\documentclass[fleqn,usenatbib]{mnras}

\usepackage{newtxtext,newtxmath}

\usepackage[T1]{fontenc}
\usepackage{cancel}
\usepackage{ulem}
\DeclareRobustCommand{\VAN}[3]{#2}
\let\VANthebibliography\thebibliography
\def\thebibliography{\DeclareRobustCommand{\VAN}[3]{##3}\VANthebibliography}

\usepackage{graphicx}	
\usepackage{amsmath}	
\usepackage[dvipsnames]{xcolor}

\title[]{Double black hole mergers in nuclear star clusters:  eccentricities, spins, masses, and the growth of massive seeds}

\author[D. Chattopadhyay et al.]{
Debatri Chattopadhyay,$^{1,2}$\thanks{E-mail: ChattopadhyayD@cardiff.ac.uk}
Jakob Stegmann,$^{1,3}$
Fabio Antonini,$^{1}$
Jordan Barber$^{1}$ and Isobel M. Romero-Shaw$^{4,5}$
\\
$^{1}$Gravity Exploration Institute, School of Physics and Astronomy, Cardiff University, Cardiff, CF24 3AA, United Kingdom\\
$^{2}$OzGrav: The ARC Center of Excellence for Gravitational Wave Discovery, Hawthorn, VIC 3122, Australia\\
$^{3}$Max Planck Institute for Astrophysics, Karl-Schwarzschild-Straße 1, 85741 Garching b. München, Germany\\
$^{4}$Department of Applied Mathematics and Theoretical Physics, Cambridge CB3 0WA, United Kingdom\\
$^{5}$Kavli Institute for Cosmology Cambridge, Madingley Road Cambridge CB3 0HA, United Kingdom
}

\date{Accepted XXX. Received YYY; in original form ZZZ}

\pubyear{2015}

\begin{document}
\label{firstpage}
\pagerange{\pageref{firstpage}--\pageref{lastpage}}
\maketitle

\begin{abstract}
We investigate the formation of 
intermediate mass black holes (IMBHs) through hierarchical mergers of stellar origin 
black holes (BHs), as well as BH mergers formed dynamically in nuclear star clusters. Using a semi-analytical approach which  
incorporates probabilistic mass-function-dependent double BH (DBH) pairing, binary-single encounters, and a mass-ratio-dependent prescription for energy dissipation in hardening binaries, we find that IMBHs with masses of
$\mathcal{O}(10^2)$~--~$\mathcal{O}(10^4)\,\rm M_\odot$ can be formed solely through hierarchical mergers in timescales of a few $100$\,Myrs to a few\,Gyrs. 
Clusters with escape velocities $\gtrsim400$\,km\,s$^{-1}$ inevitably form high-mass IMBHs.
The spin distribution of IMBHs with masses $\gtrsim 10^3$\,M$_\odot$ is strongly clustered at $\chi\sim 0.15$; while for lower masses, it peaks at $\chi\sim 0.7$.
Eccentric mergers are more frequent for equal-mass binaries containing first- and/or second-generation BHs. 
Metal-rich, young, dense clusters can produce up to 
$20\%$ of their DBH mergers with eccentricity $\geq0.1$ at $10\,\rm Hz$, and $\sim2$~--~$9\%$ of all in-cluster mergers can form at $>10$\,Hz. 
Nuclear star clusters are therefore promising environments for the formation of highly-eccentric DBH mergers, detectable with current gravitational-wave detectors. 
Clusters of extreme mass ($\sim10^8$\,M$_\odot$) and density ($\sim10^8$\,M$_\odot$pc$^{-3}$) can have about half of all of their DBH mergers with primary masses  $\geq100$\,M$_\odot$. 
The fraction of in-cluster mergers increases rapidly with increasing cluster escape velocity, being nearly unity for 
$v_{\rm esc}\gtrsim 200$\,km\,s$^{-1}$. 
Cosmological merger rate of DBHs from nuclear clusters varies $\lessapprox0.01-1$\,Gpc$^{-3}$yr$^{-1}$, where the large error bars come from uncertainties in the cluster initial conditions, number density distribution, and redshift evolution of nucleated galaxies.

\end{abstract}




\section{Introduction}
\label{sec:intro}

Since the first direct observation of gravitational waves through the merger of two black holes (BHs) \citep{LIGOScientific:2016aoc}, the LIGO-Virgo-KAGRA (LVK) gravitational-wave detector network has recorded about $85$ DBH coalescence candidates \citep{GWTC2LIGOScientific:2020ibl,GWTC3LIGOScientific:2021djp}. 
This ever-growing data-set of double BH (DBH) systems reveals that the observed BH mass spectrum ranges from $>5$~M$_\odot$ to over $\approx140$~M$_\odot$ \citep{GWTC3popLIGOScientific:2021psn}, with a smooth transition between the stellar-mass BH range (tens of M$_\odot$) to the intermediate-mass BH (IMBH) range ($\sim$10$^2$-10$^5$~M$_\odot$). Indeed, the gravitational-wave event GW190521, produced by the merger of a $\sim85$~M$_\odot$ BH with a $\sim66$~M$_\odot$ BH \citep{GW190521LIGOScientific:2020iuh}, has been cited as the first observation of the formation of an IMBH. 

The LVK observations of DBH mergers are conjectured to be originating via two main formation channels: isolated binary mergers in areas of relatively low stellar density such as fields of galaxies \citep{Belczynski2007A,Belczynski:2016obo,Stevenson:2017tfq,Giacobbo2018} and dynamically-driven DBH mergers in dense stellar systems such as star clusters \citep{Banerjee2010, RodriguezChatterjee2016, Askar2017,DiCarlo2019,ChattopadhyayStarCluster2022buz}. 
The observed BH mass, spin, and eccentricity distributions are expected to be affected by the host environment. 
For instance, due to tidal effects, isolated binaries are thought to have component spins that are aligned or nearly-aligned with the orbital angular momentum of the binary \citep{Stevenson2017MNRASS}.
Isolated binaries may also be more likely to have low spin amplitudes \citep[e.g.,][]{Bavera2020}.
On the other hand, dynamical environments are predicted to produce a population of DBHs with an isotropic spin-tilt distribution due to frequent spin-tilt-randomising interactions with other bodies \citep{RodriguezSpin:2016vmx}. 
Additionally, merger products that are retained in the dynamical environment and themselves undergo mergers should have higher dimensionless spin amplitudes $ \sim 0.7$ due to the conservation of angular momentum \citep{Doctor:2021:SpinRemnants}.
Dynamical encounters are also expected to lead to higher eccentricities in about $1$~--~$10\%$ of the mergers \citep{Wen2003,2014ApJ...781...45A,Samsing:2017oij, Gond2018,RodriguezEccen:2018pss}, unlike efficiently-circularized mergers in isolated evolution \citep{Peters:1964}. 
Finally, it is unexpected that isolated stars formed at sub-solar metallicities evolve into BHs $\gtrsim40$\,M$_\odot$ due to (pulsational) pair instability supernova ((P)PISN), which results in massive stars $\gtrsim120$~M$_\odot$ leaving no remnants (or lower-mass remnants in the case of PPISN, which leads to a peak in the BH mass distribution at $\lesssim 40$~M$_\odot$, as studied by \citealp{Belczynski2016A, Woosley:2016hmi, Spera2017}). All of the LVK observing runs have yielded detections of BHs above $>40$~M$_\odot$, within the region often referred to as the (P)PISN mass-gap. Recently, evidence has been found for spin-induced precession in merging DBHs \citep{LIGOScientific:2020kqk_ch, GWTC3LIGOScientific:2021djp,Hannam:2021pit}, as well as 
debatable evidences of
eccentric coalescences in the LVK data \citep[e.g.][]{Gayathri2020, Romero-Shaw:2022:FourEccentricMergers}, pointing towards the importance of investigating the dynamical origin of DBH mergers. 

Over the past few years, in parallel to the gravitational wave observations, there have also been radio observations by the Event Horizon Telescope (EHT) of the supermassive ($\gtrsim$10$^6$\,M$_\odot$) BH (SMBH) at the centre of M87 \citep{EventHorizonTelescopeM87:2019dse}, as well the SMBH Sagittarius A$^{\ast}$ \citep{EventHorizonTelescopeSagA:2022} at the centre of our own Milky Way. These detections reaffirm the long-standing question of the ``missing link'': IMBHs, which can connect the stellar-mass BH observed in X-ray binaries \citep{xrayTetarenko2016,xrayGenerozov2018,xrayChakraborty2020,xrayCharles2022} and gravitational-wave events
to the SMBHs at the centres of galaxies. 
Gravitational-wave detections of BHs within the (P)PISN mass gap suggest that the LVK-observed population contains BHs produced via a formation channel that subverts the mass restrictions of isolated evolution.
As such, the beginning point of an investigation into the location and formation of IMBHs can start from the study of hierarchical mergers of stellar-origin BHs in star clusters \citep{Rizzuto2021}.

The idea that a DBH merger product can subsequently keep merging with other BHs or other BH merger products 
require the remnant to be subjected to small 
(less than the cluster escape velocity)
recoil kicks, in order to be retained in the cluster \citep{Mapelli2016MNRAS_ch, Rodriguez2019PhRvD_ch}. 
Within the star cluster population, young open clusters of mass $<10^4$~M$_\odot$ and ordinary globular clusters (up to $10^6$~M$_\odot$) have low escape velocities of $\mathcal{O}$\,(1-2)km\,s$^{-1}$. 
Very massive globular clusters and nuclear star clusters with a mass range of $>10^{6}-10^{9}$~M$_\odot$ and density range of $10^{5}-10^{8}$~M$_\odot$\,pc$^{-3}$, however, have a higher likelihood of retaining subsequent generations of DBH merger remnants, leading to significant mass growth  \citep{Antonini2016ApJ,Antonini:2018auk,Fragione:2020nib}.

The modelling of such massive clusters with direct $N-$body or even Monte-Carlo simulations is extremely computationally expensive. 
Although there have been some recent trials of developing more efficient codes and utilising GPUs for this purpose \citep[e.g.][]{Wang2020MNRAS,Kamlah:2022}, most simulations require significant computational time and supercomputing facilities to produce a statistically significant dataset. 
No simulations are yet sufficiently efficient to handle $\gtrsim10^{7}$ bodies with detailed stellar evolution. 
Using a semi-analytical approach to the problem can solve most of these issues, providing a flexible, user-friendly alternative that is much faster and can assist in understanding the internal dynamics of large-$N$ systems.
Basing on the works of \cite{Henon:1972} and \cite{Breen_Heggie:2013} 
showing the macroscopic cluster properties to be insensitive to the details of their microscopic structure, we use the updated semi-analytical code $\tt{cBHBd}$ \citep{Antonini:2019ulv,Antonini:2022vib} to model massive clusters and study their DBH mergers.

In this work, we particularly focus on nuclear star clusters. These are  found at the centre of most sufficiently well resolved  low and intermediate-mass galaxies \citep{2004AJ....127..105B,2006ApJS..165...57C},  including the Milky Way \citep{2007A&A...469..125S}.
They are the densest and most massive star clusters observed in the local universe, and  are often found to host a  SMBH at their centre \citep[e.g.,][]{Georgiev:2016,Neumayer:2020}. 
Those nuclear star clusters might  be the
precursors of massive BHs in the galactic nuclei and that their might be a link between thew two types of central objects  has been suggested before \citep[e.g.,][]{2012AdAst2012E..15N, Stone2017, Atallah2023MNRAS}. Here, we consider whether a massive seed 
might be produced at the centre of a cluster through hierarchical mergers of BHs.
The key questions that we address are:
\begin{enumerate}
    \item Can we create IMBHs and SMBHs through hierarchical mergers in nuclear  and  massive globular clusters?
    \item How do the host cluster properties affect its  hierarchical mergers and hence the IMBH masses?
    \item What are the mass, spin, and eccentricity signatures of the mergers in such massive clusters, and are they detectable by present and future gravitational wave detectors?
\end{enumerate}

We discuss our methods and models in Sec.~\ref{sec:methods}. Our  results are described in Sec.~\ref{sec:results}. Sec.~\ref{sec:rates} describes our rate calculation, and, finally,  Sec.~\ref{sec:summary} sums up.

Throughout this work $G$ and $c$ refer to the gravitational constant and the speed of light, respectively.

\section{Methods}
\label{sec:methods}

In our study we use the semi-analytical fast code $\tt{cBHBd}$ developed by \cite{Antonini:2019ulv}, with updated prescriptions for BH binary sampling and three-body encounters \citep{Antonini:2022vib}. 
We have also adapted the mass sampling of binaries and triples, as is discussed further in this section. 

Within $\tt{cBHBd}$, we utilise Henon's principle \citep{Henon:1972} of steady state or balanced evolution, after the initial evolution of the star cluster. 
During this state of equilibrium, the energy per unit relaxation time created at the cluster core (by BH binaries, for a BH-rich cluster) is a constant fraction of the net energy of the cluster. 
This links the host cluster's properties to its core binary (in our case, BH) population \citep{Breen_Heggie:2013}. 

We sample the initial BH mass distribution by evolving the zero-age-main-sequence (ZAMS) stars following a \cite{Kroupa2001} initial mass function (with the maximum ZAMS mass being $150$\,M$_\odot$), using the single stellar evolution ($\tt{SSE}$) prescriptions given by \cite{HurleySSE:2000pk}\footnote{It is to be noted that we do not consider the effect of primordial binaries in the BH mass function, and the initial BH mass function is solely produced through updated $\tt{SSE}$. While binary stellar evolution \citep[i.e. $\tt{BSE}$,][]{Hurley2002MNRAS} may produce a slightly different initial BH mass spectrum, we do not consider its effect since most primordial binaries are expected to be disrupted by the core-collapse timescale. The effect of primordial binaries on these massive clusters remains a future course of study.} with metallicity-dependent wind mass loss updates of \cite{Vink:2001} and (P)PISN mass gap prescriptions from \cite{Spera2017}.
The post-stellar evolution BH mass distribution is accounted for by computing the ejection of BHs due to natal kicks \citep{Hobbs:2005}.

In $\tt{cBHBd}$, after cluster core collapse  \citep[which occurs on the order of the initial half-mass relaxation time of the cluster, scaled by NBODY models;][]{Antonini:2018auk, Antonini:2019ulv}, balanced evolution is assumed. 
It is also assumed that there is only one DBH present at any given time in the cluster, producing the required energy at the cluster-core. 
This assumption is in agreement with theoretical expectations for the massive clusters we consider \citep{1993ApJS...85..347H}.
The DBH is evolved through single BH encounters, which might result in the merger of the binary and/or its ejection and/or ejection of the single BH. 

BHs are paired following the power-law probability distribution as described in \cite{Antonini:2022vib}, with p($m_1$) $\propto m_1^{\beta_1}$ and p($q$)  $\propto q^{\beta_2}$, where $q=m_2/m_1$ (with $m_1>m_2$, such that $q\leq 1$) and $\beta_1=8 + 2\alpha$, $\beta_2=3.5 + \alpha$. Here, $\alpha$
is the power law index of the BH initial mass function. In reality, $\alpha$ should be a function of time as the BH mass function evolves through ejections and mergers. For simplicity and computational convenience, however, we fix it to its initial value\footnote{We have made a few tests where $\alpha$ was updated at each timestep and the results were similar to those with a fixed $\alpha$}.
Each binary is then encountered by a third body of mass $m_3$, drawn again from a power-law probability distribution p($m_3$) $\propto$ $m_3^{\beta_3}$, with $\beta_3=0.5 + \alpha$. 
The exponent factor $\alpha$ is obtained through fits from the initial BH mass distribution after natal kicks \citep[described in details in][]{Antonini:2022vib}. 
Since the initial BH mass spectrum is a strong function of metallicity, we extrapolate a polynomial fit for $\alpha$ in metallicity (Z) as 
\[ \alpha = c_8Z^8 + c_7Z^7 + c_6Z^6 + c_5Z^5 + c_4Z^4 + c_3Z^3 + c_2Z^2 + c_1Z + c_0 \]
where, 
c$_8= 8.5317\times10^{16}$, c$_7= 7.1772\times10^{15}$,
c$_6= 2.3818\times10^{14}$,
c$_5= 3.9582\times10^{12}$,
c$_4= 3.4364\times10^{10}$,
c$_3= 1.4564\times10^{8}$,
c$_2= 2.2885\times10^{5}$,
c$_1= 5.4322\times10^{1}$ and
c$_0= 0.1954$.
 
The semi-major axis of the binary $a\sim G\mu/\sigma^2$ is assumed to be initially in the hard-soft limit, where $\mu=m_1m_2/(m_1+m_2)$ is the reduced mass and the eccentricity $e$ for each of the $20$ resonant binary-single interaction is sampled from the thermal distribution \citep{Samsing:2017xmd}.\footnote{The average number of intermediate states for binary-single encounters is determined to be $\approx20$ by \cite{Samsing:2017xmd}, although individually, it is dependent on the target binary initial separation and initial energy state of the single, which is ignored in our case.} If $\sqrt{1-e^2} < (2G(m_1+m_2)/ac^2)^{15/4}$, there is a gravitational-wave capture merger. 

For a hard DBH, the amount of energy lost ($\Delta \mathrm{E}$) from the binary due to an encounter with a single BH is usually assumed to be $20\%$ of the initial binding energy ($\mathrm{E}$) of the binary \citep{Heggie_Hut:2003, 10.2307/j.ctvc778ff}.
This is true for equal mass systems i.e. $m_1$=$m_2$=$m_3$, averaged over all values of impact parameter. 
While this assumption is valid for most cases, inaccuracies may arise when the perturber $m_3$ is several order of magnitude smaller than the binary. 
As such, both simulation \citep{Hills_Fullerton:1980} and analytical calculation \citep{Quinlan:1996} have shown that $\Delta\mathrm{E}/\mathrm{E}\propto m_3/m_1$. 
Thus, for some of our models, we propose the functional form 
\begin{equation}
    \Delta\mathrm{E}/\mathrm{E}=0.4\frac{q_3}{(1-q_3)},
    \label{equ:delE}
\end{equation}
where $q_3=m_3/(m_1+m_2+m_3)$. 
Eq.~\eqref{equ:delE} is normalized such that $\Delta\mathrm{E}/\mathrm{E}$ reaches $0.2$, when $q_3=1/3$ so as to match the limiting condition with equal masses \citep{Heggie_Hut:2003}. 
When $m_1$=$m_2$ and $m_1>>m_3$, $\Delta\mathrm{E}/\mathrm{E}\approx0.2m_3/m_1$, which is smaller than predicted by \citet{Hills_Fullerton:1980} since that work only considered encounters with zero impact parameter. 
The semi-major axis `$a$' of the binary becomes $a\epsilon$, (where, 0.83$\lesssim\epsilon\leq1$, the boundaries obtained for extremum cases of 20\% and 0\% of binary binding energy loss) after each binary-single interaction, as the single gains (1/$\epsilon$-1) of the binding energy of the binary \citep{Antonini:2019ulv}.

Through a binary-single resonant encounter, the DBH can merge (following \citealt{Peters:1964}) and/or get ejected, or the merger remnant can be retained in the cluster, further merging with other BHs. 
The low-mass single perturber BH may also get ejected (Eqs.~3 and~4 of \citealt{Antonini:2022vib}). 
The calculation then progresses to the next BH binary. 
The computational efficiency of the code is achieved by evolving the bulk properties of the cluster (mass, half-mass density, relaxation time etc.) independently, taking the mass loss through stellar evolution and BH ejections into account (described in details in \citealt{Antonini:2019ulv} and \citealt{Antonini:2020PhRvD}).  

$\tt{cBHBd}$ does not account for binary-binary or higher-order chaotic encounters and accounts for only one DBH binary at a given time in the cluster.
The presence of many-body encounters and several concurrent DBHs has been demonstrated in detailed N-body codes like $\tt{NBODY6}$ \citep{Banerjee:2018pmh,ChattopadhyayStarCluster2022buz} and $\tt{MOCCA}$
\citep{Kamlah:2022,Hong:2020dsl}. However, these simulations refer to clusters with low masses, typically $\lesssim 10^5M_\odot$. For the very massive clusters we consider here, higher-order interactions are expected to be strongly suppressed (Pina and Gieles in-prep.) .

\subsection{Models}
\label{sec:models}

\begin{table*}\centering
\begin{tabular}{llcccccccccc}
\hline
Sl. & Model & Mass & Density & Metallicity & SN  & BH seed & Mass Loss & Natal kick & BH spin & BH pairing & Delta E \\
& &$M_\mathrm{cl,i}$ & $\rho_\mathrm{h,i}$ & Z &prescription& & & $\sigma_\mathrm{Maxw}$ & $\chi_{1,2}$& & $\Delta\mathrm{E}/\mathrm{E}$ \\
& &(M$_\odot$) & (M$_\odot$\,pc$^{-3}$) &  && & & (km\,s$^{-1}$) & & &  \\
\hline
1. & Fiducial & 2$\times$10$^7$ & 10$^7$ & 1.5$\times10^{-4}$ &rapid &0& stellar & $265$ & 0;0 & alpha & $f(\mathrm{q_3})$  \\
2. & M8D8 & 10$^8$ & 10$^8$ & 1.5$\times10^{-4}$ &rapid &0& stellar & $265$ & 0;0 & alpha & $f(\mathrm{q_3})$  \\
3. & M8D7 & 10$^8$ & 10$^7$ & 1.5$\times10^{-4}$ &rapid &0& stellar & $265$ & 0;0 & alpha & $f(\mathrm{q_3})$  \\
4. & M7D8 & 10$^7$ & 10$^8$ & 1.5$\times10^{-4}$ &rapid &0& stellar & 265 & 0;0 & alpha & $f(\mathrm{q_3})$  \\
5. & M7D7 & 10$^7$ & 10$^7$ & 1.5$\times10^{-4}$ &rapid &0& stellar & 265 & 0;0 & alpha & $f(\mathrm{q_3})$  \\
6. & M7D6 & 10$^7$ & 10$^6$ & 1.5$\times10^{-4}$ &rapid &0& stellar & 265 & 0;0 & alpha & $f(\mathrm{q_3})$  \\
7. & M7D5 & 10$^7$ & 10$^5$ & 1.5$\times10^{-4}$ &rapid &0& stellar & $265$ & 0;0 & alpha & $f(\mathrm{q_3})$  \\
8. & M6D8 & 10$^6$ & 10$^8$ & 1.5$\times10^{-4}$ &rapid &0& stellar & $265$ & 0;0 & alpha & $f(\mathrm{q_3})$  \\
9. & M6D7 & 10$^6$ & 10$^7$ & 1.5$\times10^{-4}$ &rapid &0& stellar & $265$ & 0;0 & alpha & $f(\mathrm{q_3})$  \\
10. & M6D6 & 10$^6$ & 10$^6$ & 1.5$\times10^{-4}$ &rapid &0& stellar & $265$ & 0;0 & alpha & $f(\mathrm{q_3})$  \\
11. & M6D5 & 10$^6$ & 10$^5$ & 1.5$\times10^{-4}$ &rapid &0& stellar & $265$ & 0;0 & alpha & $f(\mathrm{q_3})$  \\
12. & Z\_10 & 2$\times$10$^7$ & 10$^7$ & 1.5$\times10^{-3}$ &rapid &0& stellar & $265$ & 0;0 & alpha & $f(\mathrm{q_3})$  \\
13. & Z\_100 & 2$\times$10$^7$ & 10$^7$ & 1.5$\times10^{-2}$ &rapid &0& stellar & $265$ & 0;0 & alpha & $f(\mathrm{q_3})$  \\
14. & SN\_D & 2$\times$10$^7$ & 10$^7$ & 1.5$\times10^{-4}$ &delay &0& stellar & $265$ & 0;0 & alpha & $f(\mathrm{q_3})$  \\
15. & Sd\_50 & 2$\times$10$^7$ & 10$^7$ & 1.5$\times10^{-4}$ &rapid &50& stellar & $265$ & 0;0 & alpha & $f(\mathrm{q_3})$  \\
16. & Sd\_100 & 2$\times$10$^7$ & 10$^7$ & 1.5$\times10^{-4}$ &rapid &100& stellar & $265$ & 0;0 & alpha & $f(\mathrm{q_3})$ \\
17. & Sd\_150 & 2$\times$10$^7$ & 10$^7$ & 1.5$\times10^{-4}$ &rapid &150& stellar & $265$ & 0;0 & alpha & $f(\mathrm{q_3})$  \\
18. & Sd\_200 & 2$\times$10$^7$ & 10$^7$ & 1.5$\times10^{-4}$ &rapid &200& stellar & $265$ & 0;0 & alpha & $f(\mathrm{q_3})$  \\
19. & Ml\_ev & 2$\times$10$^7$ & 10$^7$ & 1.5$\times10^{-4}$ &rapid &0& stellar, evaporation & $265$ & 0;0 & alpha & $f(\mathrm{q_3})$  \\
20. & Ml\_0 & 2$\times$10$^7$ & 10$^7$ & 1.5$\times10^{-4}$ &rapid &0& 0 & $265$ & 0;0 & alpha & $f(\mathrm{q_3})$  \\
21. & Ml\_0$_\mathrm{M7D5}$ & 10$^7$ & 10$^5$ & 1.5$\times10^{-4}$ &rapid &0& 0 & $265$ & 0;0 & alpha & $f(\mathrm{q_3})$  \\
22. & Vk\_0 & 2$\times$10$^7$ & 10$^7$ & 1.5$\times10^{-4}$ &rapid &0& stellar & 0 & 0;0 & alpha & $f(\mathrm{q_3})$  \\
23. & Vk\_0$_\mathrm{M7D7}$ & 10$^7$ & 10$^7$ & 1.5$\times10^{-4}$ &rapid &0& stellar & 0 & 0;0 & alpha & $f(\mathrm{q_3})$  \\
24. &  Vk\_0$_\mathrm{M7D5}$ & 10$^7$ & 10$^5$ & 1.5$\times10^{-4}$ &rapid &0& stellar & 0 & 0;0 & alpha & $f(\mathrm{q_3})$  \\
25. & Vk\_0$_\mathrm{Z_{100}}$ & 2$\times$10$^7$ & 10$^7$ & 1.5$\times10^{-2}$ &rapid &0& stellar & 0 & 0;0 & alpha & $f(\mathrm{q_3})$  \\
26. & Sp\_01 & 2$\times$10$^7$ & 10$^7$ & 1.5$\times10^{-4}$ &rapid &0& stellar & $265$ & 0;1 & alpha & $f(\mathrm{q_3})$  \\
27. &  Sp\_33 & 2$\times$10$^7$ & 10$^7$ & 1.5$\times10^{-4}$ &rapid &0& stellar & $265$ & 0.3;0.3 & alpha & $f(\mathrm{q_3})$  \\
28. & Sp\_11 & 2$\times$10$^7$ & 10$^7$ & 1.5$\times10^{-4}$ &rapid &0& stellar & $265$ & 1;1 & alpha & $f(\mathrm{q_3})$  \\
29. & Sp\_LVK & 2$\times$10$^7$ & 10$^7$ & 1.5$\times10^{-4}$ &rapid &0& stellar & $265$ & LVK & alpha & $f(\mathrm{q_3})$ \\
30. & Ord\_BH & 2$\times$10$^7$ & 10$^7$ & 1.5$\times10^{-4}$ &rapid &0& stellar & $265$ & 0;0 & ordered & 0.2 \\
31. & DE & 2$\times$10$^7$ & 10$^7$ & 1.5$\times10^{-4}$ &rapid &0& stellar & $265$ & 0;0 & alpha & 0.2 \\
32. & DE$_\mathrm{M7D7}$ & 10$^7$ & 10$^7$ & 1.5$\times10^{-4}$ &rapid &0& stellar & $265$ & 0;0 & alpha & 0.2 \\
33. & DE$_\mathrm{M7D5}$ & 10$^7$ & 10$^5$ & 1.5$\times10^{-4}$ &rapid &0& stellar & $265$ & 0;0 & alpha & 0.2 \\
34. & DE$_\mathrm{Z_{100}}$ & 2$\times$10$^7$& 10$^7$ & 1.5$\times10^{-2}$ &rapid &0& stellar & $265$ & 0;0 & alpha & 0.2 \\
\hline
\end{tabular}
\caption{Initialization of the $34$ models used in this study as detailed in section~\ref{sec:models}.}
\label{tab:model}
\end{table*}

The set of $34$ models (each run $100$ times with different random seeds to account for statistical fluctuations) used for this project is tabulated in Table~\ref{tab:model} and is described as follows.

\subsubsection{Fiducial}
The base model Fiducial has an initial mass and half mass density of 2$\times$10$^7$\,M$_\odot$ and 10$^7$\,M$_\odot$\,pc$^{-3}$ respectively.
The only form of mass loss in this model (apart from BH ejections due to merger or binary-single recoils) is assumed to be due to stellar evolution \citep{HurleySSE:2000pk, Antonini:2019ulv} i.e. mass loss due to stellar winds and supernovae \citep{Lucy1970}, and natal kicks \citep{Lipunov1997}. These parameters of the Fiducial model are chosen, such that after a Hubble Time, the cluster mass and density roughly aligns with that of the Milky-way nuclear cluster \citep{Schodel2009, Schodel2020}. 
The Fiducial model metallicity is Z$=1.5\times10^{-4}$, the ``rapid'' supernova prescription is used for core-mass to BH-mass mapping \citep{Fryer:2012} and the BH natal kick is drawn from a Maxwellian distribution with $\sigma_\mathrm{Maxw}=265$\,km\,s$^{-1}$\citep{Hobbs:2005} scaled by fallback mass \citep{Janka:2013hfa}. 
The initial BH spin is assumed to be zero. 
The BH binaries and binary-singles are paired according to the metallicity-dependant $\alpha$ prescription described in Sec.~\ref{sec:methods}. 
The amount of energy lost per binary-single encounter is assumed to be a function of the masses of the third-body perturber and the binary total mass, such that $\Delta\mathrm{E}/\mathrm{E}=f(\mathrm{q_3})$, as given by equation~(\ref{equ:delE}).

\subsubsection{Other model variations}
All other models have one or two specifications changed from the Fiducial model. 
Model variations are shown in Table~\ref{tab:model}.
Different cluster initial masses and densities are explored with model serial numbers 2$-$11.
The naming of each model reflects these changes.
For example, model M8D8 has an initial mass and density of $10^8$\,M$_\odot$ and $10^8$\,M$_\odot$\,pc$^{-3}$ respectively, while M6D5 has an initial mass of $10^6$\,M$_\odot$ and an initial density of $10^5$\,M$_\odot$\,pc$^{-3}$.  
Models Z\_10 and Z\_100 have metallicities 10$\times$ and 100$\times$ that of Fiducial. 
Model SN\_D uses the ``delayed'' supernova prescription , instead of Fiducial model's ``rapid" prescription \citep{Fryer:2012}.\footnote{Z\_100 has similar order-of-magnitude metallicity as the Sun \citep{Asplund:2009}.} 
In the ``Sd'' models with serial numbers 15$-$18, we included in the cluster model a BH that is not produced through stellar evolution and with a mass which is traditionally considered above the mass limit imposed by pulsational pair instabilities. We do note that the lower edge of the (P)PISN mass gap is rather uncertain and can be pushed to further higher masses if stellar rotation is included, nevertheless, $\gtrsim100$\,M$_\odot$ is usually assumed not to be produced directly through stellar evolution from the \cite{Kroupa2001} mass function \citep{Marchant2020}. It is still possible that evolving massive stars in binaries or triples undergo mergers (pre-compact object formation), and then promptly collapses to very massive BHs \citep[e.g.][]{Stegmann2022PhRvD,ArcaSedda2023}.
These seeds can also be considered as primordial BHs; however, we note that there is tremendous uncertainty in the existence of primordial BHs, their expected mass range, and the actual process of their seeding star clusters or early galaxies \citep{Dolgov2017,Yuan2023}.  
Model Ml\_ev has added cluster mass loss due to cluster evaporation in addition to the standard mass loss via stellar evolution and BH ejections (see Fig.~1 and Section~II.B in \citet{Antonini:2020PhRvD}.
Mass loss due to stellar evolution is neglected in models Ml\_0 and Ml\_0$_\mathrm{M7D5}$. 

We explore an assumption that all BHs are born with zero natal kicks in the ``Vk\_0'' models (serial numbers 22$-$25). 
We deviate from the BH binary and triple mass selection assumption in model ``Ord\_BH'', where the two most massive BHs are selected to be in a binary, followed by the third most massive one making the triple perturber. 
The assumption of non-spinning initial BHs is varied in the ``Sp'' models (serial numbers 26$-$29), where the initial BH spin has different values. In one model we sample the initial BH spins from the 
 distribution inferred from the GW data, i.e. Fig.15 of \citet{GWTC3popLIGOScientific:2021psn}.    
The model group ``DE'' (serial numbers 31$-$34) changes the assumption of the mass-ratio dependent functional form of $\Delta\mathrm{E}/\mathrm{E}$, and replaces it by a constant value of $\Delta\mathrm{E}/\mathrm{E}=0.2$ \citep{Heggie_Hut:2003,10.2307/j.ctvc778ff}. 
The maximum integration time of the models is taken to be $13.5$\,Gyr, approximately a Hubble Time \citep[e.g.,][]{Valcin2021}.

\section{Results}
\label{sec:results} 
Specifically keeping the three primary questions of Sec.~\ref{sec:intro} in mind, we discuss the results obtained from our models some of which are summarised in Table~\ref{tab:IMBHmass}.
We divide this Section into three parts, depending on the location and mass of the DBH mergers that we discuss: (i) In-situ  (in-cluster) mergers, (ii) Mass and spin evolution of IMBHs, and (iii) Ex-situ (ejected) mergers.  

\subsection{In-situ mergers}
\label{sec:incluster}

In this Section, we discuss DBH mergers that occur inside the cluster. 
High cluster escape velocity plays a key role in retaining these DBHs, protecting them from ejection due to natal or recoil kicks. In fact, we shall see that almost all the DBH mergers formed in our models are in-cluster mergers. 
We split the discussion of the properties of the in-situ mergers in our cluster models by their masses, spins and eccentricities. 

\subsubsection{Mass}
\label{sec:mass}

\begin{figure}
\centering
    \hspace{-1cm}
	\includegraphics[width=1.1\columnwidth]{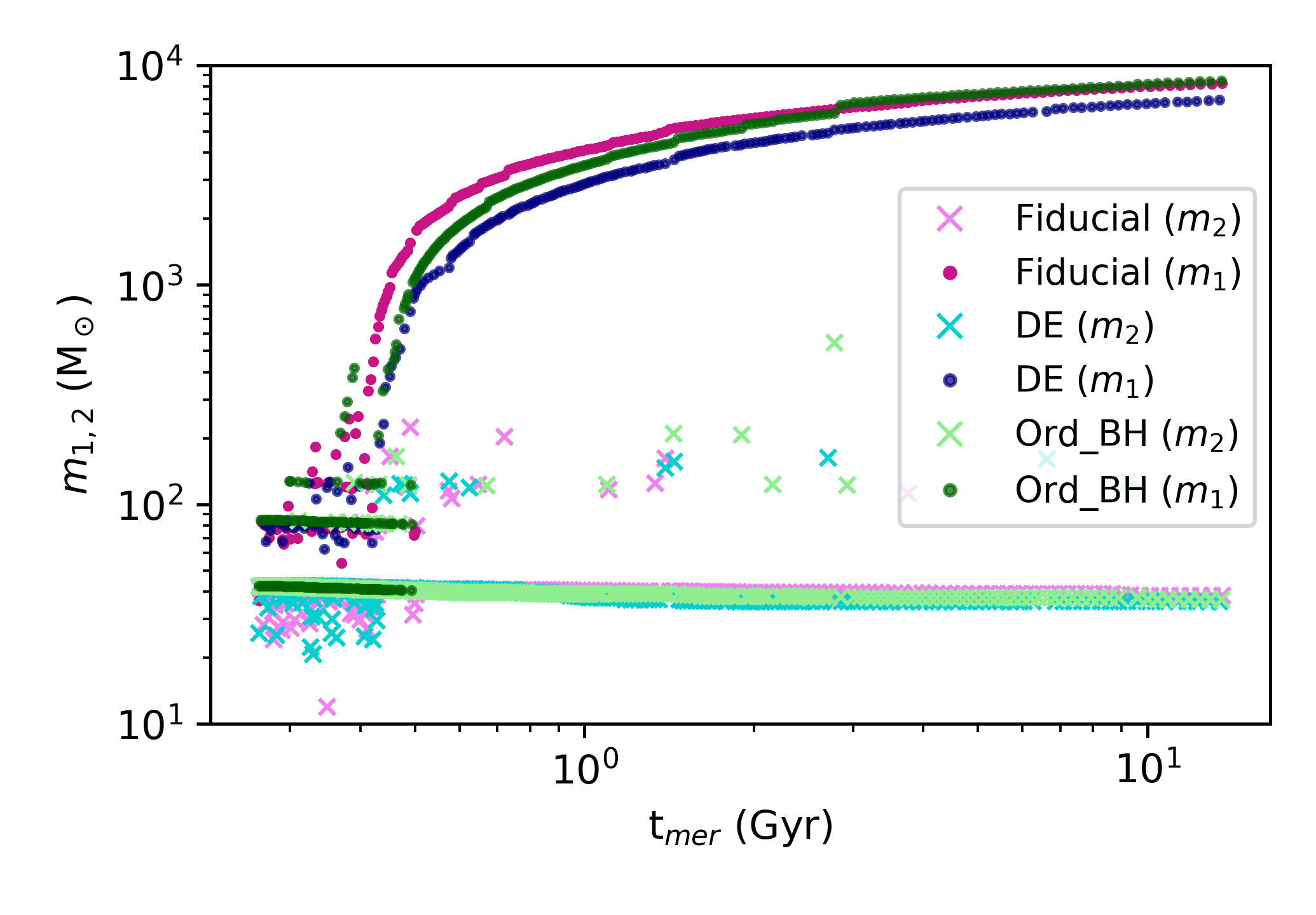}
   \caption{The BH masses of the in-cluster DBH mergers in models Fiducial (in magenta and light pink), DE (in navy and light blue) and Ord\_BH (in dark green and light green) across cluster evolution time ($t_\mathrm{mer}$). The dots of darker colours indicate the primary masses ($m_1$), while the secondary BHs ($m_2$) are depicted in the corresponding lighter shades of crosses. }
    \label{fig:M12}
\end{figure}

\begin{figure}
\centering
   \includegraphics[width=0.9\columnwidth]{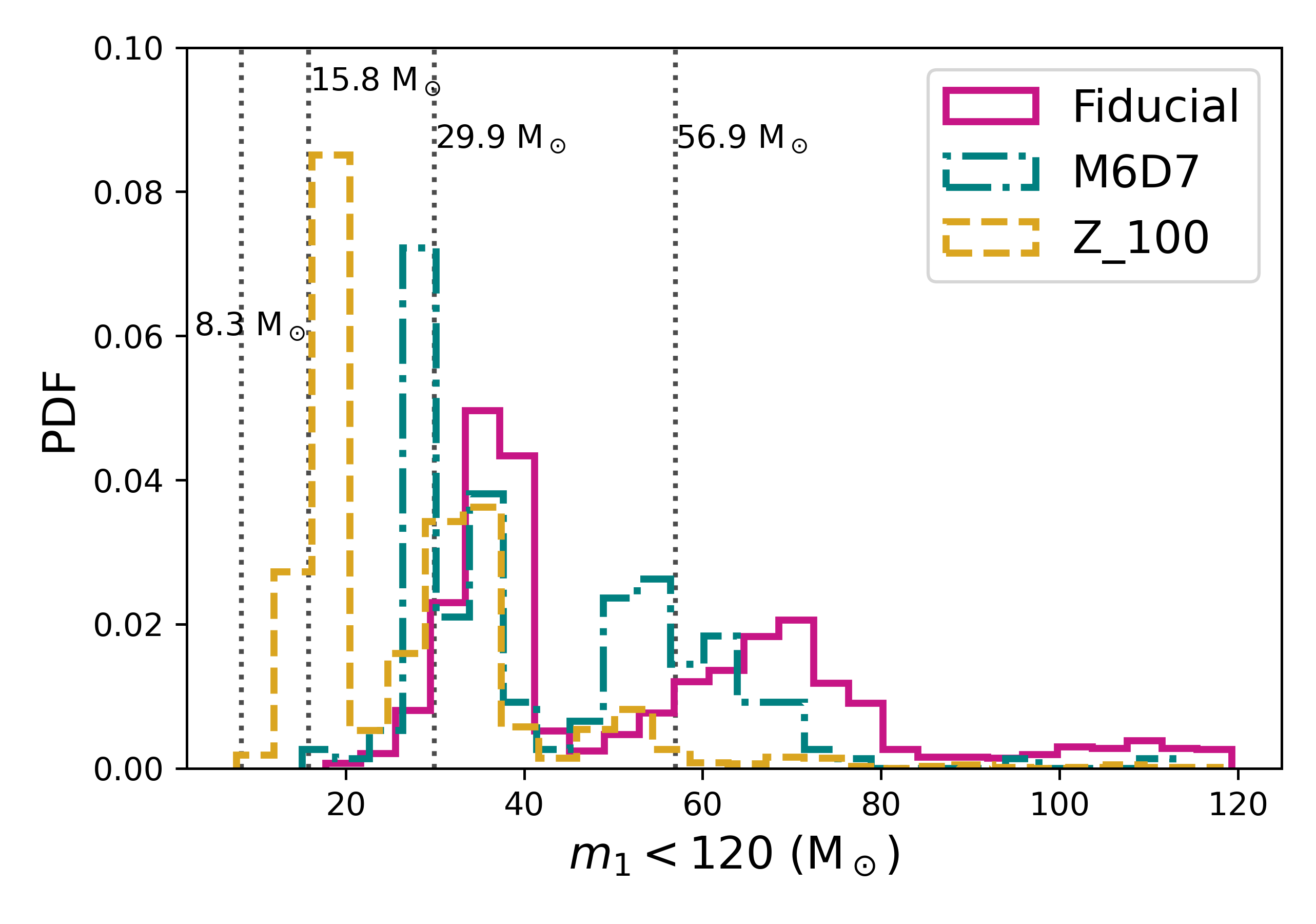}
   \caption{The primary BH mass spectrum ($<120$\,M$_\odot$) of the DBH mergers, showing structures for different merger generations shown through probability density function (PDF). The cluster escape velocity (M6D7 having a lower escape velocity than Fiducial) and initial BH mass distribution (via cluster metallicity----(Z\_100 has a metallicity \( \mathcal{O}(2) \) greater than Fiducial) determine the location of the apparent peaks in the distribution. The grey dashed lines are plotted in accordance with \protect\citet[Fig. 5.,][]{Tiwari:2021}.} 
    \label{fig:masspectrum_vaibhav}
\end{figure}

The evolution of the DBHs that merge in our cluster model Fiducial, DE and Ord\_BH is represented in Fig.~\ref{fig:M12}. 
For models Fiducial and DE, in the first few hundred million years, there are a couple of BH primaries of $\sim100$~--~$300$\,M$_\odot$ that compete until one emerges as the most massive IMBH and continues to grow in mass, following a near-logarithmic  growth curve at later times. 
This near-logarithmic growth is due to the 
secular expansion of the cluster, which causes the energy generation rate by the binary to decrease and a significant drop in the DBH merger rate.

Since there is only one binary in the cluster at any one time and the most massive BH tends to pair with the second-most massive BH, as long as the merger remnant is retained in the cluster the formation of higher-generation BHs is suppressed. Hence, in the model Ord\_BH, there can only be one $>100$\,M$_\odot$ BH at a time. 
A new $>100$\,M$_\odot$ BH, which eventually becomes the most massive IMBH of the model, is only produced after a $\approx 400$\,M$_\odot$ BH is ejected at $\approx 350$\,Myrs.

It is possible to identify the primaries in the first, second, third, and fourth DBH merger generations from the strata seen in  Fig.~\ref{fig:M12}.
At the fourth generation, the most massive IMBH starts to completely dominate in-situ mergers.
Recent studies have found multiple structures in the inferred BH mass spectrum \citep{Tiwari:2021}. There appears to be multiple peaks at $\approx$10,20,35,64\,M$_\odot$ of the primary BH mass \citep{Tiwari2023}, and
while their cause remains unknown owing to the difficulty in disentangling systematics, selection effects, unknown branching fraction between different types of environments of BH mergers and uncertainties in massive binary evolution as well as cosmological distribution and evolution of initial parameters of star clusters, different BH generations of in-cluster mergers in massive clusters could produce such features. 
We compare Fig.~5 of \cite{Tiwari:2021} to our Fig.~\ref{fig:masspectrum_vaibhav}, illustrating that a combination of clusters with different escape velocities and metallicities (as well as isolated BH merger pathways, especially for the lower-mass end of the spectrum) can potentially help to explain some of the features in the mass distribution. 

The merging DBH mass spectrum from massive clusters in which IMBHs form is remarkably different to that from young, open clusters. 
In open clusters, lower escape velocities ensure the ejection of the massive remnants formed from a DBH merger, while lower-mass first-generation BHs can still participate in dynamical mergers inside the cluster.
This means that massive clusters such as nuclear and (the most massive) globular clusters are a much more probable formation environment for hierarchical mergers, and therefore  for IMBH growth, than open clusters.

As Fig.~\ref{fig:masspectrum_vaibhav} demonstrates, the low-mass cluster M6D7 has more low-mass BH primaries participating in DBH mergers than in the Fiducial model. 
For low escape velocity clusters (especially in metal-rich environments), less massive DBH mergers take precedence; thus, small and high-metallicity clusters are more probable environments for low-mass ($\lesssim15$\,M$_\odot$) DBH mergers \citep{GWTC3popLIGOScientific:2021psn}. Due to their high mass and density, clusters with a high escape velocity such as Fiducial form a massive IMBH at ease that hegemonizes the BH merger demographics, suppressing lower-mass mergers.  

\begin{figure}
\centering
\includegraphics[width=0.9\columnwidth]{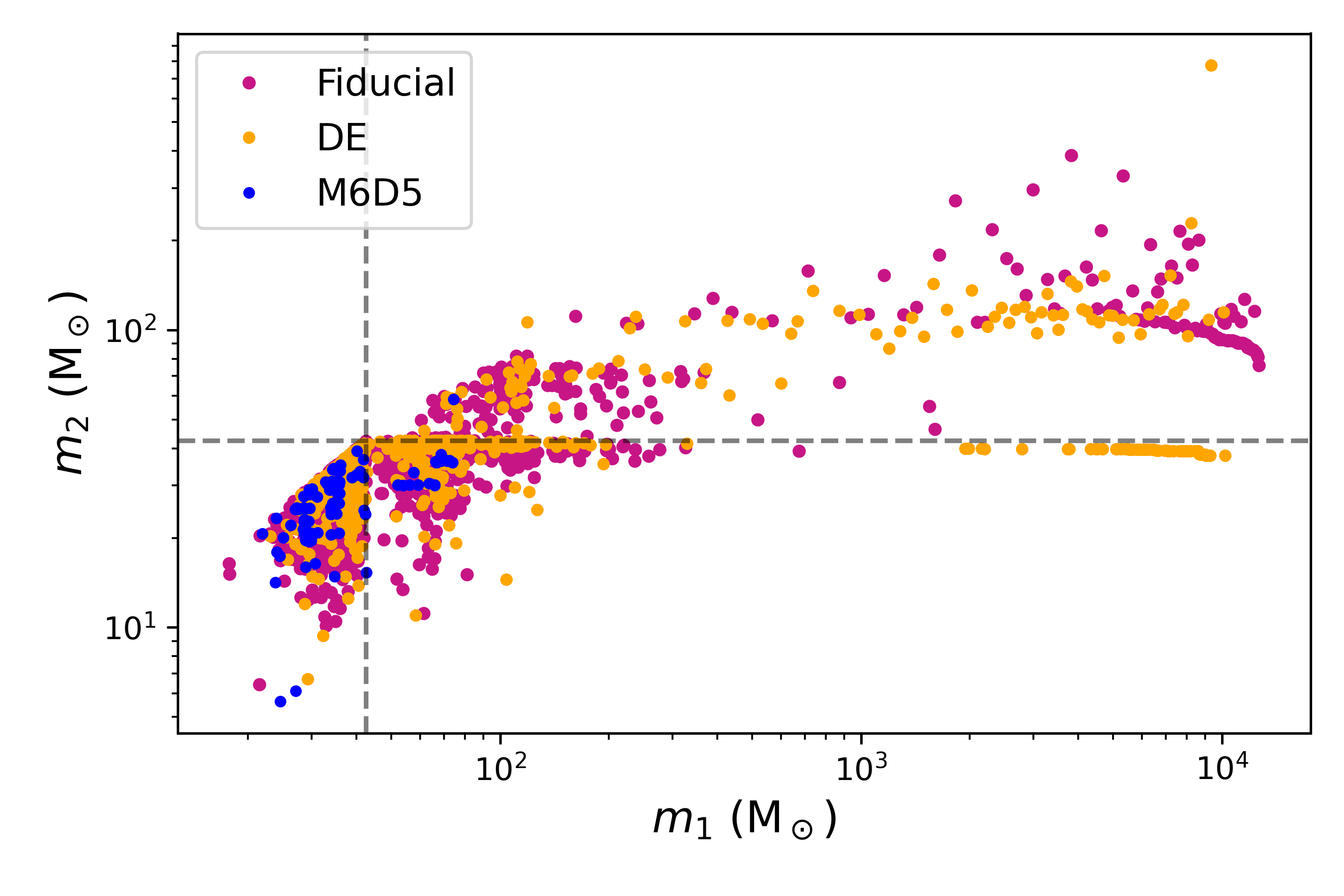}
\caption{The correlation of primary ($m_1$) and secondary masses ($m_2$) of in-cluster DBH mergers. The dashed lines show the BH maximum mass obtained from our stellar evolution model SSE, \citep{HurleySSE:2000pk}. 
}
\label{fig:secondary}
\end{figure}

We emphasise that both BH primaries and secondaries can be merger remnants of previous generations that were retained in the cluster.
In Fig.~\ref{fig:secondary}, we show the primary and secondary masses ($m_{1,2}$) of all mergers in the Fiducial, DE and M6D5 models. 
The presence of high-generation mergers in both $m_{1}$ and $m_{2}$ is clear from the build-up of BH masses above the maximum BH mass set within SSE \citep[$\sim 40$\,M$_\odot$;][]{HurleySSE:2000pk}; this build-up is more prevalent in the Fiducial and DE models than in the M6D5 model. 
The dominance of the IMBH in the in-cluster mergers is also illustrated by the plateau in secondary masses as the primary masses continue to rise. 

\subsubsection{Spin}
\label{sec:spin}
\begin{figure}
\centering
\includegraphics[width=0.9\columnwidth]{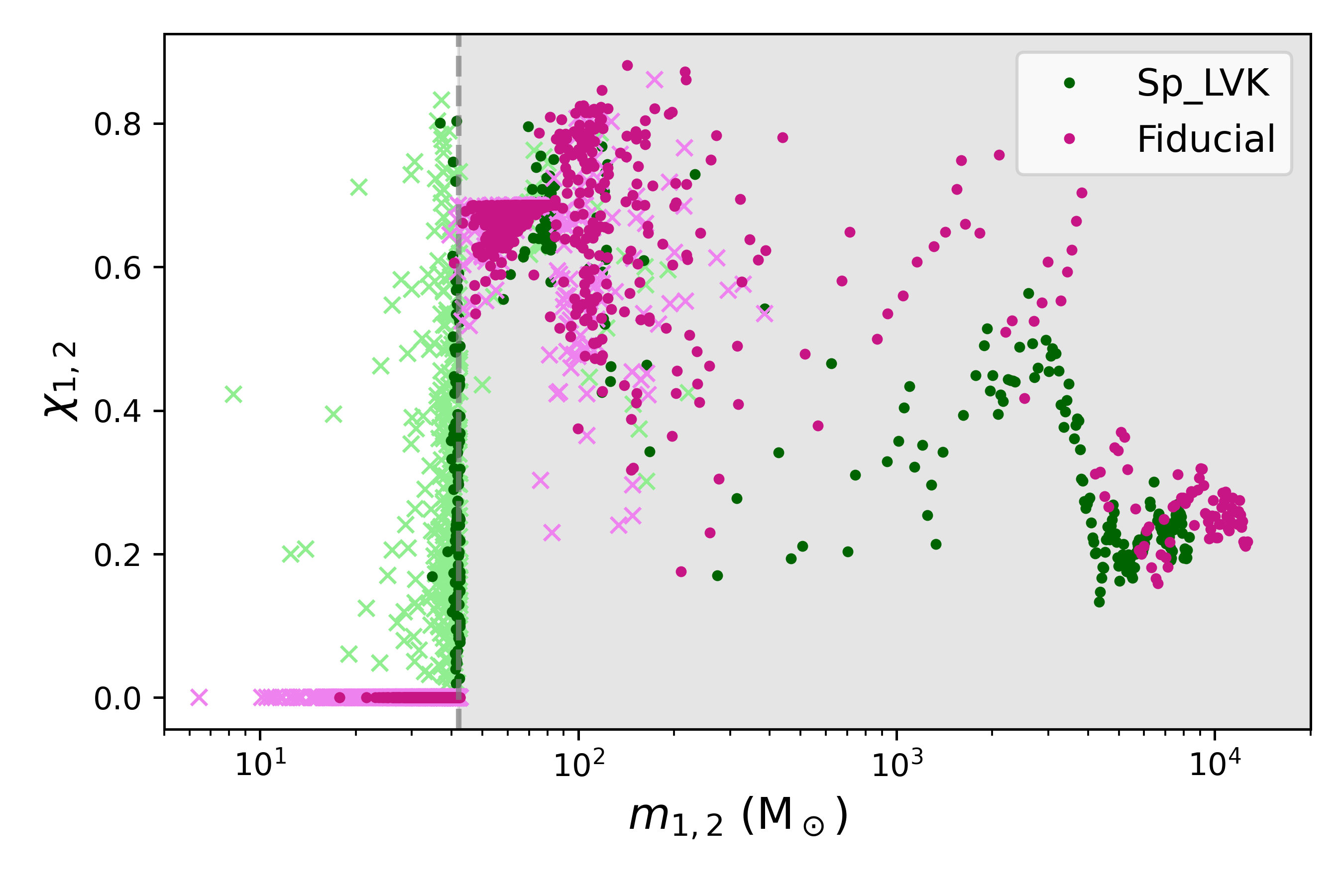}
\caption{Component spins $\chi_\mathrm{1,2}$ with respect to component masses $m_{1,2}$ of the DBH mergers in models Fiducial (magenta and light pink) and Sp\_LVK (dark green and light green). The  dark-coloured dots and light-coloured crosses identify the primary and secondary of the binaries, respectively. The left-hand region with the white background signifies the initial distribution (hence first-generation BH) and the right-hand shaded region encompasses the mass range that can only be accessed via hierarchical mergers. 
}
\label{fig:massspin}
\end{figure}
\begin{figure}
\centering
\includegraphics[width=0.9\columnwidth]{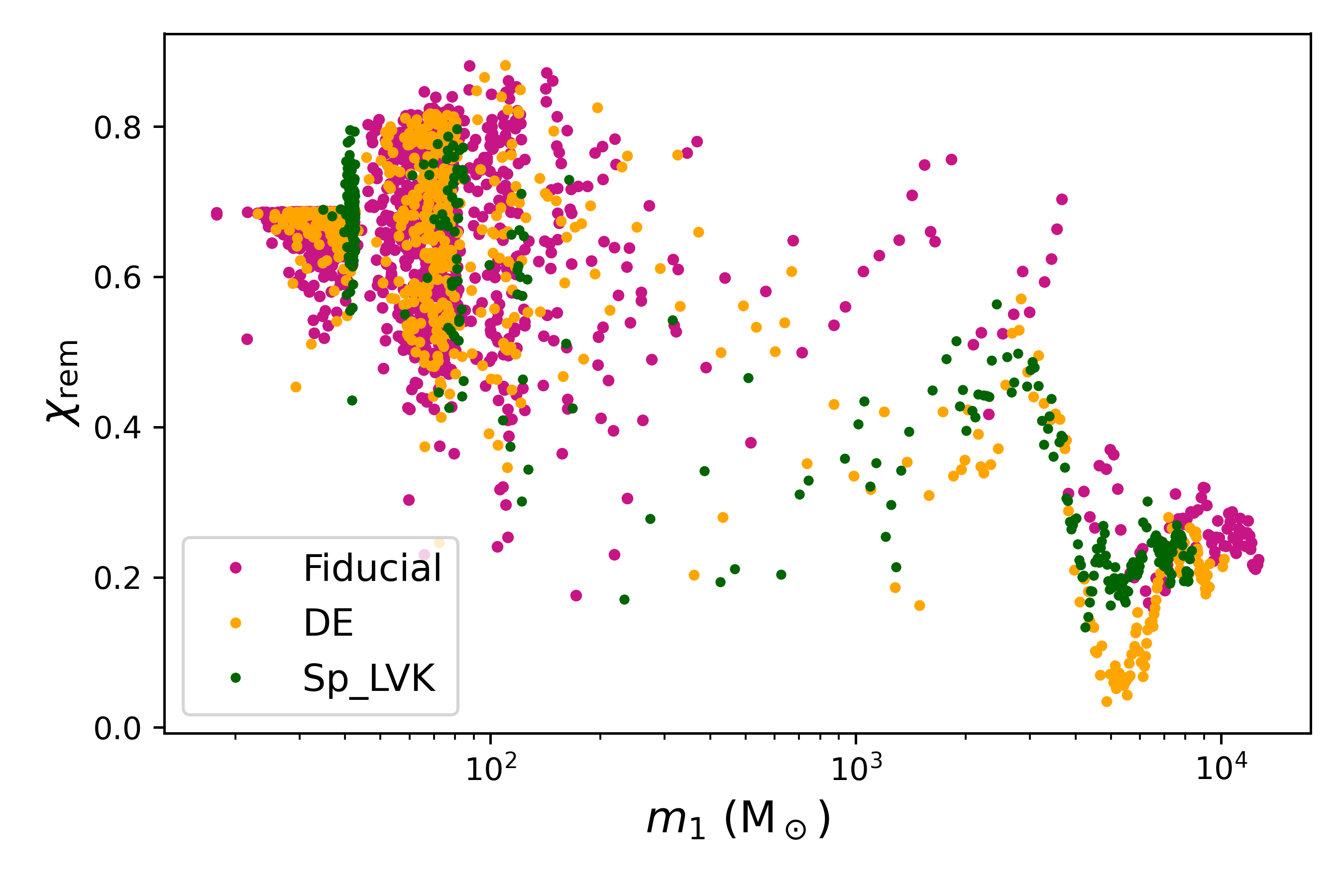}
\includegraphics[width=0.9\columnwidth]{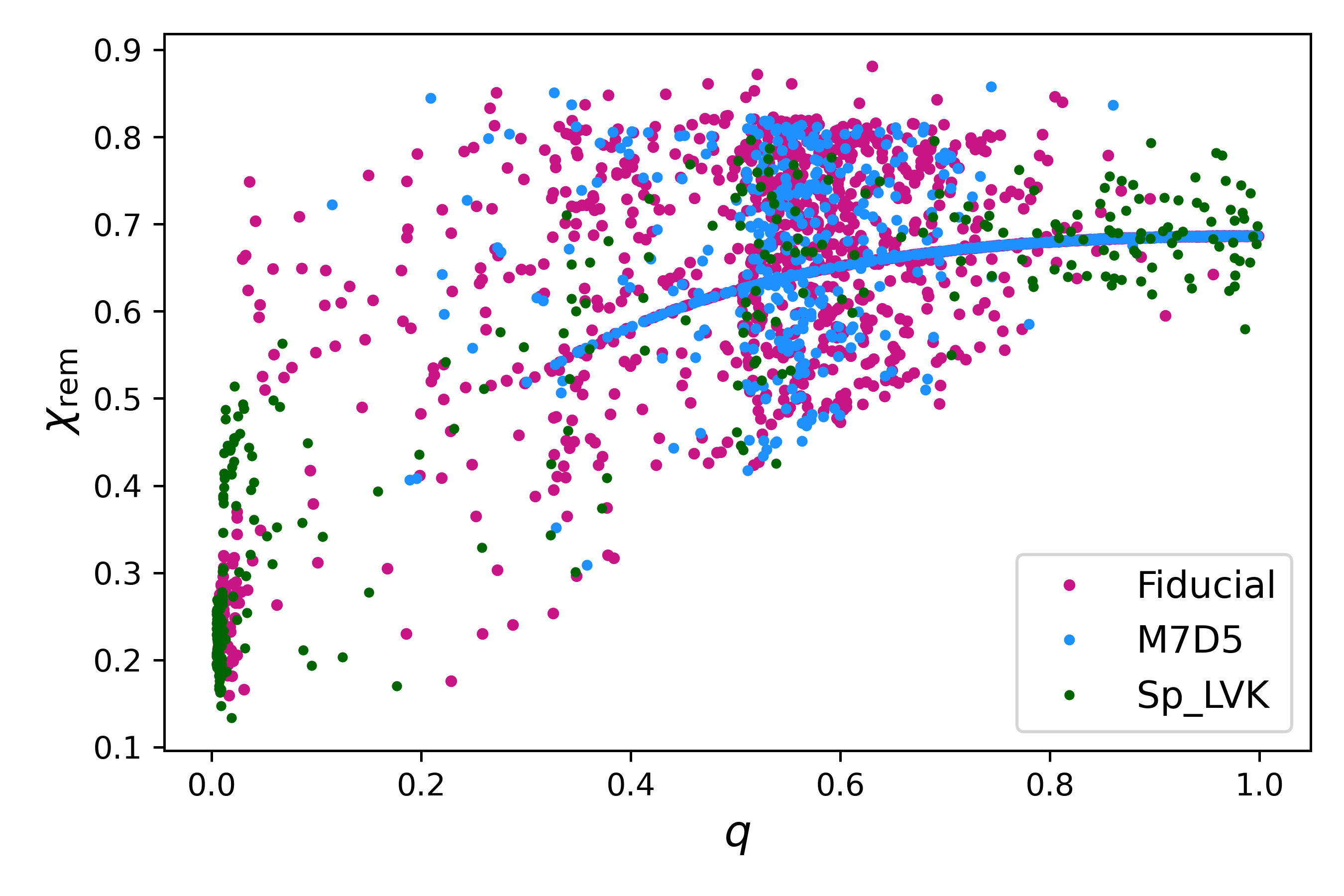}
\caption{The distribution of remnant spin $\chi_\mathrm{rem}$ with respect to the primary mass $m_1$ (upper panel) and mass ratio $q$ (lower panel). The continuous line in the middle of the lower plot for model M7D5 is the hierarchical merger of the same IMBH. For models Fiducial and Sp\_LVK, IMBH formed are one-to-two orders-of-magnitude more massive (than M7D5), making their mergers with stellar-mass BH have very small values of $q$. 
}
\label{fig:spinscor}
\end{figure}
The initial spin distribution  affects the post-merger remnant spins ($\chi_\mathrm{rem}$) to a small extent. 
While $\chi_\mathrm{rem}$ depends on the pre-merger component spins $\chi_\mathrm{1,2}$, it is also a function of the mass ratio $q=m_2/m_1$ 
of the merger. If the symmetric mass ratio is $\eta=q/(1+q)^2$, the two component BH vector spins are $\vec{\chi}_{1,2}$ and the angular momentum vector is $\vec{j}$, then the vector form of remnant spin (details in \citealt{Lousto:2010} and \citealt{Lousto:2009}) is given by 
\begin{equation}
    \vec{\chi}_\mathrm{rem}=min(1., \vec{\chi_\mathrm{t}}+\frac{q}{(1+q)^2}l\vec{j})
    \label{equ:vrec}
\end{equation}
where, 
\begin{equation}
    l=2\sqrt{3}+t_{2}\eta+t_{3}\eta^2+s_{4}\frac{(1+q)^4}{(1+q^2)^2}|\vec{\chi_\mathrm{t}}^2|
    +\frac{(s_{5}\eta+t_{0}+2)(1+q)^2}{(1+q^2)}\chi_\mathrm{p},
\end{equation}
and
\begin{equation}
    \vec{\chi}_\mathrm{t}=\frac{(q^2\vec{\chi_2}+\vec{\chi_1})}{(1+q)^2}.
\end{equation}
The parallel component of $\vec{\chi}_\mathrm{t}$ is $\chi_\mathrm{p}=\vec{\chi}_\mathrm{t}\cdot \vec{j}$ and the numerical constants are $t_0=-2.8904$, $t_2=-3.51712$, $t_3=2.5763$, $s_4=-0.1229$ and $s_5=0.4537$.

Because of the conservation of angular momentum during a merger, all models produce merger remnants with spins clustered around $\chi_\mathrm{rem}\sim0.7$.
For example, we can compare the Fiducial and Sp\_LVK models, where the former has initially non-spinning BHs and the latter has an initial BH spin distribution consistent with that observed in gravitational-wave events. 
After just the first generation of stellar-mass mergers ($\lesssim100$\,M$_\odot$), the remnant spin is already clustered around $0.7$.
Consequently, higher-generational mergers end up with similar $\chi_\mathrm{1,2}$ in both models (Fig.~\ref{fig:massspin}).
The Fiducial model only shows a few lower $\chi_\mathrm{rem}$ measurements compared to the Sp\_LVK model, even though the initial distribution of $\chi_\mathrm{1,2}$ is very different between the two models. 

Fig.~\ref{fig:spinscor} shows $\chi_{\rm rem}$ as a function of $m_1$ and $q$.
As $m_1$ increases and $q$ decreases, the spin of the IMBH goes down. This feature is present in all our models; e.g., the Sp\_LVK with its first generational rotating BHs shows hardly any difference in the final IMBH spin compared to Fiducial.
The reason why the IMBH spin goes down is because it grows by merging with smaller  BHs coming from random directions. After many mergers, the angular momentum $\vec{j}$ of Equation~\ref{equ:vrec}, averages out, leading to a net decrease in $\chi_\mathrm{rem}$.
 
Hence, $\chi_\mathrm{rem}$ reduces ($\simeq 0.15$) with increasing primary mass, as illustrated in the upper panel of Fig.~\ref{fig:spinscor}. 
The effect of $q$ on $\chi_\mathrm{rem}$ is further illustrated in the lower panel of Fig.~\ref{fig:spinscor}, showing more symmetric masses indeed produce more rapidly-spinning remnants. 
In model M7D5, where the IMBHs are of low masses $\lesssim200$\,M$_\odot$ and never reach $>500$\,M$_\odot$, the $\chi_\mathrm{rem}$ distribution extends to higher values, as the mass ratio remains confined to relatively high values, within the range $0.1$~--~$1$.
In contrast, the massive $\sim10^4$\,M$_\odot$ IMBH in the Fiducial model produces mass ratios as low as $0.001$, resulting in lower $\chi_\mathrm{rem} = 0.13$. 
The result is a double-peaked distribution of $\chi_{\rm rem}$. IMBH remnants with masses $\lesssim 10^3 M_\odot$ have spins clustered near $\chi_{\rm rem}= 0.7$, while the largest IMBHs formed in our models have  $\chi_{\rm rem}\simeq 0.15$ (see Table~\ref{tab:IMBHmass}).

\subsubsection{Eccentricity}
\label{sec:eccentricity}

While most DBHs formed in clusters are circularized by the time they merge, a small fraction of them will  still have a significant eccentricity ($e>0.1$) when they reach the frequency band of current detectors  \citep[e.g.,][]{2014ApJ...781...45A,Samsing:2017xmd,RodriguezEccen:2018pss,Rodriguez18a}. 
It has been argued that eccentric mergers are the most robust signature of DBH formation via the dynamical channel, and that a sub-population of eccentric binaries could help to resolve the branching fraction between isolated and dynamical DBH formation channels \citep[e.g.,][]{Lower18, Romero-Shaw2021,Zevin:2021,Romero-Shaw:2022:FourEccentricMergers,DallAmico2023}. 

\begin{figure}
\centering
	\includegraphics[width=0.9\columnwidth]{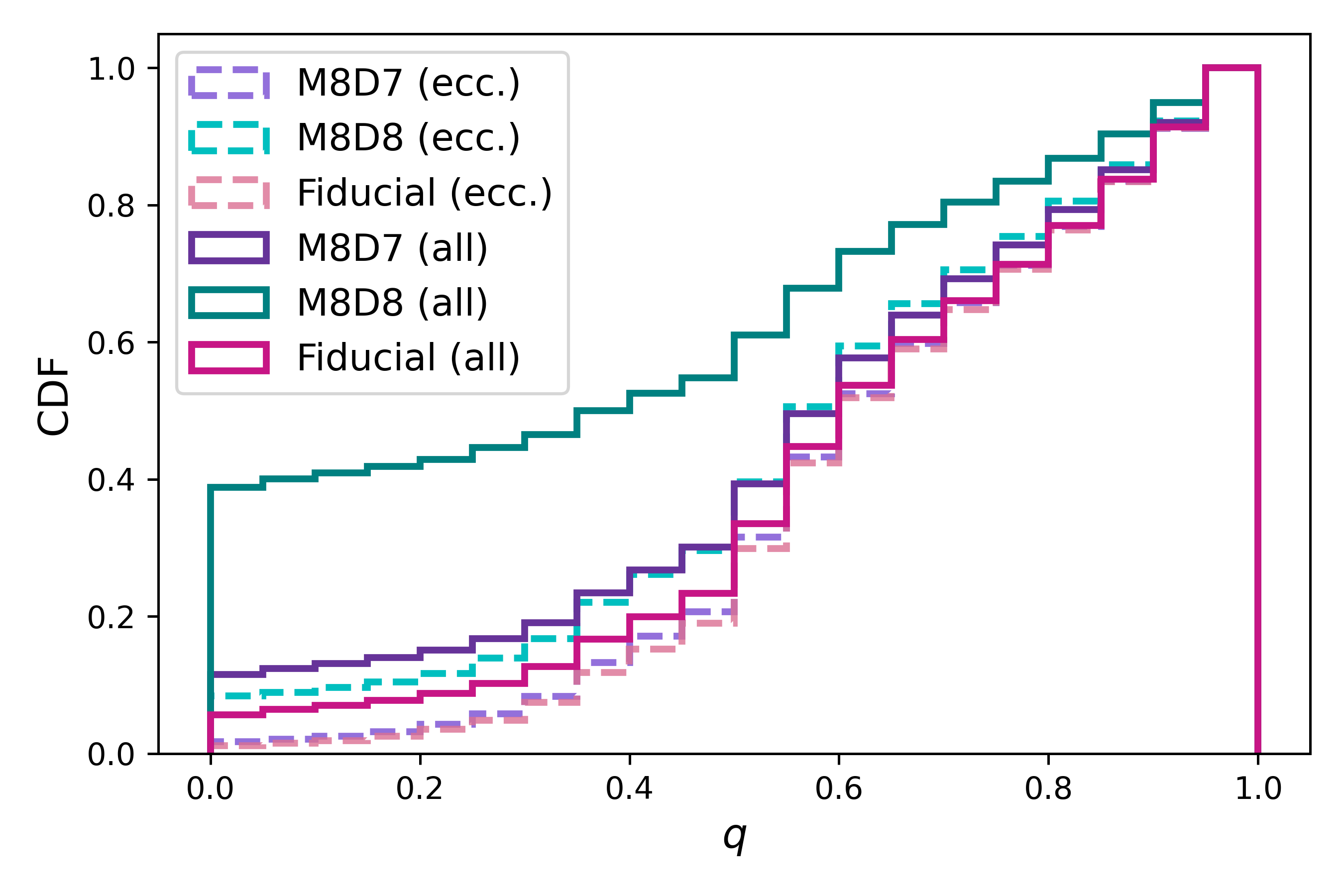}
    \includegraphics[width=0.9\columnwidth]{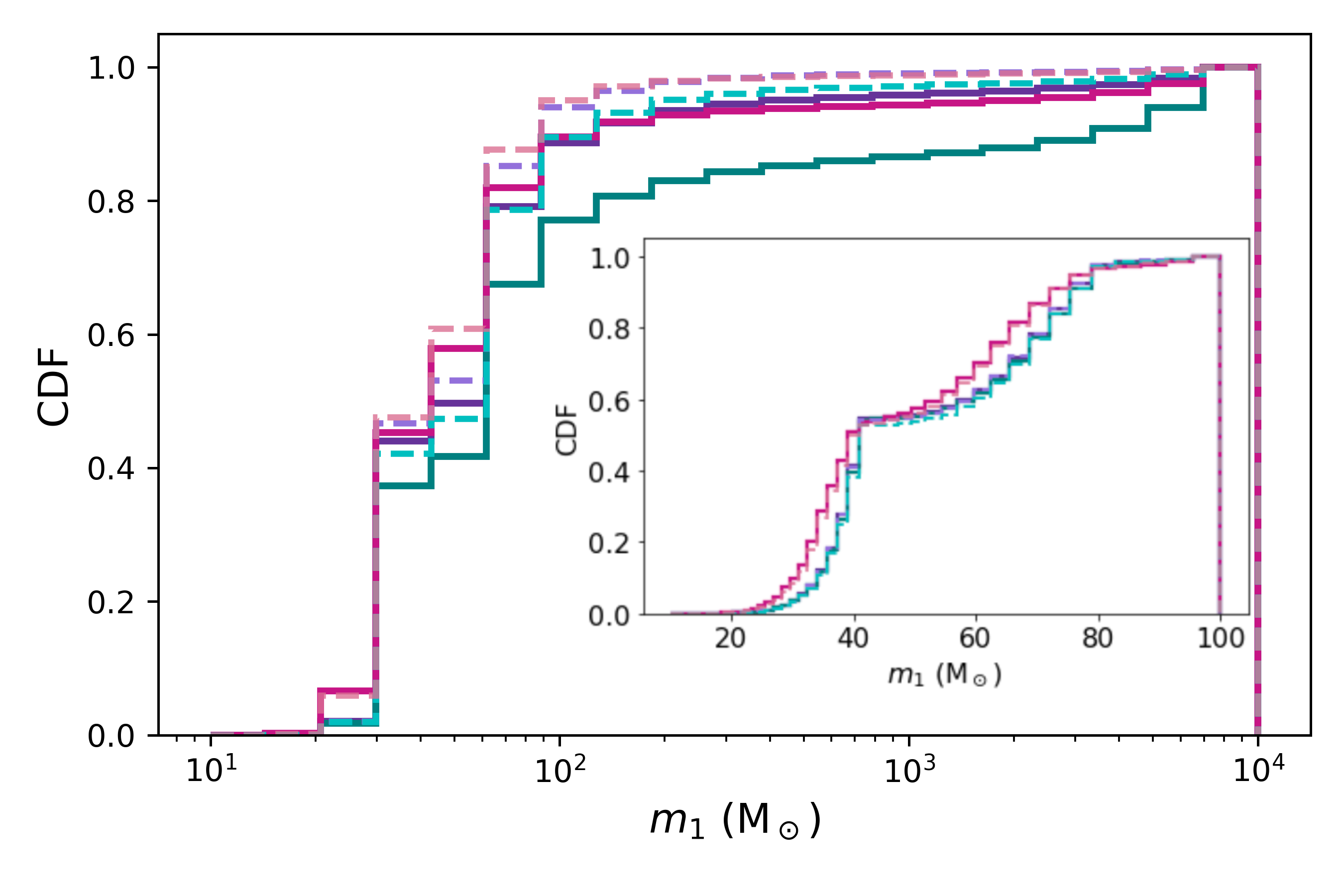} 
    \caption{Cumulative density functions (CDFs) of eccentric mergers (dashed-line histograms) to all (both in-situ) DBH mergers (solid-line histograms) across mass ratio $q$ (top panel) and primary mass $m_1$ (bottom panel). 
    Three different models are compared. The labels are identical for the two plots. The lower panel also shows the zoomed-in CDF of $m_1$ for primary masses $\leq100$\,M$_\odot$.}
    \label{fig:eccComp}
\end{figure}

\cite{Samsing:2017xmd} (see Fig.\,2) shows that while binary-single hardening can potentially harden a binary to ejection before it merges or to merge within the cluster, inclusion of 2.5PN terms in the orbital evolution can also lead to gravitation-wave capture during the binary-single resonant encounter. Eq.~2 of \cite{Antonini:2022vib}, which states the condition of a gravitational-wave capture, can be rewritten as 
\begin{equation}
    \sqrt{1-e^2} < h\left( \frac{2Gm_1(1+q)}{c^2a}\right)^{5/14},
    \label{equ:ecc_merger}
\end{equation}
where $e$, $m_1$, $q$ and $a$ are the BH binary eccentricity, primary mass,  mass ratio, separation respectively, while $h$ is a normalisation constant of the order of unity. By this condition, higher $e$, $m_1$, $q$ and smaller $a$ ensure gravitational-wave capture. 

\begin{figure}
\centering
	\includegraphics[width=0.9\columnwidth]{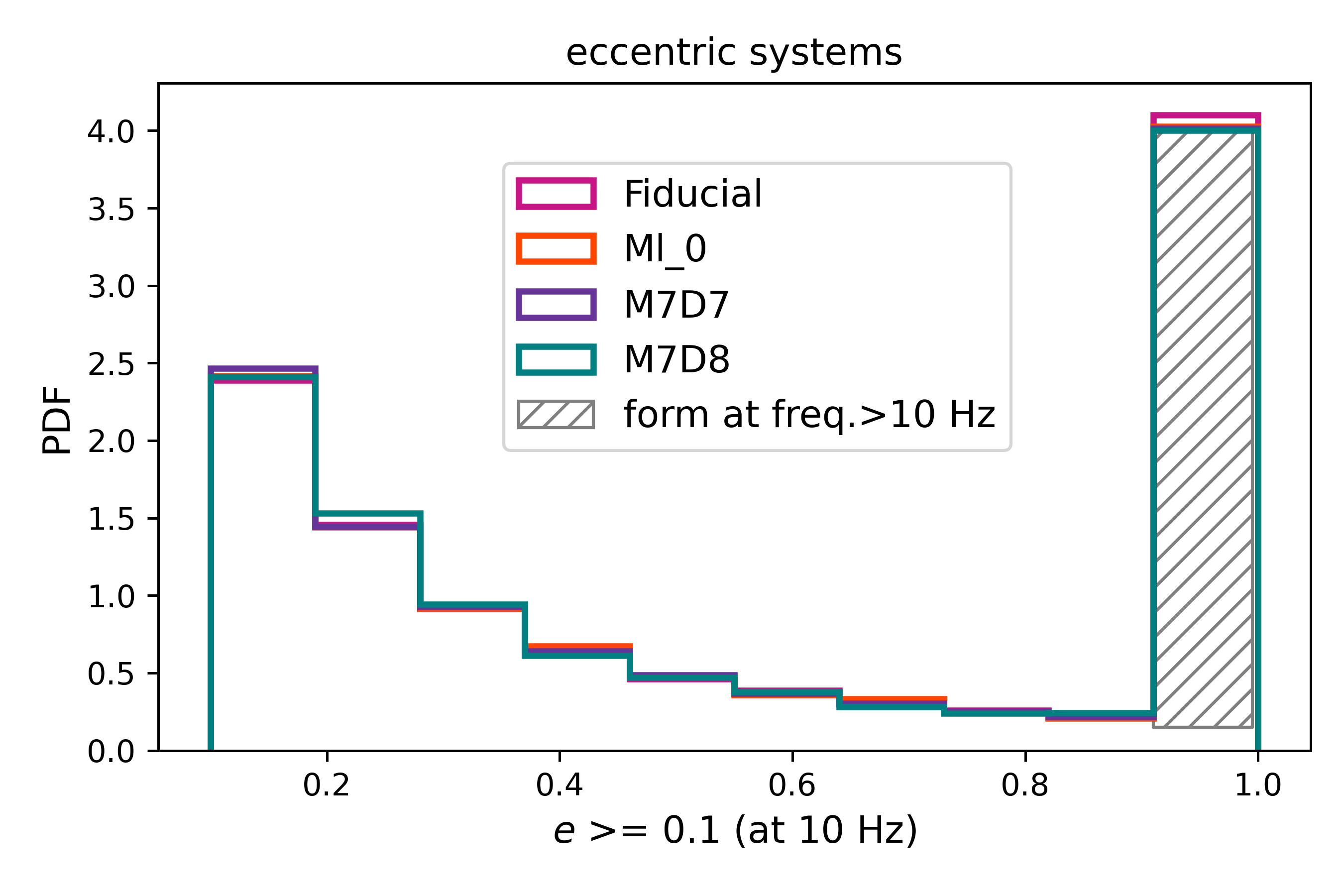}
    \caption{Eccentricity distribution of eccentric mergers with $e \geq 0.1$ at a gravitational-wave frequency of $10$~Hz (calculated using Equation~\ref{equ:frequency}) for four different models. Binaries that become bound with gravitational-wave frequency $\geq10$\,Hz are represented in the shaded region. Typically, eccentric binaries form at lower ($<10$\,Hz) frequencies with higher eccentricities and evolve to lower eccentricity at $<10$\,Hz, unless they form in the LVK band.}
    \label{fig:eccDist}
\end{figure}

The peak frequency $f$ of a DBH of total mass $M$, orbital separation $a$ and eccentricity $e$ is calculated by \cite{Wen2003} as 
\begin{equation}
    f=\frac{1}{\pi}\sqrt{\frac{GM}{a^3}}\frac{(1+e)^{1.1954}}{(1-e^2)^{1.5}}.
    \label{equ:frequency}
\end{equation}
We define a BH binary merger to be eccentric if, at a gravitational-wave frequency of $10$\,Hz (corresponding roughly to the low-frequency limit of the LVK band), $e\geq0.1$. 
All eccentric mergers are expected to be in-situ mergers, as ex-situ systems that are ejected from the influence of dynamical activity have larger time delays and circularize by the time they merge \citep[e.g.,][]{ChattopadhyayStarCluster2022buz}.  We find that about $17\%$ of all mergers in the Fiducial model are eccentric, and about one-third of these mergers become bound within the LVK band. 
This sub-group of eccentric binaries that form at  frequencies $\geq 10$\,Hz at the source frame, will be called ``high-frequency mergers'' from here onward. All of the high-frequency mergers and about $90-95\%$ of the eccentric mergers are gravitational-wave captures. 

Younger clusters also provide lower-mass BHs and, when no IMBH is present to dominate the cluster dynamics, eccentric mergers become commonplace between nearly-equal-mass BHs. 
The top panel of Fig.~\ref{fig:eccComp} shows the mass ratio $q$ distribution of the eccentric versus all mergers (marked ``all" in the figure)
, illustrating that indeed eccentric mergers occur preferentially in more equal mass systems. 
Since a lower value for both $m_1$ and $m_2$ is preferred in captures, low mass primaries are more frequent in eccentric mergers than IMBHs.
The lower panel of Fig.~\ref{fig:eccComp} demonstrates the trend for eccentric mergers to have less massive $m_1$.
In the inset plot, we compare the distribution for masses $\leq100$\,M$_\odot$; there is barely any difference between the distributions for eccentric DBH mergers and the total population of DBH mergers, demonstrating that the difference arises due to circularised mergers involving the IMBH. 
The $e$ distribution at a gravitational-wave frequency of $10$\,Hz for eccentric binaries, together with the high-frequency mergers, is shown in Fig.~\ref{fig:eccDist}, illustrating that the eccentricity distribution itself is nearly indistinguishable from model-to-model. 

Eccentric DBHs are also expected to have  shorter time delays between formation and merger, as a consequence of (i) gravitational-wave captures occurring early in the cluster's life when the velocity dispersion $\sigma$ is large, causing the hard-soft binary separation limit (and hence the semi-major axes of the merging DBHs) $a$ to be small ($a\propto1/\sigma^2$); 
and (ii) high $e$, significantly reducing the merger time of DBHs \citep{Peters:1964}. 
In practice, robustly measuring the eccentricity of systems that form outside the LVK band but still retain $e\geq0.1$ at detection is challenging; this is largely due to a lack of waveform models containing the effects of both orbital eccentricity and spin-induced orbital precession, which can lead to confusion between these two parameters \citep{RomeroShaw2020ApJ}, especially when only a few orbital cycles are visible in-band \citep{Romero-Shaw2023}.
Very high-$e$ systems that form inside the LVK band will produce gravitational-wave bursts, which are more likely to be visible in unmodelled burst searches than with pipelines that search based on templates of circularised compact binary coalescences \citep{Gond2018,Loutrel2020,Romero-Shaw2022arXiv}. 

\begin{figure}
\centering
	\includegraphics[width=0.9\columnwidth]{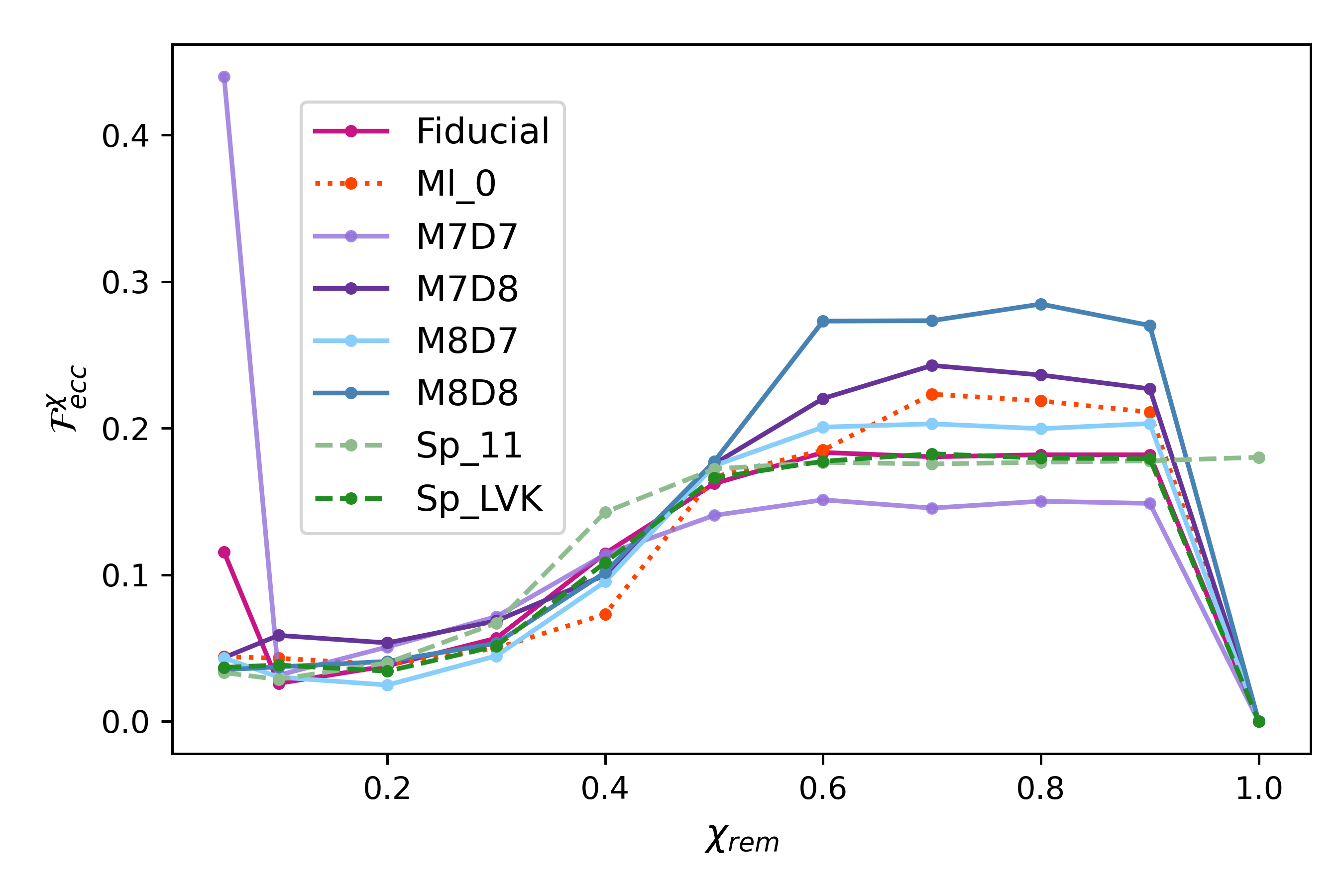}
    \includegraphics[width=0.9\columnwidth]{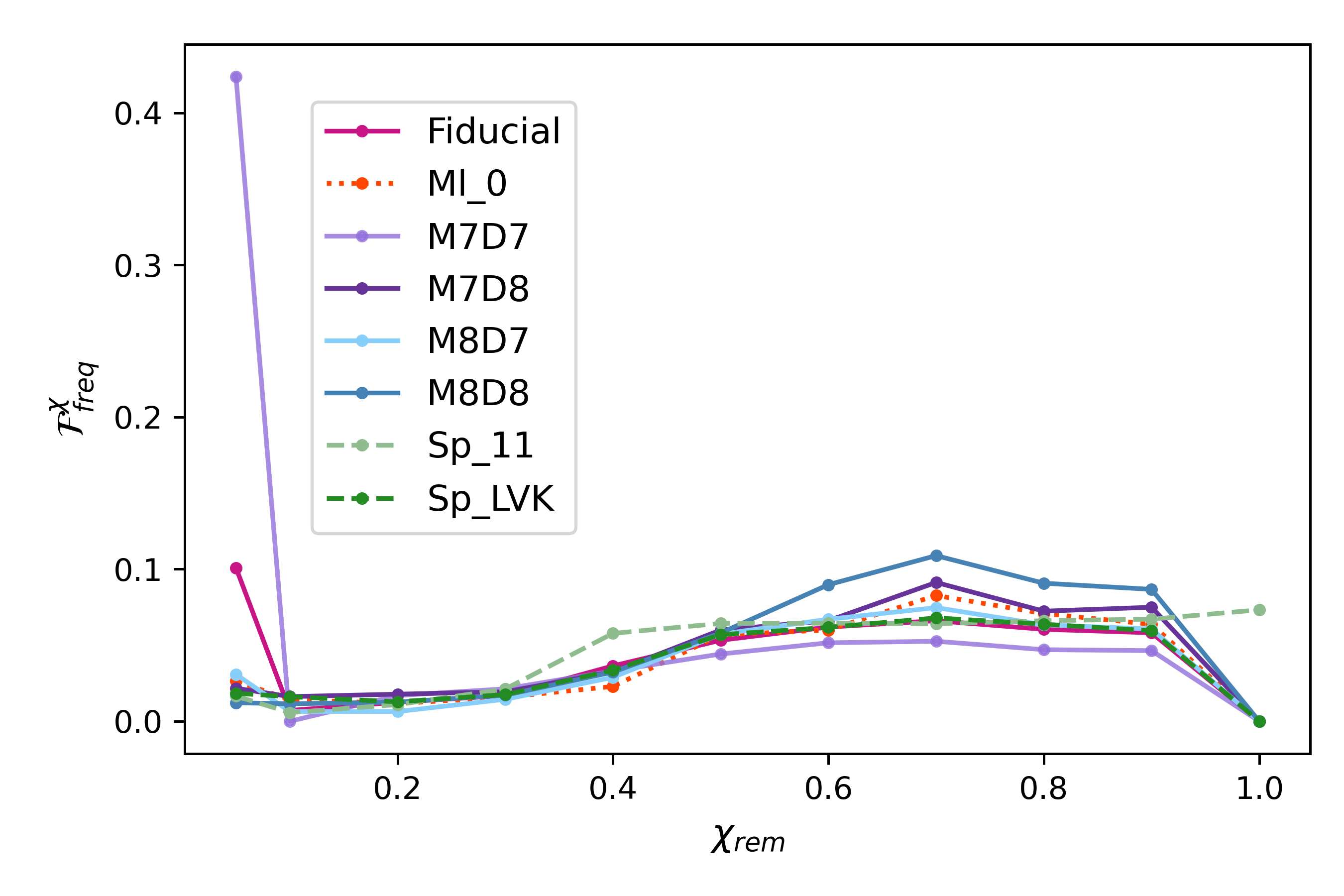}
 
    \caption{Fraction of eccentric binaries (top panel) and high-frequency mergers (bottom panel) as a function of remnant spins $\chi_\mathrm{rem}$. For each bin in $\chi_\mathrm{rem}$, $\mathcal{F_\mathrm{ecc}^{\chi}}$ (or $\mathcal{F_\mathrm{freq}^{\chi}}$) gives the fraction of  DBH mergers in that bin  with an eccentricity $e>0.1$ at a frequency $>10\rm Hz$ (or when the binary birth frequency is $>10$\,Hz).  
    The distribution peaks around $\chi_{\rm rem}\approx0.6$~--~$0.9$, accounting for most eccentric mergers. An aggregate of $100$ realizations per model is taken to improve over the otherwise small-number statistics influencing the distribution of $\mathcal{F_\mathrm{ecc}^{\chi}}$ (and $\mathcal{F_\mathrm{freq}^{\chi}}$).  The very extreme ends of the distribution still suffer from small number statistics in some cases. 
    }
    \label{fig:eccspin}
\end{figure}

Since gravitational-wave captures are more likely to occur in near-symmetric 
mass DBHs frequenting young massive clusters, eccentric DBHs are naturally biased towards first and second generation mergers. Thus we would expect to see a correlation between spin and eccentricity.
This correlation is illustrated in
Fig.~\ref{fig:eccspin} where we show for each bin of $\chi_{\rm rem}$, the fraction of mergers that have  
an eccentricity $e>0.1$ at $10$Hz (top panel). In the figure,  $\mathcal{F^{\chi}_\mathrm{ecc}}=N_{\rm tot}/N_{\rm ecc}$ with $N_{\rm tot}$ the total number of mergers in a given  bin and $N_{\rm ecc}$ the corresponding number of eccentric mergers. 
This  distribution shows that eccentric  
binaries are relatively more common in mergers that result in a  remnant with a high spin $0.6\lesssim\chi_\mathrm{rem}\lesssim0.9$. In this range of remnant spins, we expect that about $20\%$ of the binaries are still eccentric within the $10$\,Hz frequency window.

The value of $\mathcal{F^{\chi}_\mathrm{ecc}}$ varies from model to model, with densest clusters (M7D8, M8D8) and more massive clusters (M8D8, M8D7)
producing more eccentric mergers, since dense and massive clusters have a higher $v_\mathrm{esc}$ and hence $\sigma$. Model Ml\_0, which has no mass-loss apart from BH ejections, remains denser and more massive than Fiducial and hence has a higher $\mathcal{F^{\chi}_\mathrm{ecc}}$. 
The initial spin distribution appears to have no strong impact on $\mathcal{F_\mathrm{ecc}^{\chi}}$; the distributions from models Sp\_11 and Sp\_LVK  follow the distribution from the Fiducial model closely in Fig.~\ref{fig:eccspin}.
Indeed, the general nature of the $\mathcal{F^{\chi}_\mathrm{ecc}}$ curve does not change, implying that irrespective of models, eccentric mergers are mainly first- or sometimes second-generation mergers. 

The general eccentric merger fraction $\mathcal{F_\mathrm{ecc}}$ is tabulated in the penultimate column of Tab.~\ref{tab:IMBHmass}, demonstrating that denser clusters have more eccentric mergers. 
Metallicity also appears to play a role in determining $\mathcal{F_\mathrm{ecc}}$. 
Metal-rich model Z\_100, in which nearly $20\%$ of the DBH mergers are eccentric, has a narrower initial BH mass spectrum due to the increased metallicity; this increases the efficiency of eccentric DBH formation, since the condition of lower- and nearly-equal mass BH binaries is satisfied relatively easily and is aided by the delayed formation of IMBHs \citep[this is similar to the upper limit of eccentric mergers that][finds; albeit for Lidov-Kozai mechanism in globular clusters]{Antonini2016}. 
This implies that eccentric mergers are more frequent in the metal-rich dynamical environments of the local universe, opening exciting avenues for future runs of current ground-based detectors.

\begin{figure}
\centering
	\includegraphics[width=0.9\columnwidth]{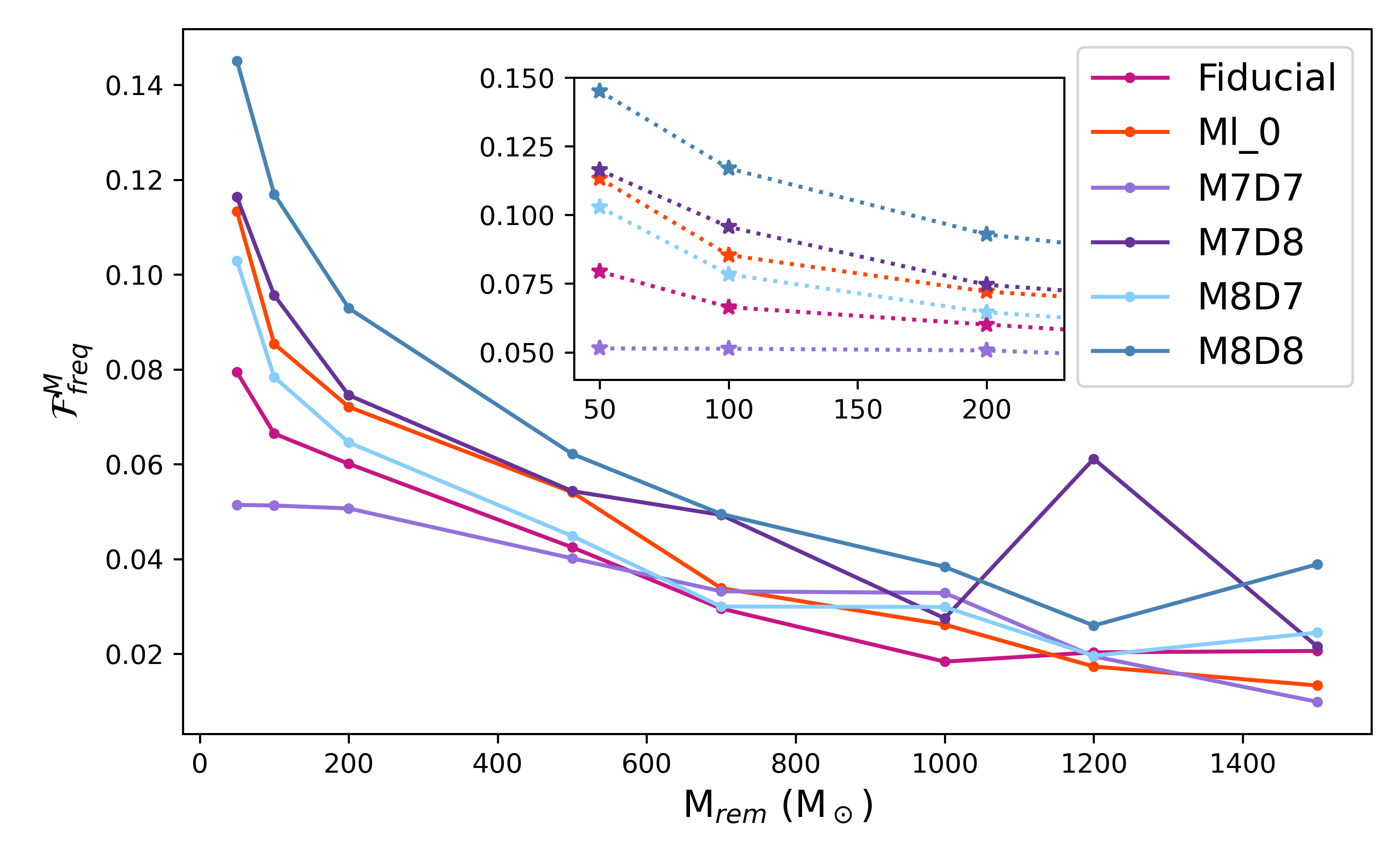}
 \includegraphics[width=0.9\columnwidth]{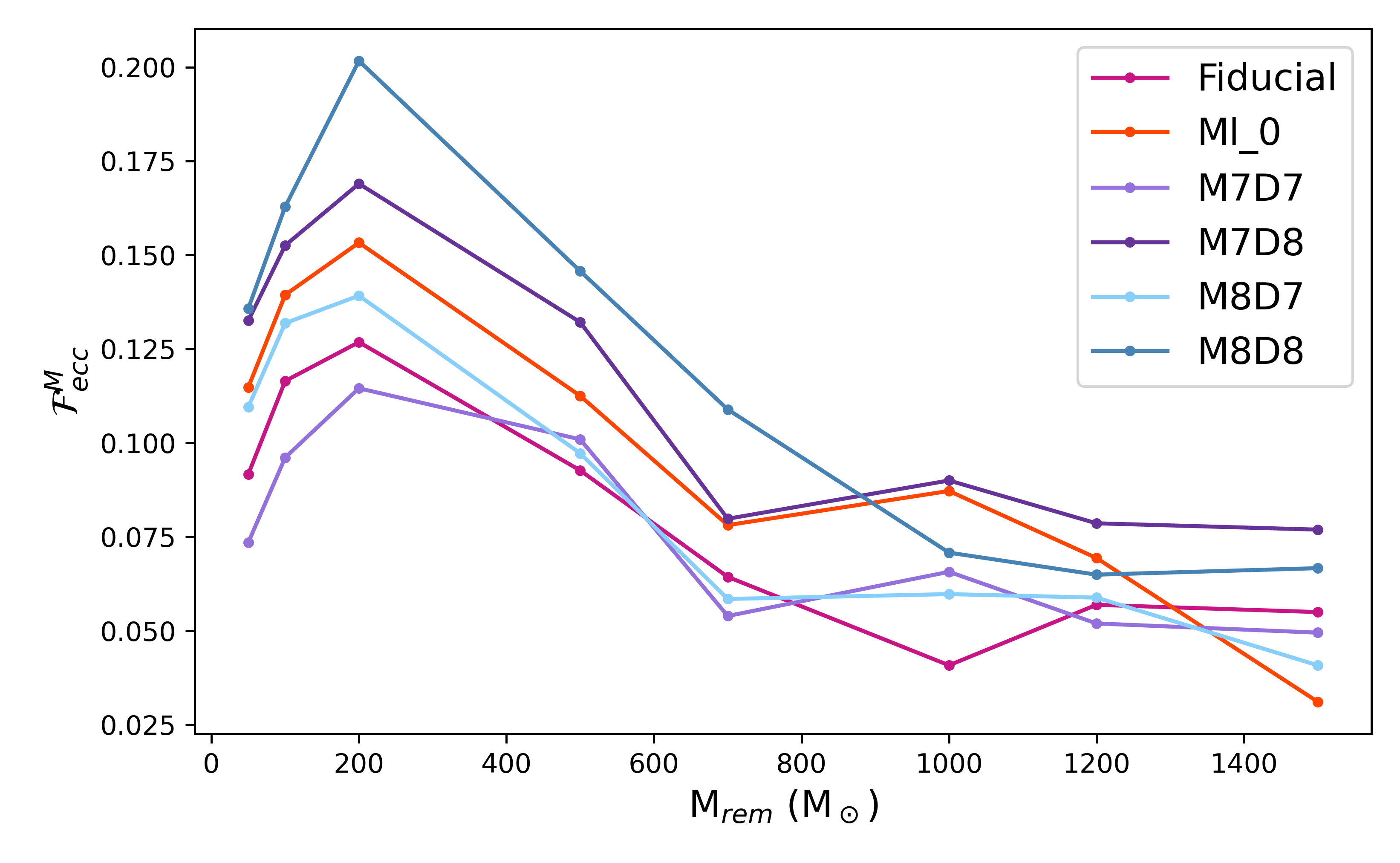}
    \caption{The fraction of high-frequency (upper panel) and eccentric (lower panel) mergers binned $\mathcal{F_\mathrm{freq}^{M}}$ (and $\mathcal{F_\mathrm{ecc}^{M}}$)  by remnant mass $M_\mathrm{rem}$, emphasizing the precedence of less massive BHs in high-frequency mergers. For eccentric mergers, $M_\mathrm{rem}$ slightly increases till 200\,M$_\odot$ and then decreases.
    As with net eccentric mergers, even with $100$ realisations of each model, the tail end of the distribution is  impacted by low-number statistics. 
    }
    \label{fig:highFreqFrac}
\end{figure}

High-frequency mergers--- those binaries that form with gravitational-wave frequencies $>10$\,Hz---obey the same Eq.~\ref{equ:ecc_merger}.
While having very similar $e>0.99$ at formation, they have even smaller masses. 
For the Fiducial model, the median $m_1$ for all eccentric mergers is$\sim51$\,M$_\odot$. 
However, for high-frequency mergers the median $m_1 \sim 41$\,M$_\odot$ is about $0.8$ times lower. 
The preference for lower masses in these high-frequency, high-eccentricity mergers
is also apparent when we compare the mass-binned high-frequency merger fraction ($\mathcal{F_\mathrm{freq}^{M}}$) across models in Fig.~\ref{fig:highFreqFrac}, where we demonstrate the  decline of high-frequency mergers as the remnant mass bin value increases. The fraction of high-frequency mergers $\mathcal{F_\mathrm{freq}}$ scales from model to model as $\mathcal{F_\mathrm{ecc}}$, ranging from $\approx2-8$\% of all mergers, with Fiducial having $\mathcal{F_\mathrm{freq}}\approx6$\% (final column of Table~\ref{tab:IMBHmass}). 

We note, finally, that the fraction of eccentric and high-frequency mergers  is a decreasing function of time. 
As the cluster expands and the IMBH grows in mass,
the time $t_3$ between individual binary-single encounters  increases \citep[see Equation 20 in][]{Antonini:2019ulv}.
The increased value of $t_3$ means a higher probability of  in-cluster inspirals and a lowered fraction  of eccentric GW captures. 

\subsection{Mass and spin evolution of IMBH} 
\label{sec:IMBH}

In this Section, we concentrate on the most massive IMBH formed in our cluster models. Primarily focusing on the mass of the IMBH, we discuss the contributing factors, such as global cluster properties, assumptions associated with stellar evolution, and BH binary/triple pairing. We also correlate the IMBH spin to its mass, as well as to the host cluster properties.  

\subsubsection{Cluster initial mass and density}
\label{sec:clusters_mass_den}
\begin{figure}
\centering	\includegraphics[width=0.9\columnwidth]{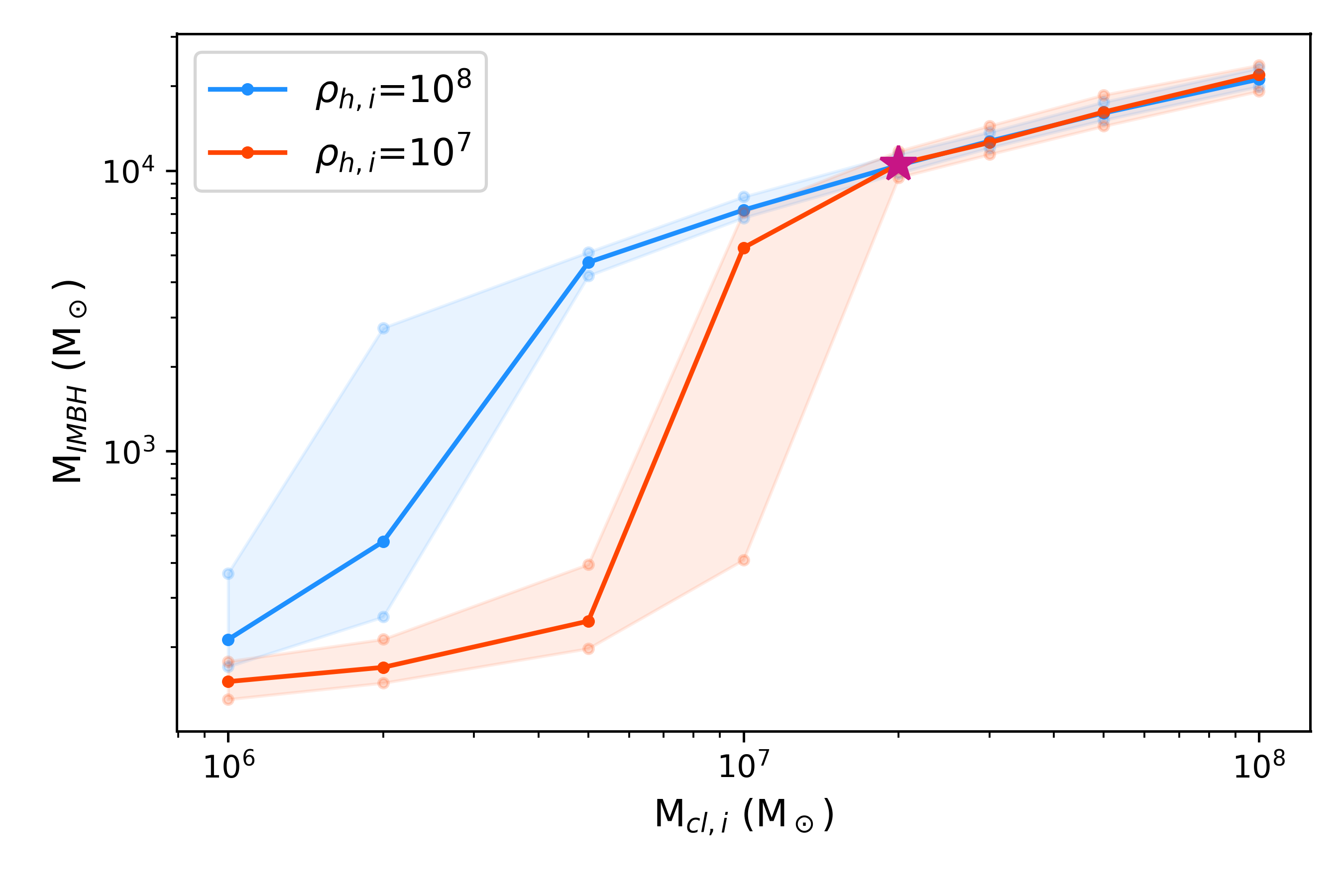}
\includegraphics[width=0.9\columnwidth]{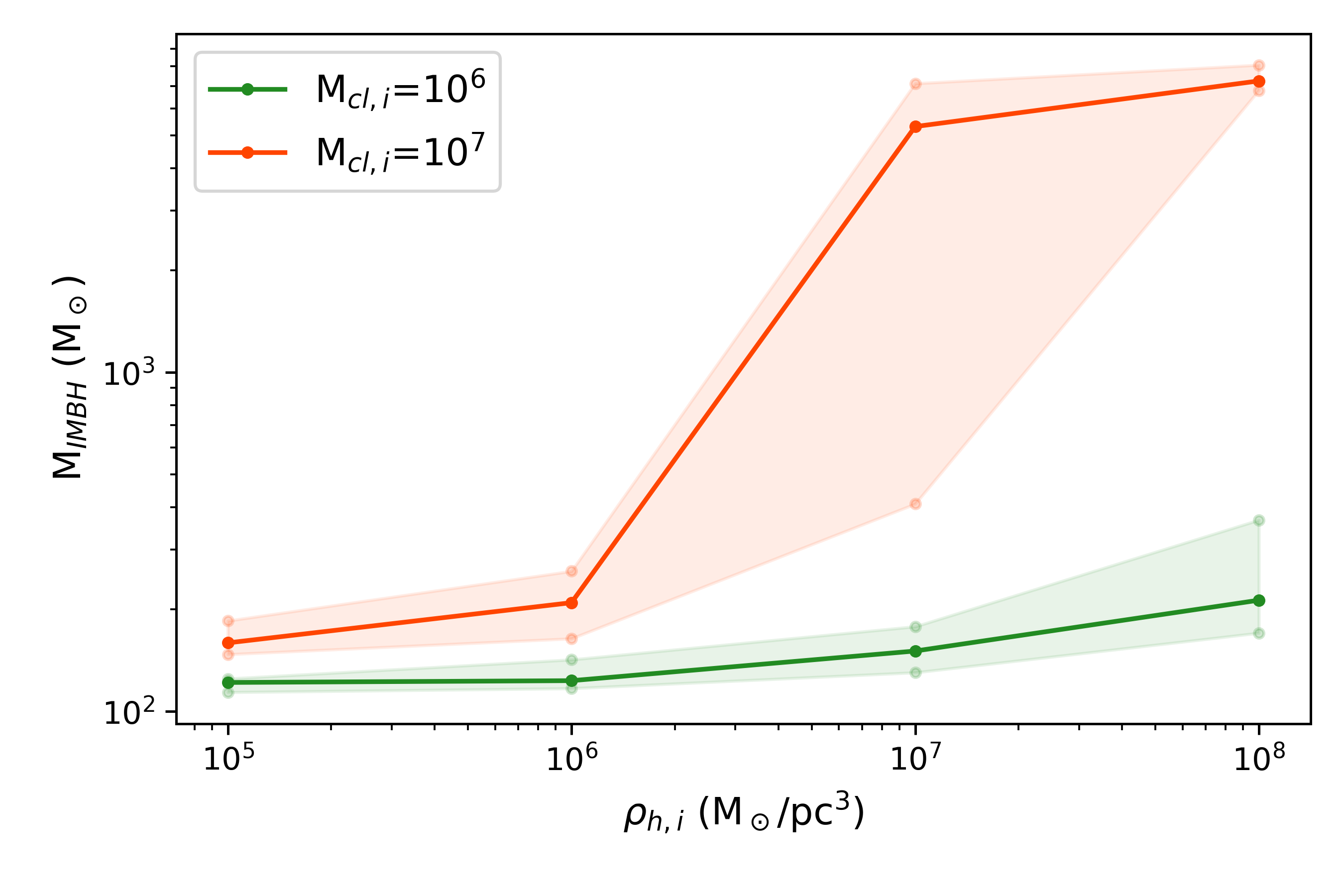}
\includegraphics[width=0.9\columnwidth]{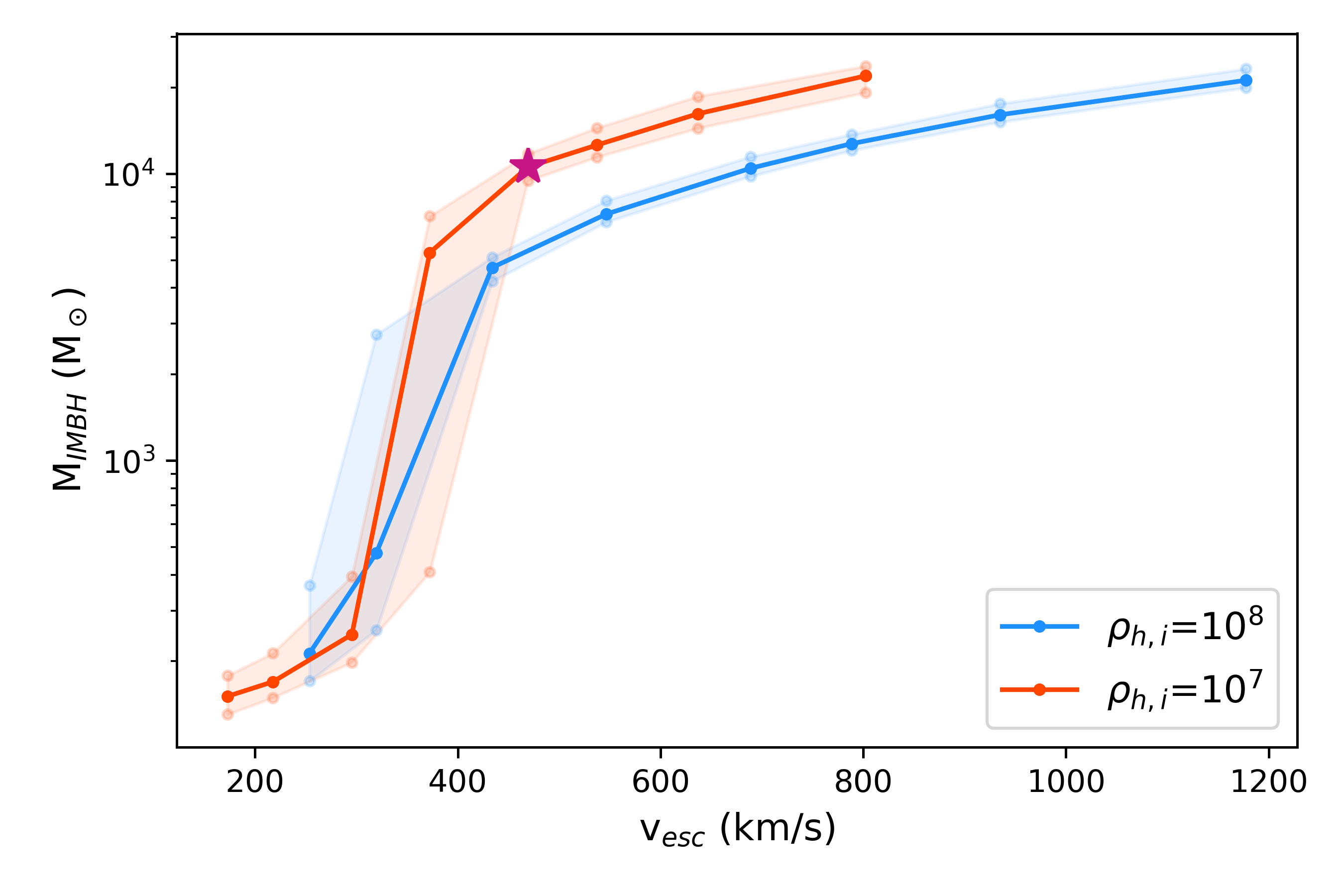}
    \caption{Final mass of the IMBH ($M_\mathrm{IMBH}$) after a Hubble time of cluster evolution with respect to cluster initial mass (upper panel), initial density (middle panel) and escape velocity (lower panel). Solid lines represent the median values ($M_\mathrm{IMBH}^{50}$ of Table~\ref{tab:IMBHmass}) and the shaded region shows the boundary region between the $90^\mathrm{th}$~--~$10^\mathrm{th}$ percentiles ($M_\mathrm{IMBH}^{10}$ and $M_\mathrm{IMBH}^{90}$ of Table~\ref{tab:IMBHmass}) for $100$ realizations of each model. 
    The magenta star symbol in the upper and lower plots represents the Fiducial model.}
    \label{fig:Mimbh}
\end{figure}

The maximum IMBH mass reachable through only hierarchical mergers is strongly affected by the initial mass and density of the host cluster. 
Models Sl. no. 2$-$11 of  Table~\ref{tab:IMBHmass} (columns denoted by ``$M_\mathrm{IMBH}^{50}$'', ``$M_\mathrm{IMBH}^{10}$'', ``$M_\mathrm{IMBH}^{90}$'' showing the IMBH mass $50^\mathrm{th}$, $10^\mathrm{th}$ and $90^\mathrm{th}$ percentiles respectively of 100 realizations of each model) clearly demonstrate this relationship. 
While this can be predicted from Eq.~(4) of \citet[][]{Antonini:2018auk}, we observe that the maximum BH mass reachable in a cluster of initial mass $M_{cl,i}$ (in M$_\odot$) and half-mass density $\rho_{h,i}$ (in M$_\odot$\,pc$^{-3}$) is lower by up to an order of magnitude in our models. Although \cite{Antonini:2018auk} estimates the upper limit on the maximum IMBH mass while ignoring recoil kicks, the incorporation of binary-single encounters that can potentially eject BHs (the single and/or the binary) lowers our maximum obtained IMBH mass.

Fig.~\ref{fig:Mimbh} shows the variation of the maximum IMBH mass ($M_\mathrm{IMBH}$) in our models as a function of the host cluster initial mass, half-mass density and escape velocity. 
We see from the upper panel of Fig.~\ref{fig:Mimbh} that an increase in initial cluster mass leads to the formation of more massive IMBHs. However, we also note that (i) there appears to be a transition in $M_\mathrm{IMBH}$ with respect to cluster initial mass ($M_{cl,i}\sim10^7$~M$_\odot$ for $\rho_{h,i}=10^7$~M$_\odot$\,pc$^{-3}$ and $M_{cl,i}\sim3 \times 10^6$M$_\odot$ for $\rho_{h,i}=10^8$~M$_\odot$\,pc$^{-3}$) where the median $M_\mathrm{IMBH}$ makes an order-of-magnitude jump, and (ii) around the same transitory phase, the width in the $M_\mathrm{IMBH}$ spectrum, i.e., the difference between the $90^\mathrm{th}$ and $10^\mathrm{th}$ percentile, calculated from $100$ realisations of each model, is rather broad. Comparing the upper to the middle panel of Fig.~\ref{fig:Mimbh}, we observe that this transition occurs at lower $M_{cl,i}$ for higher $\rho_{h,i}$: denser clusters. This behaviour is explained by the combination of three velocities---cluster escape velocity ($v_\mathrm{esc}$), BH natal kick ($v_\mathrm{kick}$) and post-merger gravitational wave recoil kick ($v_\mathrm{rec}$) magnitudes. 

\begin{figure}
\centering
	\includegraphics[width=0.8\columnwidth]{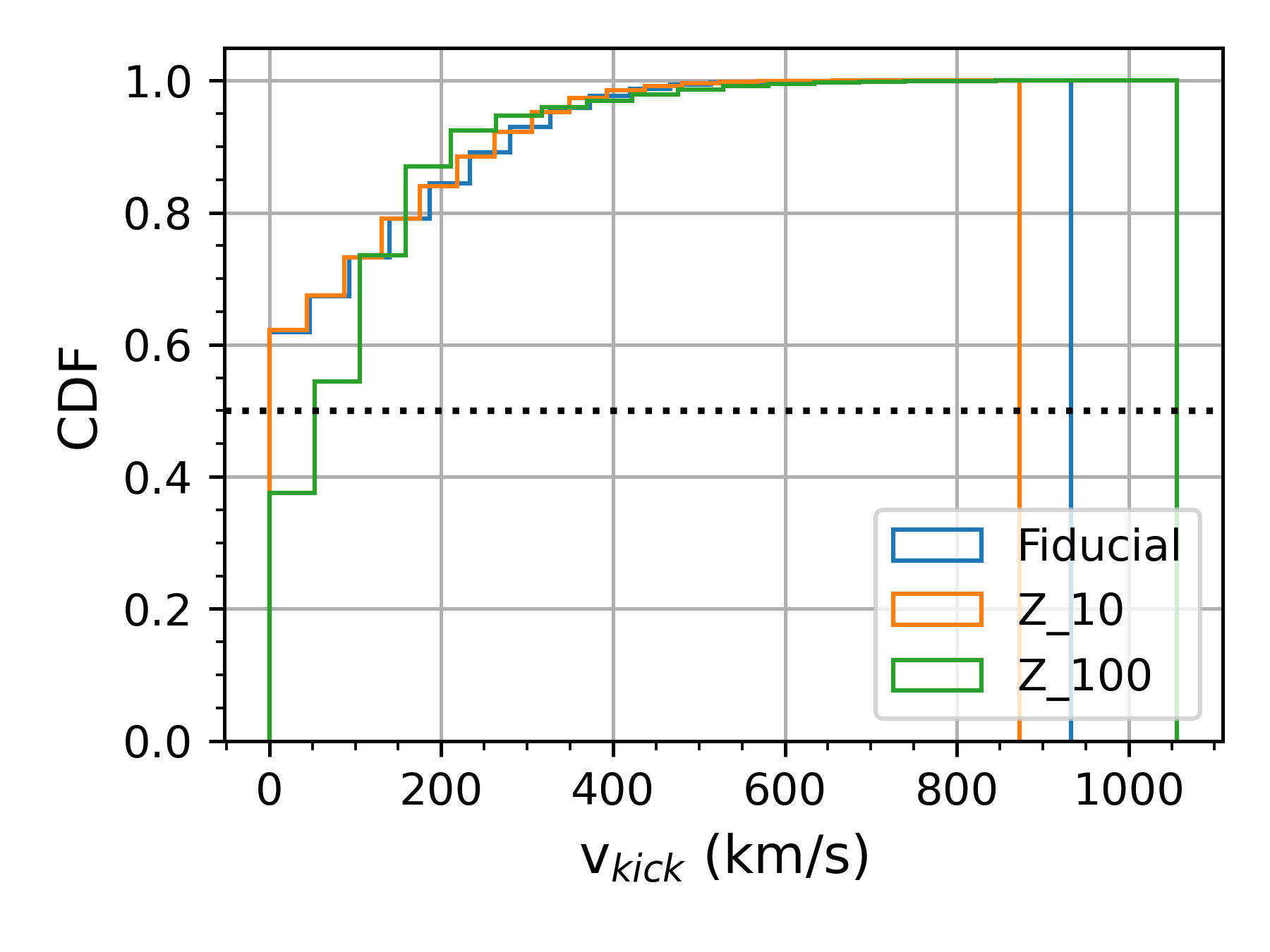}
    \caption{BH natal kick distribution for models Fiducial, Z\_10 and Z\_100, with metallicities $1.5 \times 10^{-4}$, $1.5 \times 10^{-3}$ and $1.5 \times 10^{-2}$ respectively. The dotted line emphasises where CDF$=0.5$.}
    \label{fig:vkick}
\end{figure}

\begin{figure}
\centering
	\includegraphics[width=0.9\columnwidth]{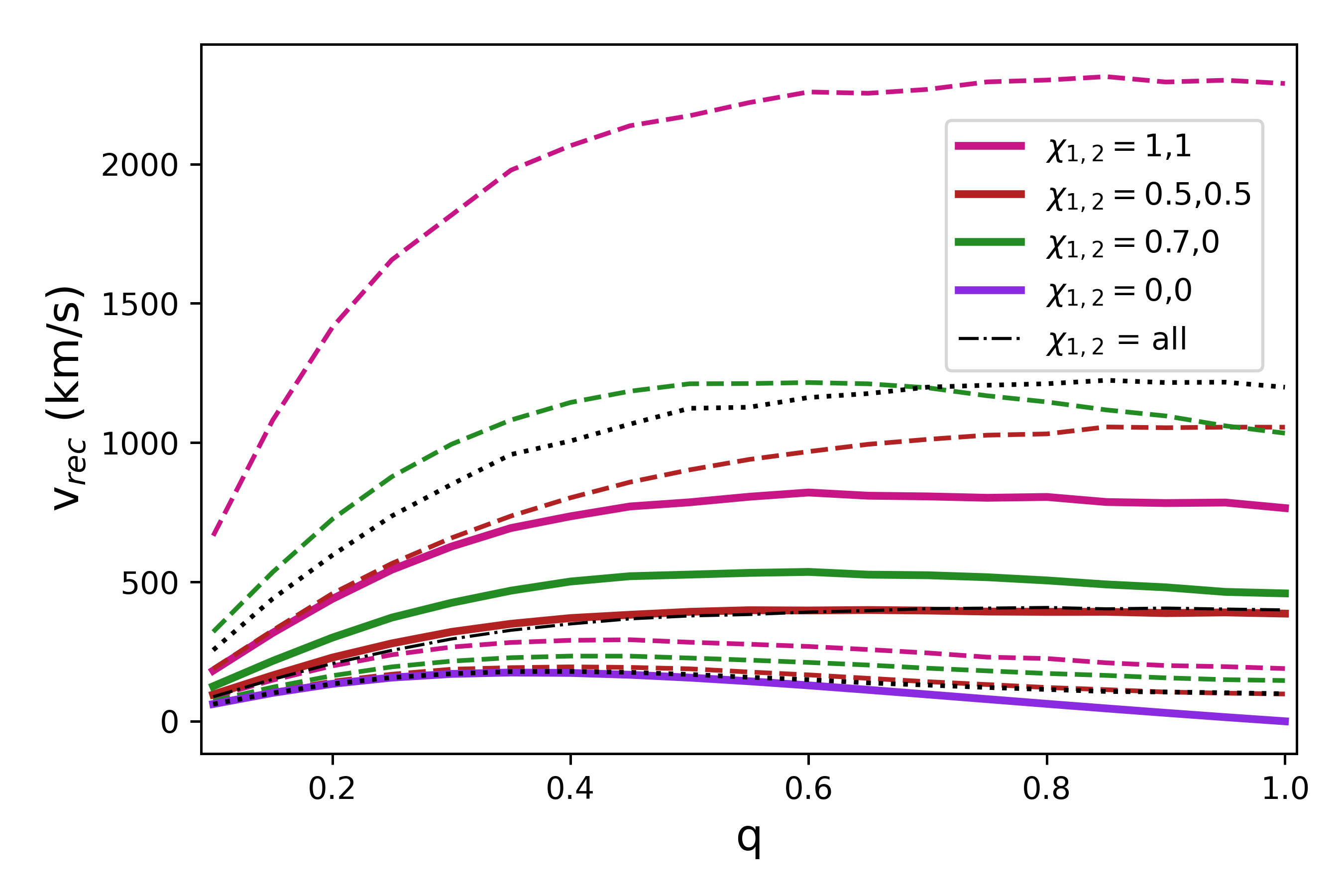}
    \caption{Post-merger gravitational wave recoil $v_\mathrm{rec}$ vs binary mass ratio $q$ for different binary spin magnitudes $\chi_{1,2}$ integrated over all possible spin-orbit angles \citep{Lousto:2010}. The solid lines show the median $v_\mathrm{rec}$ magnitude for all possible spin angles, and the dashed  lines depict the corresponding $90^\mathrm{th}$-$10^\mathrm{th}$ percentile boundaries. The black lines characterize the same for all possible values and orientations of $\chi_{1,2}$.
    }
    \label{fig:vreq}
\end{figure}

The $v_\mathrm{kick}$ distribution of BHs from our single stellar evolution prescription \citep{HurleySSE:2000pk} is depicted in Fig.~\ref{fig:vkick}. For clusters with initial $v_\mathrm{esc}$ $\gtrsim 400$\,km\,s$^{-1}$, nearly all BHs are retained post-stellar evolution (losing only about $0.6\%$ of the BH mass generated through stellar evolution), compared to host clusters with $v_\mathrm{esc}$ $\lesssim 100$\,km\,s$^{-1}$ that retains  $\approx70\%$ of their BHs (losing about $18\%$ of the BH mass through natal kicks).\footnote{For comparison, post-stellar evolution (with stellar evolution parameters as described in Sec.~\ref{sec:methods}) $\approx10-20\%$ of the initial mass of the cluster is expected to be held in BHs.}

For comparison with \cite{Antonini2016ApJ}, we see that $\lesssim40\%$ of metal-poor cluster BHs receive natal kicks greater than $50$\,km\,s$^{-1}$, and it is only in metal-rich (Z\_100) environments that more than $60\%$ of BHs receive a natal kick of $\gtrsim50$\,km\,s$^{-1}$. Since Z\_100 is a rather high metallicity even for the local universe, the nuclear cluster DBH merger rate is unlikely to dominate over the merger rate from other dynamical environments, such as globular clusters, or isolated evolution. 

The post-merger recoil kick $v_\mathrm{rec}$ calculated with different three-dimensional spin magnitudes  \citep{Lousto:2010} for all possible spin orientations is shown in Fig.~\ref{fig:vreq}. The median curve for all possible spin magnitudes for $q\gtrsim0.5$ is about $400$\,km\,s$^{-1}$ and its $10^\mathrm{th}$ percentile is at about $\approx200$\,km\,s$^{-1}$. This indicates that, after an in-cluster DBH merger, the remnant is nearly always retained in clusters with $v_\mathrm{esc}\gtrsim400$\,km\,s$^{-1}$, and is likely to be ejected for clusters with $v_\mathrm{esc}\lesssim100$\,km\,s$^{-1}$. For clusters with 200\,km\,s$^{-1}\lesssim v_\mathrm{esc}\lesssim 400$\,km\,s$^{-1}$ range, the retention fraction varies. It must be also noted that the $v_\mathrm{esc}$ for models shown in the lower panel of Fig.~\ref{fig:Mimbh} refers to the initial value, while cluster evolution (e.g., expansion, mass loss) tends to reduce $v_\mathrm{esc}$ both in reality and in our simulations. 
Hence, it can be concluded that clusters with initial $v_\mathrm{esc} \gtrsim 400$\,km\,s$^{-1}$ with very high BH retention fraction are more likely to form IMBHs with masses $\sim10^4-10^3$ \,M$_\odot$, whereas clusters with $v_\mathrm{esc} \lesssim 200$ km\,s$^{-1}$ lose most of their BHs through stellar evolution birth kicks as well as GW recoil kicks post-merger, prohibiting further growth of IMBHs greater than a few $100$\,M$_\odot$. Our results are  lower in final IMBH mass estimate than \cite{Miller2002}, whose analytical limit of IMBH mass in a  $\approx10^6$\,M$_\odot$ cluster is about $\mathcal{O}(3)$\,M$_\odot$, an order of magnitude higher than ``M6" models in Table~\ref{tab:IMBHmass}. Underestimating binary-single ejections and the overall more simplistic model used in \cite{Miller2002} is a possible cause of the discrepancy. It is also interesting to note that for $\chi_\mathrm{1,2}=0;0$, i.e. first generation mergers, remnants will always be retained for $v_\mathrm{esc}\gtrsim100$\,km\,s$^{-1}$. 

The Poisson oscillation $\sigma_\mathrm{P}$ for $M_\mathrm{IMBH}^{50}$ can vary from $0.04$ to $0.65$. Clusters with $v_\mathrm{esc}\lesssim200$\,km\,s$^{-1}$ or $v_\mathrm{esc}\gtrsim400$\,km\,s$^{-1}$ corresponds to the lower values $\sigma_\mathrm{P}$, while $v_\mathrm{esc}$ in the mid-transitional region have higher $\sigma_\mathrm{P}$. Fiducial model has $\sigma_\mathrm{P}=0.15$.   \footnote{$\sigma_\mathrm{P}$=$\sigma$/$M_\mathrm{IMBH}^{50}$ where,
$\sigma=\sqrt{\frac{1}{n}\sum_{i=1}^{n}({\mathrm{M}_\mathrm{IMBH}[\mathrm{i}] - \mathrm{M}_\mathrm{IMBH}^{50})^2}}$.} 
That $v_\mathrm{esc}$ plays the key role in determining whether an IMBH  will form is shown by \cite{Antonini:2018auk, Fragione:2020nib, Mapelli:2021MNRAS} and by our Fig.~\ref{fig:Mimbh}. 

\begin{figure}
\centering
	\includegraphics[width=0.9\columnwidth]{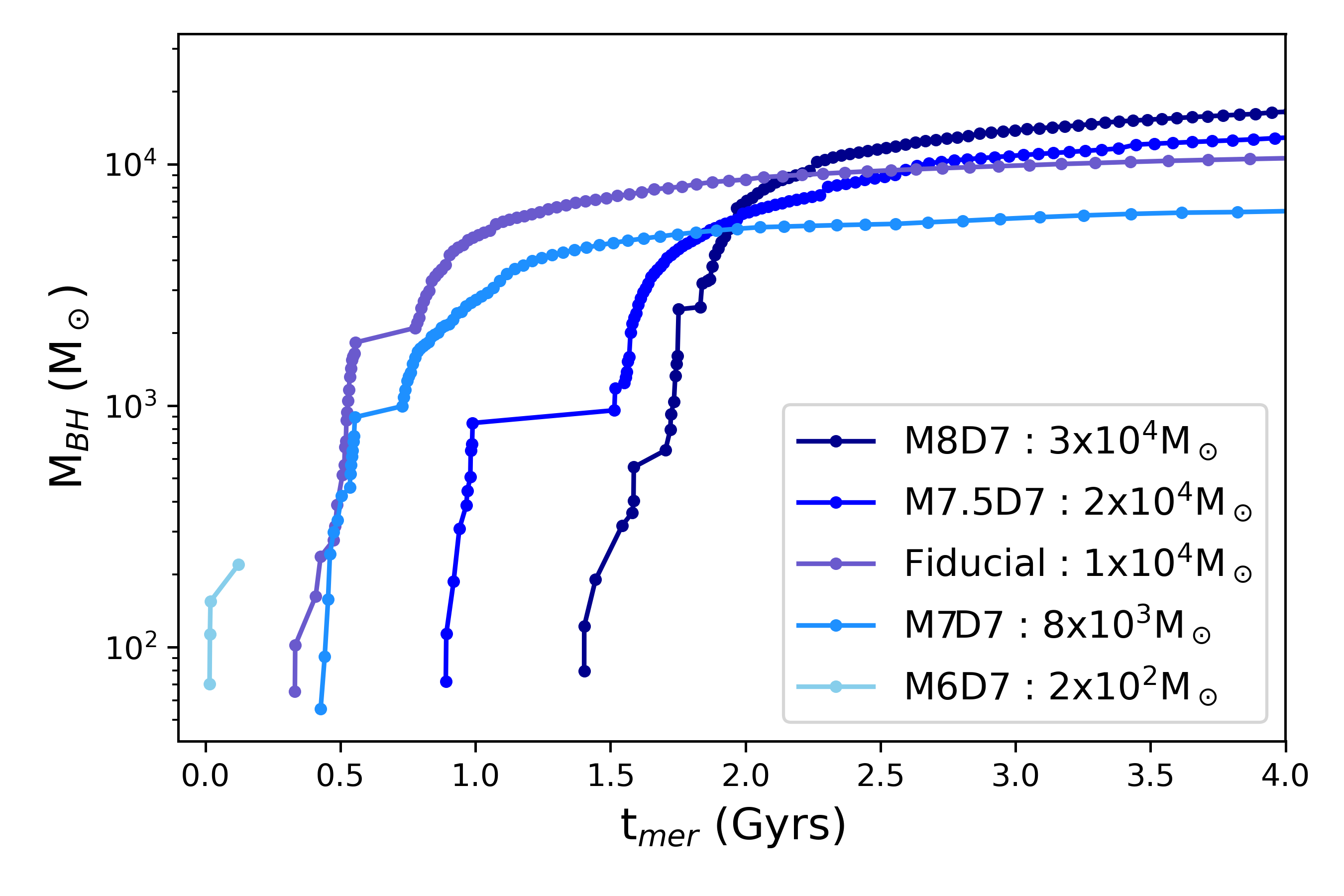}
	\includegraphics[width=0.9\columnwidth]{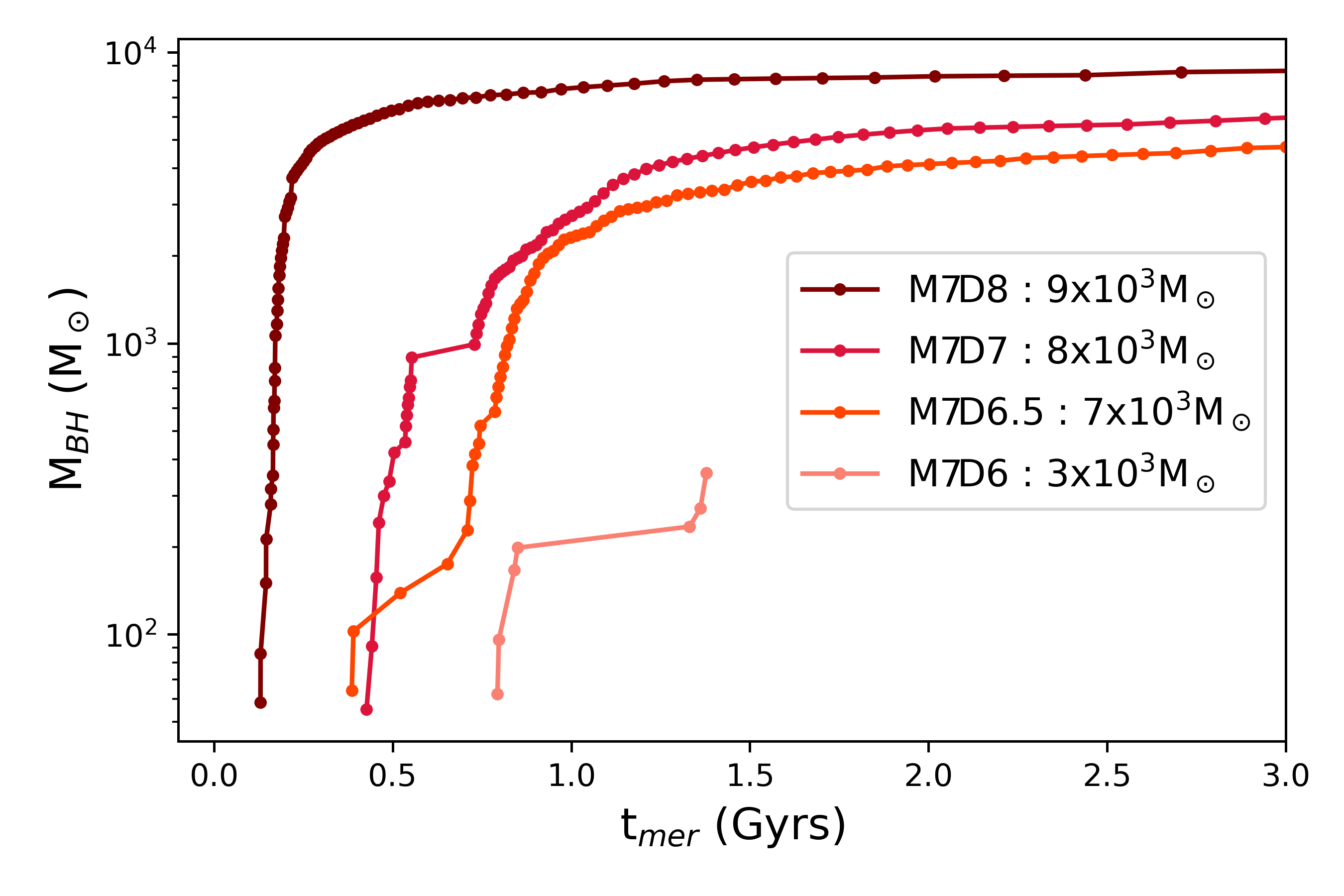}
    \caption{Mass growth of IMBH through hierarchical mergers across cluster evolution time ($t_\mathrm{mer}$). The upper panel shows clusters with same initial density $10^7$\,M$_\odot$\,pc$^{-3}$ and different masses $10^8$\,M$_\odot$ (M8D7), $5\times10^7$\,M$_\odot$ (M7.5D7), $2\times10^7$\,M$_\odot$ (Fiducial), $10^7$\,M$_\odot$ (M7D7) and $10^6$\,M$_\odot$ (M6D7). The lower panel illustrates clusters of same initial mass $10^7$\,M$_\odot$ and different densities $10^8$\,M$_\odot$\,pc$^{-3}$ (M7D8), $10^7$\,M$_\odot$\,pc$^{-3}$ (M7D7), $5\times10^6$\,M$_\odot$\,pc$^{-3}$ (M7D6.5) and $10^6$\,M$_\odot$\,pc$^{-3}$ (M7D6). The legend also notes the maximum IMBH mass reached in a Hubble time by each model. }
    \label{fig:MDHeir}
\end{figure}

The rapidity with which DBH mergers 
occur and a massive BH remnant grows through hierarchical mergers depends on the host cluster's initial mass and density. This can be seen by inspection of the ninth and tenth columns of Table~\ref{tab:IMBHmass}; these record the median time required to create a BH of $100$\,M$_\odot$ ($t_{100}$) and a BH of $1000$\,M$_\odot$ ($t_{1000}$) respectively, and show that more massive (more dense) clusters require more time (less time) to create an IMBH of $1000$\,M$_\odot$, with all other parameters remaining constant. 
The first DBH in our models is expected to form after cluster core collapse, when the dense core, formed through mass segregation, harbours interacting BHs. Equation 9 and 10 of \cite{Antonini:2019ulv} show the dependence of the  core-collapse time ($t_\mathrm{cc}$) on the cluster initial mass and density, such that $t_\mathrm{cc} \propto \mathrm{M}_{cl,i}/(\rho_{h,i})^{1/2}$. While $M_{cl,i}$ is the most dominant factor in determining how rapidly the hierarchical mergers commence in a cluster, $\rho_{h,i}$ does play a  role too. The upper panel of Fig.~\ref{fig:MDHeir} shows the growth history of what becomes the most massive IMBH, reflecting this strong dependence on the initial cluster mass. M8D7 has its initial BH mergers occurring around $1.4$\,Gyrs, and M6D7---a cluster two orders of magnitude lower in its initial mass---around $0.01$\,Gyrs. It is interesting to note that although M8D7 begins its hierarchical mergers later than the lower-mass clusters, it overtakes the others to form the most massive IMBH of all models compared in this plot. A cluster too massive, computed without relativistic treatments for its evolution, can have a $t_\mathrm{cc}$ greater than a Hubble time and is therefore unsuitable for hierarchical mergers. Several models of \cite{Fragione:2020nib}, under the assumptions of the mass spectrum factor $\psi=6$ (Eq.~(9) and~(10) of \cite{Antonini:2019ulv} and with N$_\mathrm{rh}=3.21,$ ln$\Lambda=10$, $\langle m_{all}\rangle=0.6$), yield a $t_\mathrm{cc}$ greater than a Hubble time. Of course, $\psi$ can be lower than its traditionally-assumed value of $1$, further increasing $t_\mathrm{cc}$. The lower panel of Fig.~\ref{fig:MDHeir} shows the hierarchical history of the cluster IMBH with respect to cluster density, showing  denser clusters to have a more rapid IMBH growth. Indeed, in a cluster of too-low initial density, an IMBH never grows over a few hundred solar masses.  

\begin{figure}
\centering
\includegraphics[width=0.9\columnwidth]{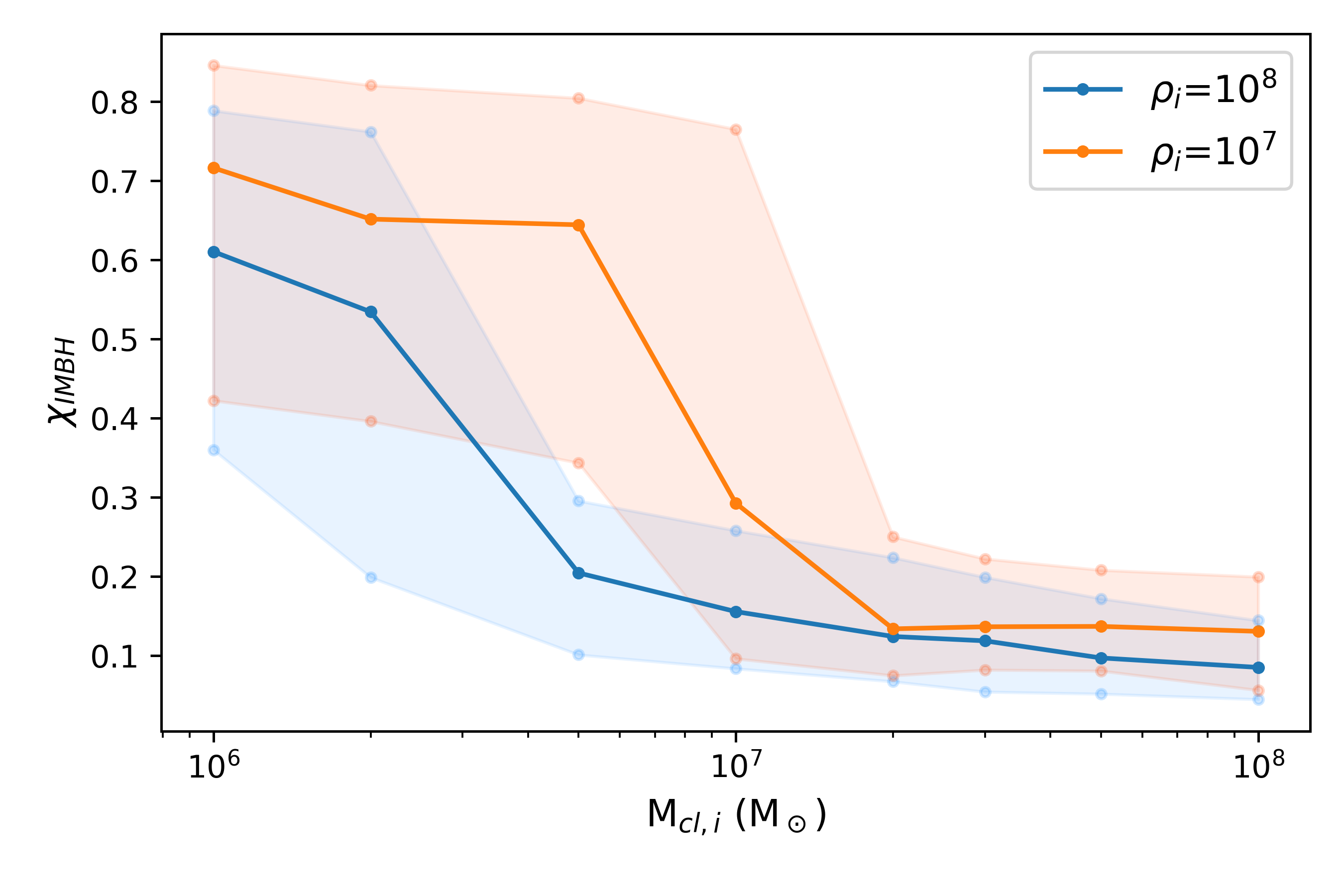}
    \caption{Final spin of the IMBH ($\chi_\mathrm{IMBH}$) vs. host cluster initial mass ($M_{cl,i}$) for two initial density grids of $10^7$\,M$_\odot$\,pc$^{-3}$ (blue) and $10^8$\,M$_\odot$\,pc$^{-3}$ (orange). The solid lines denote 50$^\mathrm{th}$ percentile and the shaded region depicts 10$^\mathrm{th}$-90$^\mathrm{th}$ percentile region, since each model has 100 realizations.}
    \label{fig:Mimbhspin}
\end{figure}

It is of no surprise that the spin of the IMBH, at the end of the cluster simulation ($13.5$\,Gyrs; a Hubble Time), shows an inverse correlation to the host cluster's initial mass, as depicted in Fig.~\ref{fig:Mimbhspin}. Low-mass host clusters only have a couple of generations of  hierarchical mergers, with the remnant spin reaching around $0.7$, as discussed in Sec.~\ref{sec:spin}. It is only in massive clusters with over $100$ mergers (column ``n$^\mathrm{th}$ G'' of Table~\ref{tab:IMBHmass}) that the repeated  low mass ratio mergers cause the spin of the IMBH to become lower. This also translates as the higher-mass IMBHs have lower spins than their less-massive counterparts (columns ``$M_\mathrm{IMBH}^{50}$'' and ``$\chi_\mathrm{IMBH}^{50}$'' of Table~\ref{tab:IMBHmass}). Although our suite of models never reach the SMBH threshold mass of 10$^5$\,M$_\odot$, with our maximum IMBH measuring $\mathcal{O}$(4)\,M$_\odot$, it is intriguing that X-ray astronomy has shown SMBH spins to have a slight anti-correlation to their masses \citep{Reynolds:2013rva,Reynolds:2021}, although the spins of the SMBHs are mostly very high $\gtrsim0.5$ \citep{Piotrovich2022P} \citep[however, there may be bias towards observing those with high spin][]{Bonson2016}. There are occasional studies showing that SMBHs may have a retrograde spin effect \citep[e.g.][]{Wang2019}---where there is a possibility of the BH spin being lowered through anti-alignment with the accretion disc ---there lacks conclusive observational evidence of retrograde spin \citep{Garofalo2013} in SMBHs \citep{Reynolds:2013rva}.

We acknowledge that the (likely most significant) impact of gas and accretion onto the massive BHs spin is not taken into account in our study. However, if we make the assumption that the observed SMBHs at the centre of the galaxy has formed through --- a) in-situ hierarchical mergers in the galaxy's nuclear cluster (note that the 10$^3$\,M$_\odot$ IMBH seed is formed within the first or a couple of Gyrs of cluster evolution) and b) gas accretion (we ignore the possibility of IMBH mass growth through infalling of massive globular clusters and/or inter-galactic mergers), it would appear as if the SMBHs still followed the intrinsic mass-spin distribution obtained through hierarchical mergers.

It is also interesting to note that Sagittarius A$^{\ast}$ of the Milky Way ($\mathcal{O}(6)$\,M$_\odot$) is estimated to have a spin $\lesssim0.1$ \citep{Fragione:2020khu}, and our Fiducial model, where the cluster mass and density after $13.5$\,Gyr evolution is similar to that of the Milky-Way nuclear cluster as the current time, has an IMBH of$\chi_\mathrm{IMBH}^{50}=0.13$. M87, with its very high mass ($\mathcal{O}(9)$\,M$_\odot$), also has a lower spin measurement of $\approx0.2-0.3$ \citep{Nokhrina2019}. We reiterate here that we are not drawing any conclusions with regards to SMBH mass-spin correlation here, as our study is constrained to much lower masses. Instead, we draw attention to this phenomenon and highlight the opportunity for future studies focusing on the evolution of SMBH from IMBHs through hierarchical mergers and accretion.

\subsubsection{Metallicity}
\label{sec:metallicity}

The metallicity (Z) impacts the mass of the cluster IMBH through determining the width of the initial BH mass spectrum. 
While the highest metallicity model Z\_100 with $Z=0.0158$ (this metallicity value is approximately similar to solar metallicity, $Z_\odot\approx0.0142$ \sout{close to solar metallicity}; \citealp{Asplund:2009}) shows a narrow BH mass spectrum, with median BH mass of only about 7.3~M$_\odot$ and maxima of $\approx 30.3$~M$_\odot$, the Fiducial model of the lowest $Z=0.00158$ has a much wider spectrum with median around $17.7$~M$_\odot$ and a maximum of $\approx 42.5$~M$_\odot$, owing to diminished stellar winds at lower metallicities \citep{Vink:2001, Belczynski:2010}. Mass segregation and hence cluster core-collapse time is shortened at lower metallicities due to the broader mass spectrum, resulting in a more rapid onset of massive DBH mergers that quickly build up a hierarchically-formed IMBH. Table~\ref{tab:IMBHmass} shows that the Fiducial model takes $t_\mathrm{1000}\approx0.6$\,Gyr to form an $1000$\,M$_\odot$ IMBH, compared to Z\_100, which takes only $t_\mathrm{1000}\approx1$\,Gyr---the metal-rich cluster takes about $40\%$ more time to create an IMBH of the same size. 
The dearth of initial BHs of higher masses to form the first binaries (and yield subsequently more massive systems through hierarchical mergers), combined with the slower pace of hierarchical mergers, causes the Z\_100 model to have a median IMBH mass of $\sim6.5\times10^3$\,M$_\odot$, only $0.6$ times that of the Fiducial model.

\subsubsection{Delayed supernovae prescription}

The ``delayed'' prescription of \cite{Fryer:2012} is used in the SN\_D model. Ordinarily, the ``rapid'' prescription of \cite{Fryer:2012} enforces a mass-gap between 2-5\,M$_\odot$ between neutron stars and BHs, while the ``delayed'' model allows for BHs of masses between $\approx$2.5-5\,M$_\odot$ as well. The effect of metallicity through stellar winds is the most important parameter in determining BH masses, and though the slightly more numerous lower mass BHs in SN\_D receive higher natal kicks (due to the dependence on fallback mass), the mass function of SN\_D and Fiducial are not significantly different. However, due to the correlation between BH natal kick and fallback mass, about 99.7\% of BHs in SN\_D have $v_\mathrm{kick}<400$\,km\,s$^{-1}$, compared to 96.8\% in Fiducial. Due to having a little more number of massive BHs retained initially, the IMBH mass in the SN\_D model is increased very slightly, only by a few $100$\,M$_\odot$.

\subsubsection{BH seed}
\label{sec:seed_models}

\begin{figure}
\centering
\includegraphics[width=0.9\columnwidth]{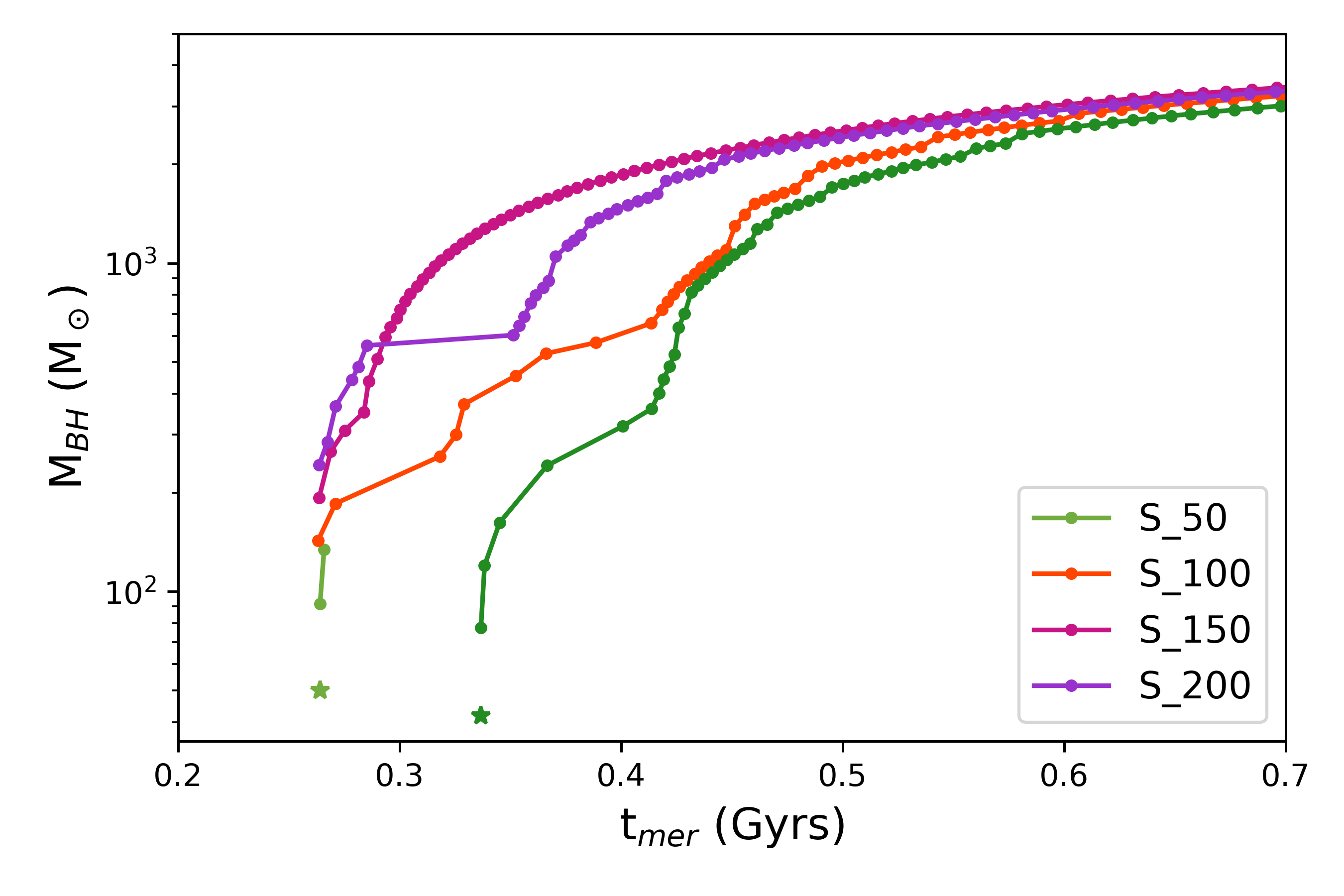}
\includegraphics[width=0.9\columnwidth]{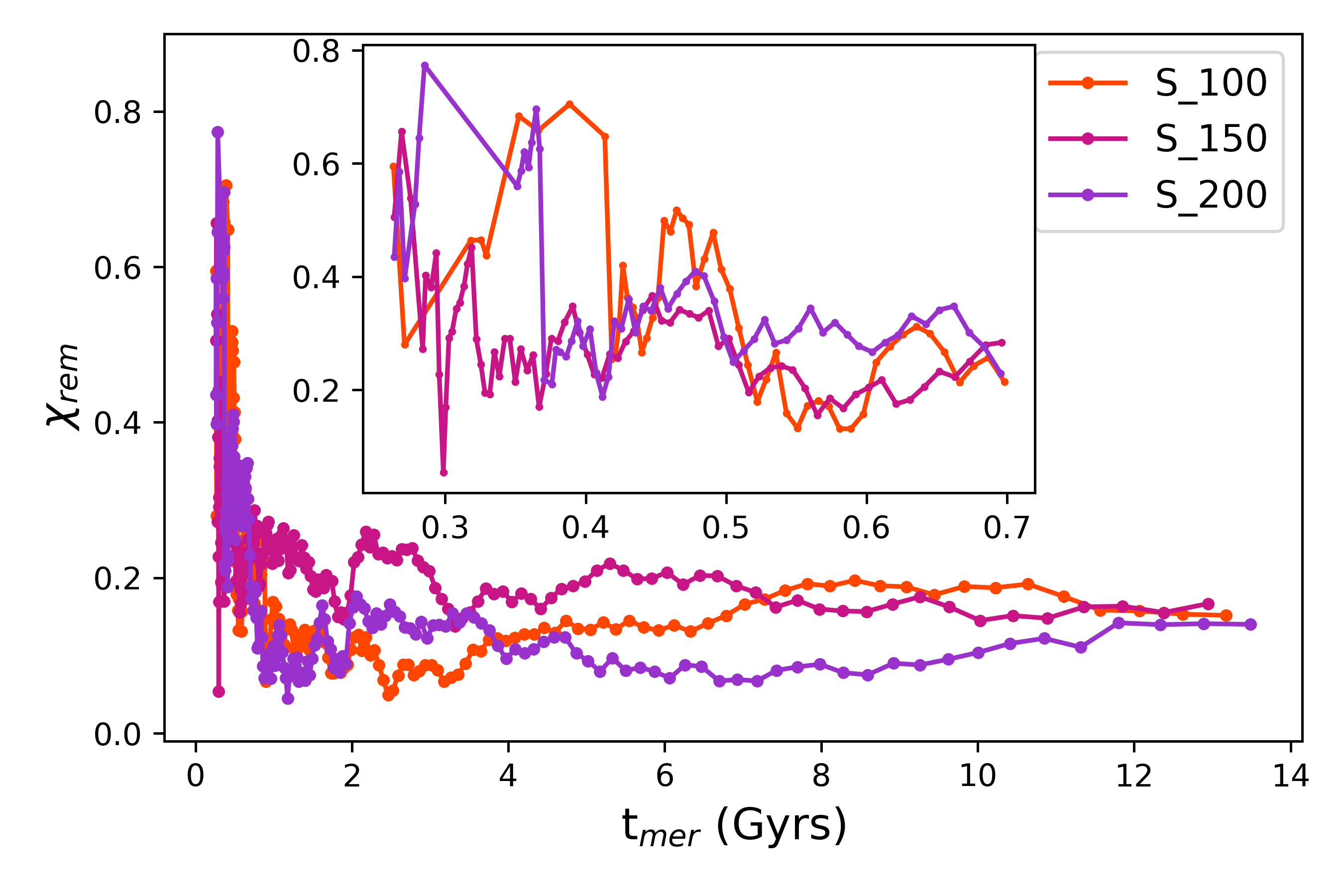}
    \caption{Hierarchical growth of BH seed in models Sd\_50 (green), Sd\_100 (orange), Sd\_150 (magenta) and Sd\_200 (purple). Out of $100$ realizations of each model, only one is chosen per model for illustrative purposes. 
    The upper panel shows the mass growth of the IMBH, where light green depicts the $50$\,M$_\odot$ seed (star mark) and its two consecutive mergers, and dark green shows the slightly lower-mass stellar-remnant BH of the Sd\_50 model, which becomes the most massive IMBH. The lower panel displays the evolution of remnant spin $\chi_\mathrm{rem}$ of models Sd\_100, Sd\_150 and Sd\_200. 
    } 
    \label{fig:seed}
\end{figure}

The effect of adding an initial seed BH of masses beyond that of the stellar evolution prescription is explored through models Sd\_50, Sd\_100, Sd\_150, Sd\_200, where we include at cluster initialisation BH of $50$\,M$_\odot$, $100$\,M$_\odot$, $150$\,M$_\odot$ and $200$\,M$_\odot$ respectively. We term these additional BHs as `seeds'' since they are not directly produced through stellar evolution. Massive stars with Helium cores above $\approx50$\,M$_\odot$ are expected to produce a remnant $\approx40$\,M$_\odot$ and those with Helium cores larger than $60$\,M$_\odot$ are expected to fully disrupt due to thermonuclear eruptions and leave behind no remnant \citep{Woosley:2016hmi, Belczynski2016A, Spera2017, Farmer:2019jed}. This apparent gap in the BH mass spectrum (created through stellar evolution) is often termed as the (pulsational) pair instability or (P)PISN mass gap. It should be remembered with caution that the exact location of the (P)PISN gap in the BH mass spectrum is uncertain \citep[e.g.,][]{Farmer2019, Belczynski:2020:0521, Sakstein2020PhRvL, Woosley2021, Vink2021MNRAS, Spera2022Galax}, and hence a $50$\,M$_\odot$ seed BH may even be a stellar evolution remnant that evolved under special circumstances of, say, high stellar rotation \citep{Marchant:2020haw}, or the unlikely event of suppressed stellar winds at high metallicity \citep{Belczynski:2019fed}. Moreover, very massive stars (initial mass $>$200-1000\,M$_\odot$), can directly collapse to form IMBHs \citep{Belkus2007, Yungelson2008, Sabhahit2023}. Clusters with high primordial binary fractions may also have runaway stellar collisions and very efficient mass accretion from companions, creating stars as massive as $\gtrsim200-600$\,M$_\odot$ which easily form IMBHs through direct collapse \citep{DiCarlo2021,Gonzalez2021ApJ}.  

The hierarchical evolution of the initial seeds of models Sd\_50, Sd\_100, Sd\_150, and Sd\_200 are compared in the upper panel of Fig.~\ref{fig:seed}. While for the last three models with seed mass $\geq 100$\,M$_\odot$, the BH seeds themselves develop to become the final IMBH of the cluster, this may not be the case for Sd\_50. 
The $50$\,M$_\odot$ seed grows through merger to enter its second generation merger as a 91\,M$_\odot$ BH, which merges with a $42$\,M$_\odot$ BH and receives a recoil kick of $\approx950$\,km\,s$^{-1}$ (almost double the cluster escape velocity) and is hence ejected.
Meanwhile, an originally $41.8$\,M$_\odot$ stellar-origin BH grows to become the $\approx9\times10^3$\,M$_\odot$ IMBH in this model. However, in other realisations, the $50$\,M$\odot$ seed survives, and so while $M_\mathrm{IMBH}^{50}$ and $M_\mathrm{IMBH}^{10}$ of the Sd\_50 and Fiducial models are  similar, $M_\mathrm{IMBH}^{90}$ of the Sd\_50 model is a few $100$\,M$_\odot$ more massive. 

There are also instances of the more massive seeds getting ejected after a few generations. Ordinarily, in a non-seeded model such as the Fiducial model, the mass ratio $q$ will be around $1$ for the first generation and then gradually become lower through hierarchical mergers. For seeded models, specifically Sd\_150 and Sd\_200, the starting point of $q$ is significantly lower, around $0.3$ and $0.2$ respectively. Fig.~\ref{fig:vreq} shows that $v_\mathrm{rec}$ actually increases between $q=0.2$ and $0.5$, which may cause more second- or third- generation mergers of seeded models to be ejected compared to the Fiducial model where $q$ may be sufficiently higher in second- and third-generation mergers. Fig.~\ref{fig:vreq} also illustrates that the median and $90^\mathrm{th}$ percentile values for non-zero BH component spin magnitudes peaks around $q=0.4-0.6$ and remains nearly constant. 
For the seeded models, at the time of the first few hierarchical mergers, $v_\mathrm{esc}\approx 400$\,km\,s$^{-1}$. If the seed has undergone a merger or two, the remnant obtains a spin magnitude $\chi_1\approx0.5-0.7$ (see Fig.~\ref{fig:seed}, lower panel), while its non-merger remnant companion has spin magnitude $\chi_2\approx0.0$. Looking at the corresponding curves of Fig.~\ref{fig:vreq} we see that the $10^\mathrm{th}$ percentile peak of $v_\mathrm{rec}$ is at $q\approx0.4$, and the $v_\mathrm{rec}$ median reaches about a constant value from $q\gtrsim0.4$. Indeed, this also supports the observation that in models Sd\_150 and Sd\_200, we do obtain the $M_\mathrm{IMBH}^{10}$ smaller than Sd\_50. Indeed, in cluster realizations of the model with the most massive seed, Sd\_200, we do obtain the $M_\mathrm{IMBH}$ slightly less massive than in the Fiducial and Sd\_50 models. 
 
Even a $200$\,M$_\odot$ seed is not completely protected from ejection post-merger. The best way to ensure that the IMBH growth occurs solely through the seed BH, the cluster (with initial $v_\mathrm{esc}\gtrsim400$~--~$500$) must have the seed BH at least about $10\times$ massive than the upper end of its BH initial mass function (i.e. the seed should be $\geq400$\,M$_\odot$ in our case) such that $q\lesssim0.1$ since first merger. This choice restricts $v_\mathrm{rec}$ to its lowest magnitude region in the parameter space.\footnote{Gravitational wave merger recoil kicks are also illustrated in \citealp{LeTiec_spinkick:2009yg} (Fig.1,2) and for eccentric cases in \citealp{Sopuerta_spinkickecc:2007} (Fig.1,2,3), with respect to the symmetric mass ratio $\eta=q/(1+q)^2$. For $\eta=0.2$ (corresponding to $q=0.4$), the recoil kick shows a clear peak.} This case of retention vs non-retention of seed also illustrates the importance of having multiple realizations of each model, as statistical variations can change the fate of the BH seed.

\subsubsection{Host cluster mass evolution}
\label{sec:mlev_models}

Our Fiducial model has mass loss only through stellar evolution and BH recoil ejection (Eq.~(16) of \citealt{Antonini:2019ulv}). The cluster mass for the Fiducial model after a Hubble time is about $10^7$\,M$_\odot$, half of its initial mass.   

We use model Ml\_ev as a variation which allows mass loss due to evaporative expansion \citep{Antonini:2020PhRvD}, but the difference in final cluster mass is negligibly small ($\approx 9.8 \times 10^6$\,M$_\odot$). Ml\_ev has very similar median IMBH mass but the $10^\mathrm{th}$ percentile is a few $100$\,M$_\odot$ less than that of the Fiducial model. 

In the Ml\_0 model, stellar mass loss is stopped, with the only BH ejections due to binary-single encounters and gravitational-wave recoils 
being sources of cluster mass reduction. However, since both Ml\_0 and Fiducial have rather high $v_\mathrm{esc}$ to begin with, IMBH masses are very similar for the two models. A lower mass density modification, represented by the Ml\_0$_\mathrm{M7D5}$ model, does result in more massive IMBHs compared to M7D5. The consequence of no stellar mass loss is also reflected in higher cluster density at Hubble Time, and in slightly shorter seed formation time.

\subsubsection{BH natal kick}
\label{sec:vkick_models}

In view of Fig.~\ref{fig:vkick}, it can be seen that host clusters with escape velocities $\gtrsim400$\,km\,s$^{-1}$ will not be significantly affected by reducing the natal kick magnitude of the BHs; such clusters tend to retain over $90\%$ of the BHs immediately after stellar evolution, making them participate in the cluster dynamics. Consequently, the Vk\_0 model with mass and density settings as Fiducial and initial $v_\mathrm{esc}\approx 450$\,km\,s$^{-1}$ shows very similar IMBH masses as Fiducial (Table~\ref{tab:IMBHmass}). Due to the broadening of the initial BH mass spectrum by keeping the less massive BHs, reducing the BH natal kick only slightly alters $M_\mathrm{IMBH}^{10}$ and $M_\mathrm{IMBH}^{90}$ to lower values.

The effect of no BH natal kick is insignificant even on the cluster model with lower escape velocities, Vk\_0$_\mathrm{M7D5}$, which has the same mass density settings as model M7D5 with $v_\mathrm{esc}\approx150$\,km\,s$^{-1}$. In the intermediate cluster model, Vk\_0$_\mathrm{M7D7}$ with $v_\mathrm{esc}\approx350$\,km\,s$^{-1}$, $M_\mathrm{IMBH}^{50}$ is lowered to about $20\%$ that of M7D7. This is due to the retention of more lower mass BHs and the cluster $v_\mathrm{esc}$ being in the transitional region, as explained in Sec.~\ref{sec:clusters_mass_den} and shown in the lower panel of Fig.~\ref{fig:Mimbh}.
We also run model Vk\_0$_\mathrm{Z_{100}}$ with solar metallicity and zero BH birth kick; its change in the IMBH mass is only marginal compared to Z\_100. 
 
We hence conclude that the BH natal kick prescription is not an important factor in deciding the hierarchical IMBH growth for clusters, as long as cluster $v_\mathrm{esc}$  is either sufficiently high ($\gtrsim400$\,km\,s$^{-1}$) or sufficiently low ($\lesssim150$\,km\,s$^{-1}$) because the lower-mass BHs that are retained get eventually ejected by either merger recoils or binary single encounters. It is only in the intermediate  region of $v_\mathrm{esc}$ (Fig.~\ref{fig:Mimbh}, lower panel) that altering $v_\mathrm{kick}$ may significantly affect $M_\mathrm{IMBH}^{50}$.

\subsubsection{initial BH spin}
\label{sec:sp_models}

\label{subsec:initialSpin}
\begin{figure}
\centering
	\includegraphics[width=0.9\columnwidth]{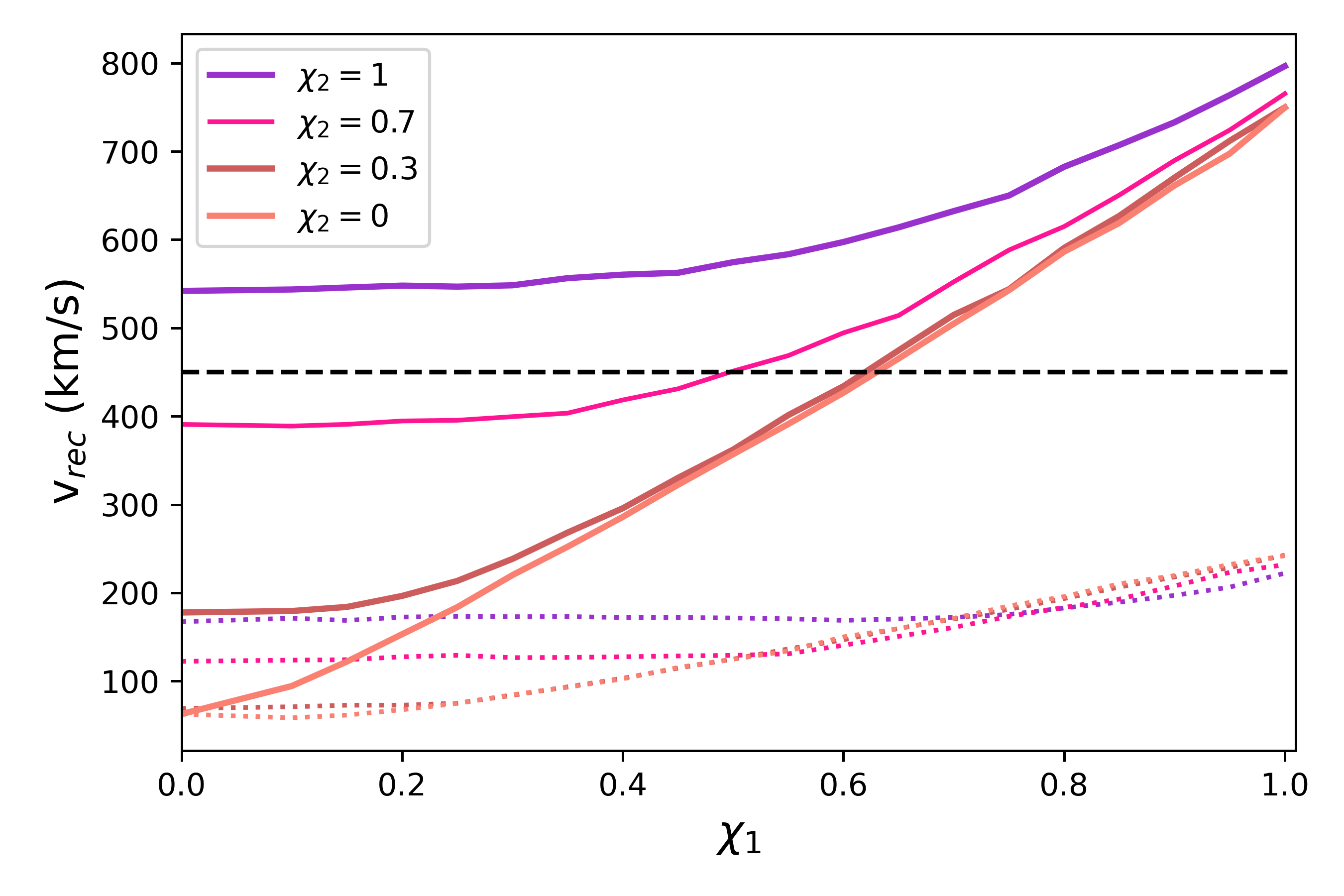}
    \caption{Gravitational wave recoils $v_\mathrm{rec}$ vs primary spin $\chi_1$ for different secondary spin magnitudes $\chi_2=1$ (purple); $0.7$ (pink); $0.3$ (red); $0$ (orange), at a fixed mass ratio $q=0.8$ which roughly estimates the first generation mergers. The solid lines show the median for random orientations of $\chi_{1,2}$, while dotted lines show the $10^\mathrm{th}$ percentiles.
    The black dashed line marks $v_\mathrm{rec}=450$\,km\,s$^{-1}$, approximately the value of the escape velocity during first generation mergers for the Fiducial, Sp\_01, Sp\_33, Sp\_11 and Sp\_LVK models.}
    \label{fig:vrecChi}
\end{figure}

Models Sp\_01, Sp\_33, Sp\_11 and Sp\_LVK explore the effect of initial BH spin on the cluster IMBH mass. While the Fiducial model has initial BH spins set to $0$, Sp\_01, Sp\_33 and Sp\_11 set the initial spin combinations (for primary:secondary) to $0:1$ , $0.3:0.3$ and $1:1$ respectively. Sp\_LVK has its initial BH spins drawn from the spin distribution inferred from current observations of DBH coalescences through gravitational waves.  

We initiate our stellar evolution with only single stars and BH progenitor Helium stars, the latter of which (with their effective angular momentum transfer from core to envelope) are expected to become non-spinning BHs. In binaries, however, the companion of a compact object (neutron star or BH) can get tidally spun up \citep{Qin2018, Bavera2020, Chattopadhyay2021MNRAS, Chattopadhyay2022, Broekgaarden2022spin, Ma2023}. Sp\_01 hence allows the lower-mass secondary component to have higher spin. Although there are conflicting results on the efficacy of tidally spinning up BHs through dynamics \citep{LeTiec2021L,Chia2021}, models Sp\_33 and Sp\_11 can be thought of as intermediate and extremal cases of the effect of initial BH spin on hierarchical mergers.  Given that we start our cluster models with only single stars and expect close dynamical encounters to onset after the core forms, the Fiducial model is the most realistic. 
The variation of $v_\mathrm{rec}$ with respect to the primary spin $\chi_1$ for different secondary spins $\chi_2$ with a fixed mass ratio $q=0.8$ (roughly representing the initial in-cluster mergers) is shown in Fig.~\ref{fig:vrecChi}. We concentrate particularly on the region with $v_\mathrm{rec}<450$\,km\,s$^{-1}$, which approximately corresponds to the $v_\mathrm{esc}$ for the Fiducial and ``Sp'' models at the time of the first DBH mergers. 
At the onset, high-spin BHs easily obtain large kicks, prohibiting the growth of the IMBH and resulting in a suppression of the value of $M_\mathrm{IMBH}^{10}$ of Table~\ref{tab:IMBHmass} (models Sp\_11 and Sp\_LVK have $M_\mathrm{IMBH}^{10}$ about $0.04\times$ and $0.58\times$ of Fiducial). For the model realizations where the initial mergers chance to remain in the cluster, after a couple of mergers with $\chi_1\approx0.7$ now, the $v_\mathrm{rec}$ become very similar to each other for all models, thereby resulting in very similar values of $M_\mathrm{IMBH}^{50}$.
We conclude that the choice of initial spins has a secondary effect on the hierarchical growth  of an IMBH in a star cluster.

\subsubsection{BH ordered pairing}
\label{sec:ordbh_models}

In the Ord\_BH model, we change the (initial BH mass spectrum dependent) power-law probability distribution in pairing the BHs in binaries and triples (as described in Sec.~\ref{sec:methods} and \citealp{Antonini:2022vib}) to complete ordered pairing. In other words, the most massive BH in the Ord\_BH model is paired with the second-most massive one, followed by the third-most massive as the single perturber. 

For a binary of mass $m_{1}+m_2$ and a single of mass $m_3$, the recoil kick of the binary from this binary-single encounter is 

\begin{equation}
\begin{split}
    v_\mathrm{bin} & \sim \sqrt{(1/\epsilon -1)G\frac{{m_{1}m_{2}}}{{m_{1}+m_{2}+m_{3}}}\frac{q_3}{a}} \\
    & = \sqrt{(1/\epsilon -1)G
    \frac{m_{1}m_{2}}{(m_{1}+m_{2})(1+\frac{m_{1}+m_{2}}{m_3})}
    \frac{1}{a}}  ,
    \label{equ:vbin}
\end{split}
\end{equation}
where $q_3=$$m_3$/($m_1$+$m_2$), $a$ is the semi-major axis, $G$ is the gravitational constant, and $(1/\epsilon -1)$ is a function of $q_3$ which is always $\leq0.2$ for all models but the set labelled ``DE'', where it is constant at $0.2$. Equation~\ref{equ:vbin} reveals that an increase in $m_3$ increases $v_\mathrm{bin}$. All other variables remaining constant, With $m_1$, $v_\mathrm{bin}$ reaches a local maximum and then decreases (although the variation is obviously lower than that with respect to $m_3$). The expression for $v_\mathrm{bin}$ is symmetric in $m_1$ and $m_2$.  

In the Fiducial model, the power-law probability distribution ensures the binary-single encounter is composed of massive BHs from the mass spectrum.
Unlike in the Ord\_BH model, there is no guarantee that the three most massive BHs will be the objects engaging in the binary-single encounter. 
This results in substantially decreasing $M_\mathrm{IMBH}^{10}$ in the Ord\_BH model to only about $7\%$ of that of the Fiducial model: in some of the cluster realisations,  the third BH can become massive enough to increase $v_\mathrm{bin}$. 
$M_\mathrm{IMBH}^{50}$ and $M_\mathrm{IMBH}^{90}$ are also reduced significantly in the Ord\_BH model compared to the Fiducial model.

\subsubsection{Tertiary mass dependent energy loss}
\label{sec:m3_DE_models}

The set of models labelled with the ``DE'' moniker contains those with constant tertiary-induced binary energy loss fraction $\Delta\mathrm{E}/\mathrm{E}=0.2$.
In all other models this value is the maximum, valid only in equal-mass interactions. 
One outcome of this is that the IMBH mass is reduced by a few thousand M$_\odot$ in the DE models in comparison to those produced in the Fiducial model. 
Since there is  more energy absorption per interaction in the DE models than the Fiducial model, the IMBH formation timescales ($t_{100}$ and $t_{1000}$) are slightly shorter. 
It takes longer for a binary in the Fiducial model to reach the gravitational-wave driven regime from dynamically-driven regime, since $t_3$ in Equation 20 of \citealt{Antonini:2019ulv} is lowered. 
The energy loss per binary-single interaction, fixed at $20\%$ in the DE models, is lower in the Fiducial model, in the case of $m_3$ << $m_1$.\footnote{A binary of masses `$m_{1,2}$', semi-major axis `$a$' and generating energy `$\dot{E}$' at the cluster core is expected to enter gravitational-wave driven regime when eccentricity $e\gtrsim$ $1.3\Bigl[\frac{G^4(m_1m_2)^2(m_1+m_2)}{c^5\dot{E}}\Bigr]^{1/7}a^{-5/7}$. For DE models, the semi-major axis is larger than that in the Fiducial model due to the perturbing single absorbing more energy (the maximum of $20\%$, valid in other models only for equal-mass systems) from the binary, causing the binary separation to shrink more per interaction. However, $\dot{E}$ is still set by Henon's principle, so only depends on the cluster global properties of mass and density.} 
For a bound period of time (a Hubble time, in our case), the Fiducial model cluster hosts fewer mergers than the DE models, by a factor of $\approx0.8$. If we compare the ratio of mergers with $m_1>100$\,M$_\odot$ to the total number of mergers ($\mathcal{F}_{100}$, listed in Table~\ref{tab:IMBHmass}), the DE models still have about $20\%$ more IMBH-regime (primary mass $\geq100$\,M$_\odot$) mergers than the Fiducial model. 

However, it is clear from Eq.~\eqref{equ:vbin} that the DE models will also have a larger $v_\mathrm{bin}$, as well as larger tertiary kick $v_3$ \citep{Antonini:2022vib}. 
Hence, the DE models eject more binaries (and tertiaries) through binary-single encounters, reducing the mass growth of IMBHs in the long run. 
This lowering of IMBH mass is most strongly affected in the intermediate $v_\mathrm{esc}$ cluster DE$_\mathrm{M7D7}$, where $M_\mathrm{IMBH}^{50}$ is only $12.5\%$ that of M7D7. Even lower-density cluster DE$_\mathrm{M7D5}$ and metal rich DE$_\mathrm{Z\_100}$ have less massive IMBHs compared to those that arise in M7D5 and Z\_100. 
It may therefore be argued that the functional form of $\Delta\mathrm{E}/\mathrm{E}$
which takes into account the mass of the tertiary $m_3$ is indeed an important parameter to be taken into account in the fast codes of rapid cluster evolution models.

\subsubsection{IMBH relative rates}
The fraction of mergers that occur in the IMBH regime, i.e. having a primary mass of $\geq100$\,M$_\odot$, can be expressed as the ratio of the total number of mergers with $m_1\geq120$\,M$_\odot$ to the total number of mergers per cluster; this is shown in the column marked $\mathcal{F}_{100}$ of Table~\ref{tab:IMBHmass}. 
Denser and more massive clusters have a much higher $\mathcal{F}_{100}$, and within our models, it varies largely $9 \times 10^{-5}\lesssim\mathcal{F}_{100}\lesssim0.53$. 
In the Fiducial model, $\sim16\%$ of all mergers have a $\geq100$\,M$_\odot$ primary, while in the M8D8 model $\approx53$\% of all mergers are in the IMBH regime.

\subsection{Ex-situ mergers}
\label{sec:ejectedmer}
\begin{figure}
\centering
	\includegraphics[width=0.9\columnwidth]{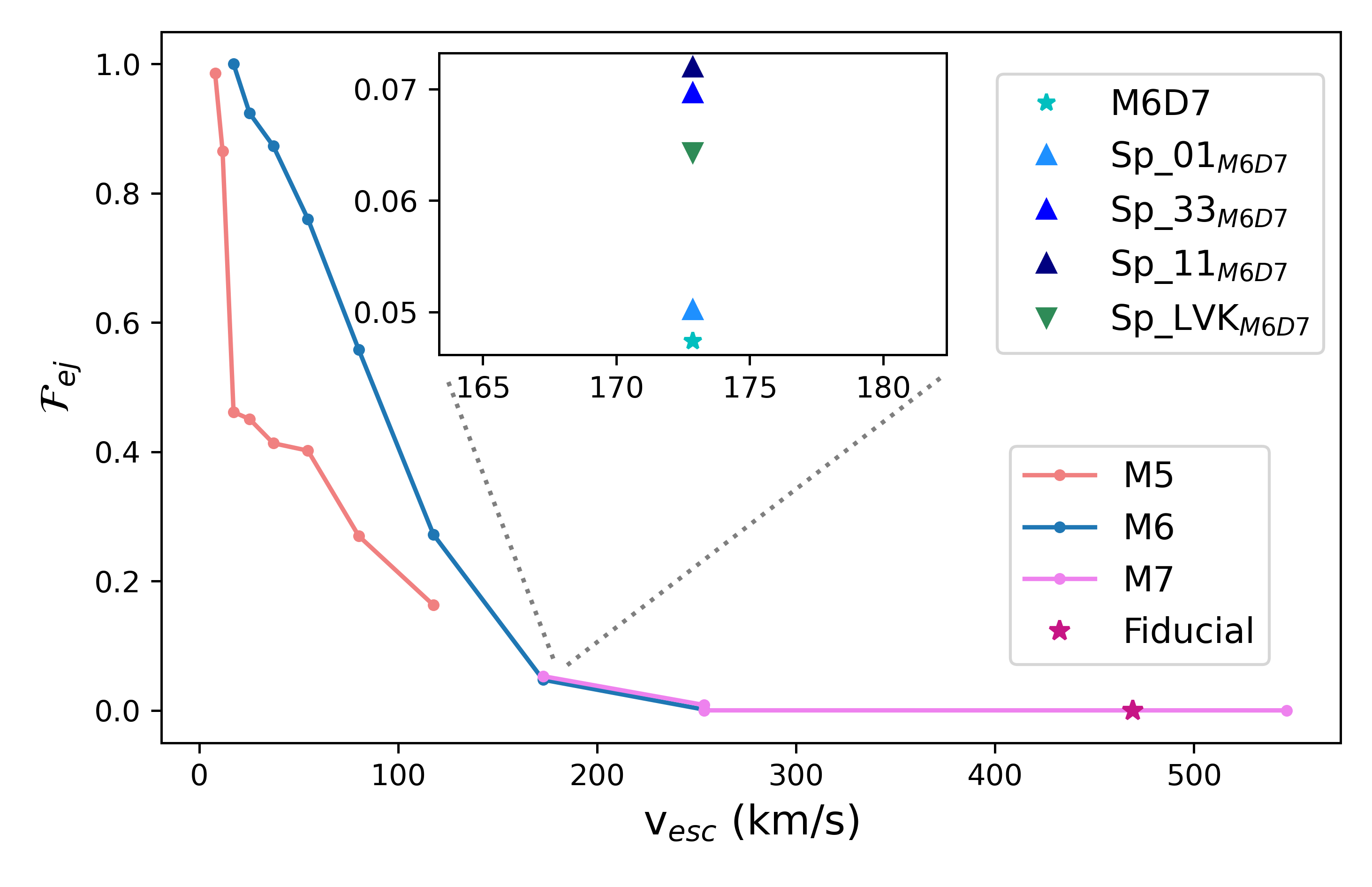}
    \caption{Fraction of ejected binary mergers out of all mergers ($\mathcal{F_\mathrm{ej}}$) with respect to escape velocity for different cluster models. `M5' (coral), `M6' (blue) and `M7' (pink) denote clusters with initial masses $10^5$\,M$_\odot$, $10^6$\,M$_\odot$ and $10^7$\,M$_\odot$ respectively, but with different initial half-mass radii. The Fiducial model is identified with a star. The zoomed-in inset plot shows M6D7 (initial mass and density $10^6$\,M$_\odot$ and $10^7$\,M$_\odot$\,pc$^{-3}$ respectively) with non-spinning initial BHs and clusters with same initial mass-density but initial BH spins of $0:1$ (Sp\_01$_\mathrm{M6D7}$), $0.3:0.3$ (Sp\_33$_\mathrm{M6D7}$), $1:1$ (Sp\_11$_\mathrm{M6D7}$) and the spin distribution inferred by the LVK from gravitational-wave observations (Sp\_LVK$_\mathrm{M6D7}$).}
    \label{fig:ejected}
\end{figure}

\begin{figure*}\centering
    \includegraphics[width=0.33\textwidth]{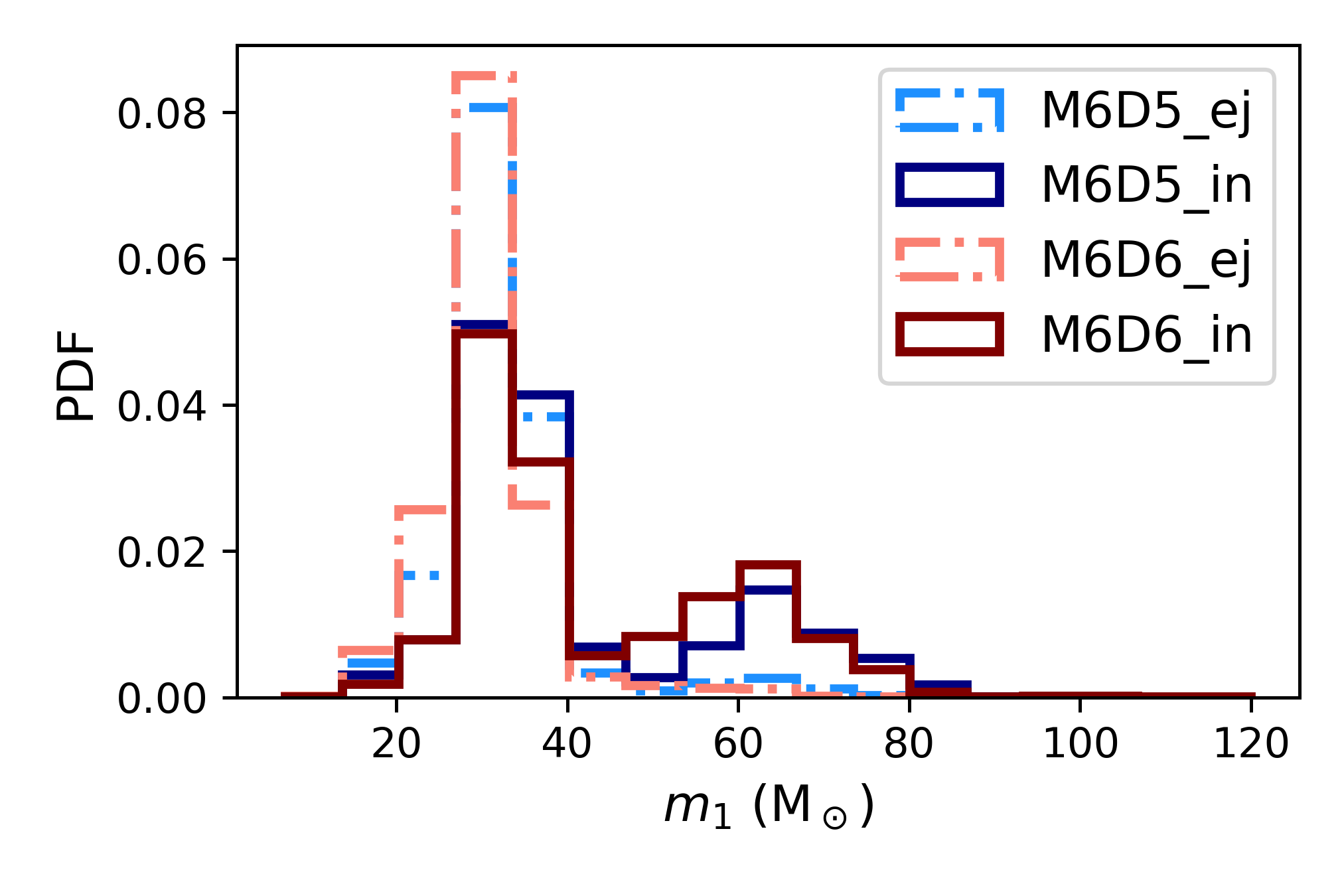}
    \includegraphics[width=0.33\textwidth]{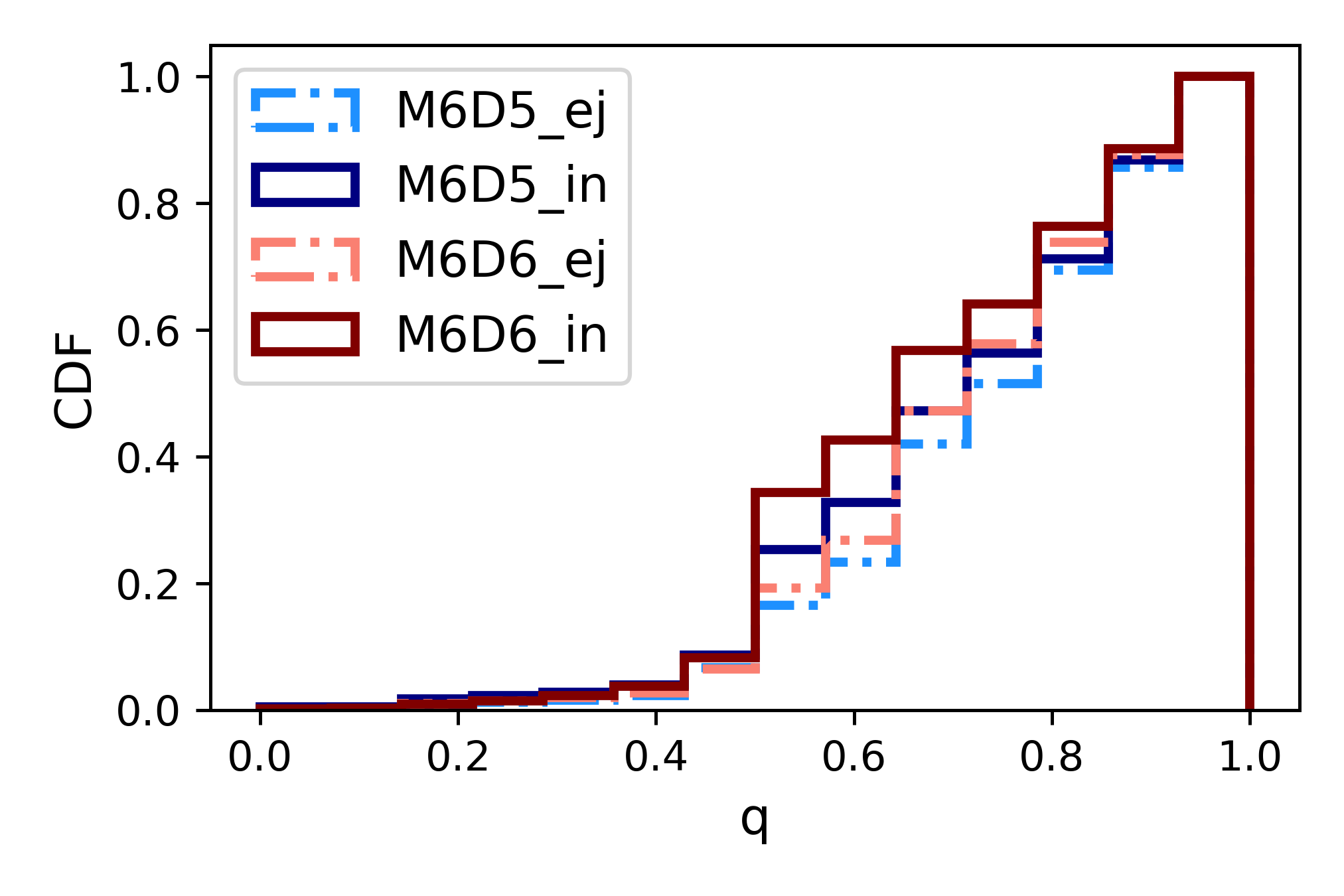}
    \includegraphics[width=0.33\textwidth]{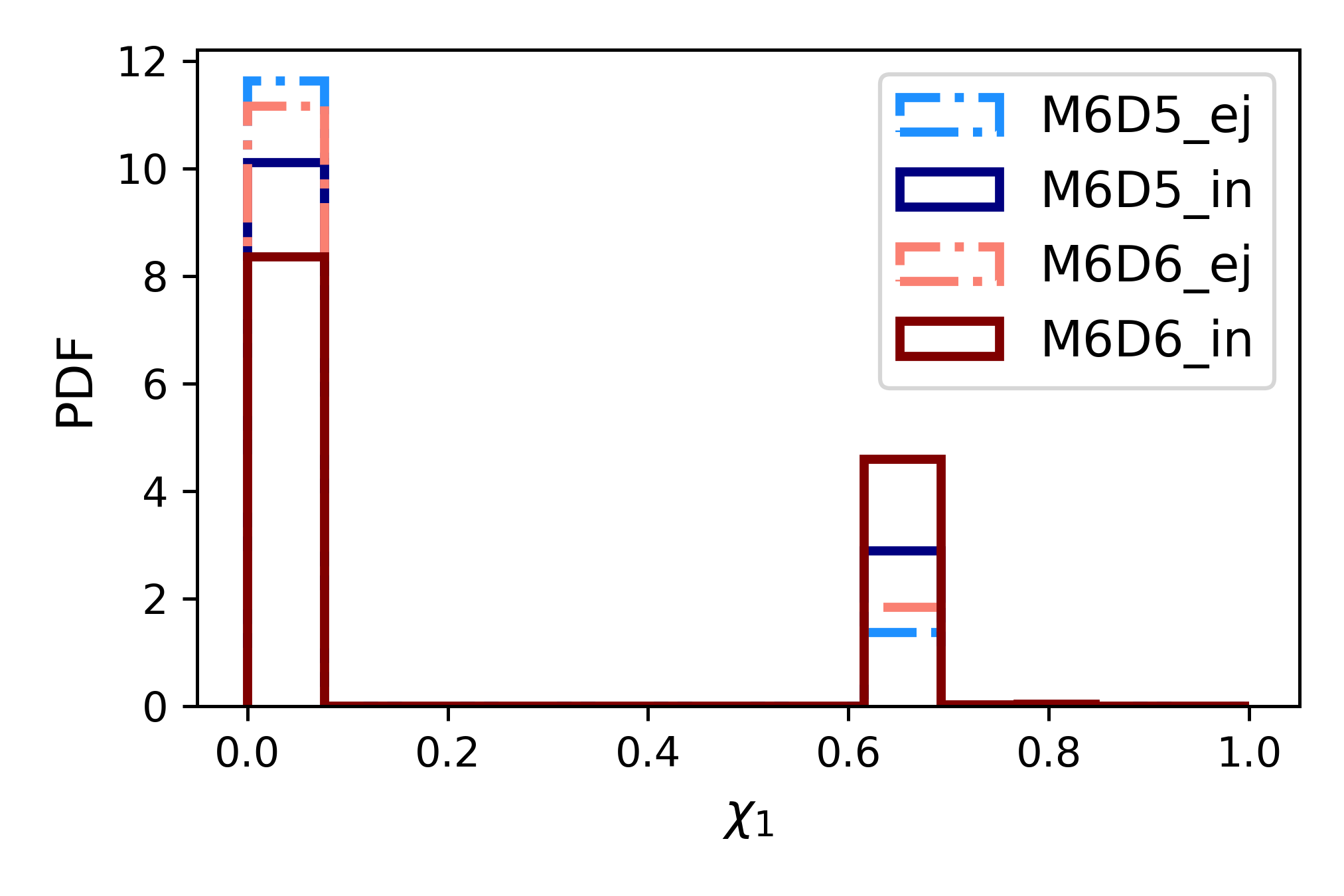}
     \includegraphics[width=0.33\textwidth]{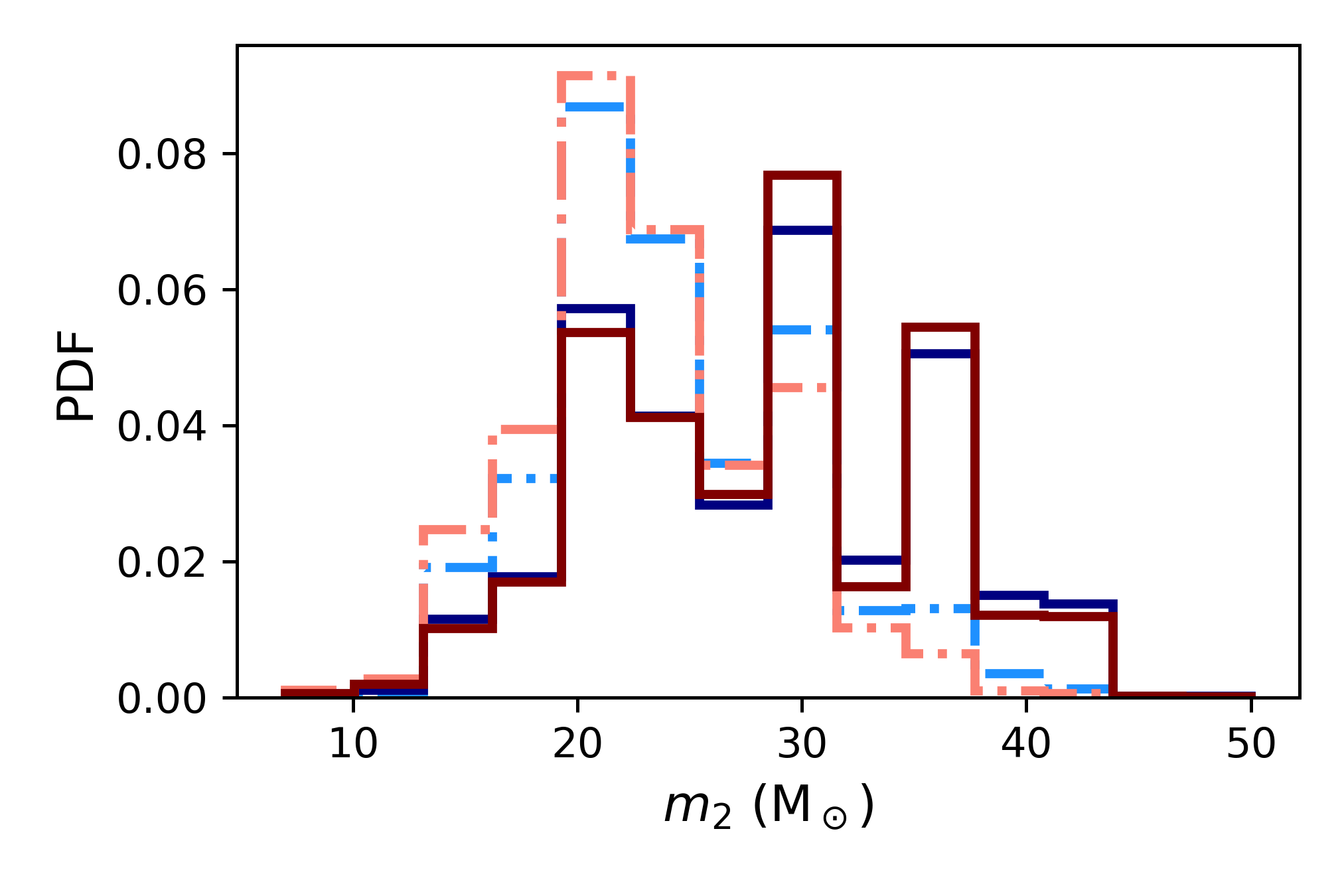}
    \includegraphics[width=0.33\textwidth]{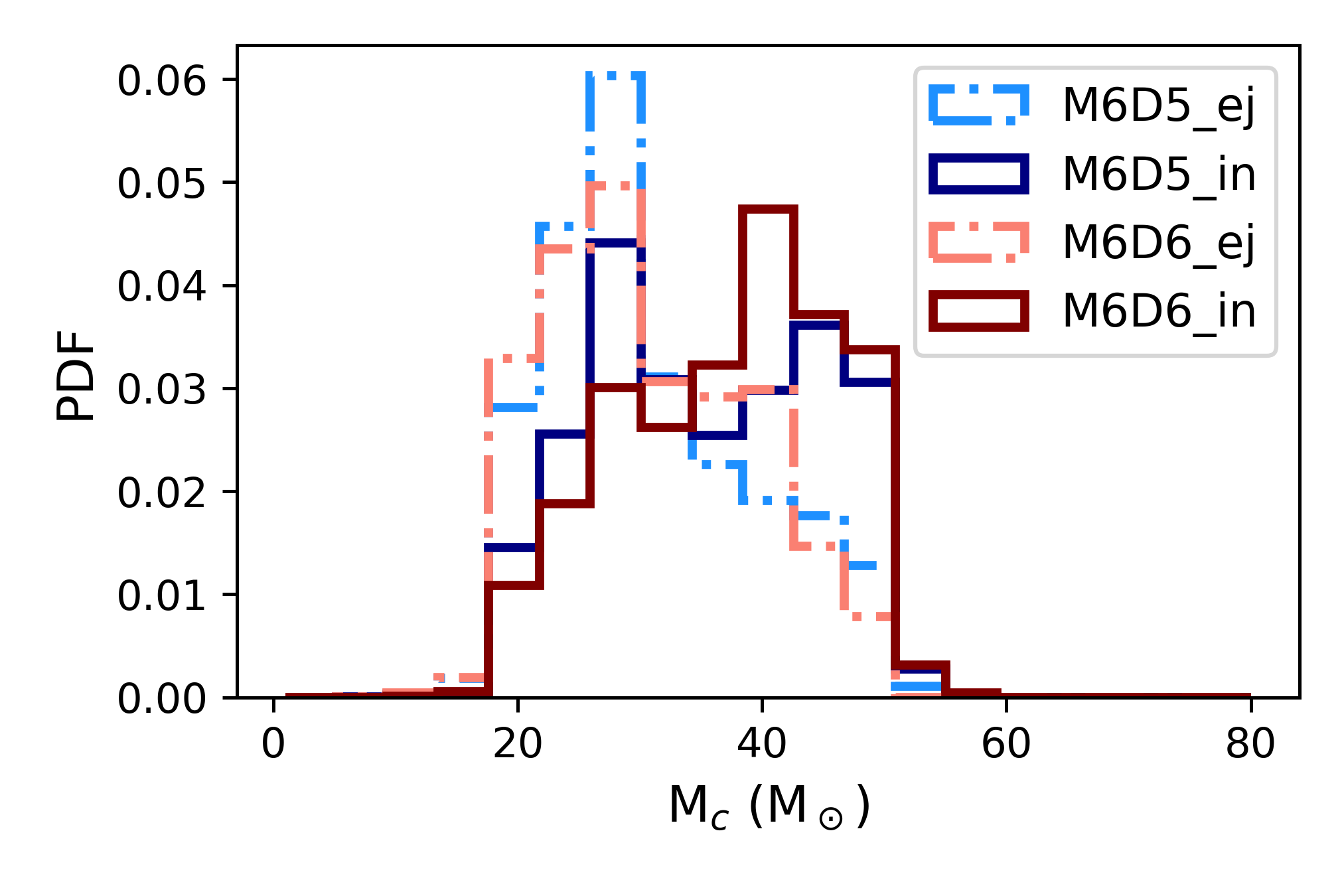}
    \includegraphics[width=0.33\textwidth]{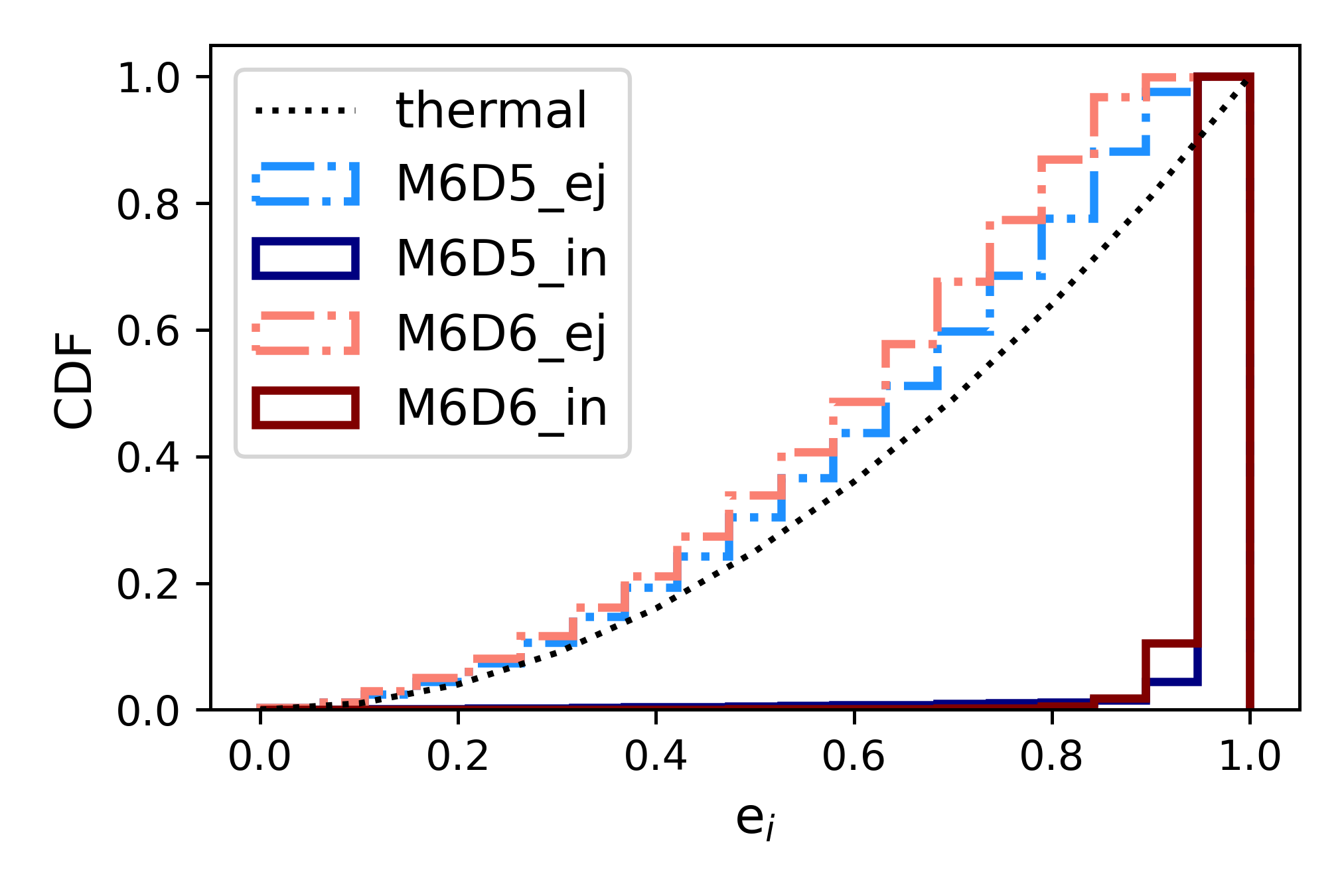}
    \caption{Comparison of DBH parameters in ejected (\_ej) vs in-cluster (\_in) mergers. The left panel shows both the primary (upper plot) and secondary (lower plot) in ejected mergers are less massive, making the chirp mass 
    (M$_{c}$) smaller for ejected DBHs (lower middle panel). Equal mass-ratio $q$ is slightly more preferred in ejected systems (upper middle panel), as are first generation mergers (upper right panel). Away from further dynamical binary-single encounters, the initial (at formation) eccentricity $e_i$ distribution of ejected mergers are marginally more circular than the thermal distribution, while in-cluster DBH mergers typically have much steeper $e_i$. 
    }
   \label{fig:ejgrid}
\end{figure*}

Binaries can get ejected and merge ex-situ, i.e., outside the cluster. 
A  binary is expelled through binary-single interaction when  $v_\mathrm{bin}>v_\mathrm{esc}$, where $v_\mathrm{bin}$ is obtained from Equation~\ref{equ:vbin}. Under the ejection condition, we can expect in the ex-situ mergers less massive binaries, triples with a more massive third body interloper, smaller binary semi-major axes, and lower $\epsilon$. 
In the DE set of models, $(1/\epsilon -1)=0.2$ always (making $\epsilon=0.83$), while in all other models including the Fiducial model, $(1/\epsilon -1)$ is mass-dependent, and is lowered as the mass ratio between the interloper and the binary becomes more asymmetric with the formation of an IMBH. 

The fraction of ejected DBH mergers to in-cluster mergers ($\mathcal{F_\mathrm{ej}}$) for varying initial cluster escape velocity is shown in Fig.~\ref{fig:ejected}. 
Cluster $v_\mathrm{esc}$ plays the key role in deciding $\mathcal{F_\mathrm{ej}}$, with nearly all of the mergers being ex-situ for sufficiently low $v_\mathrm{esc}<10$\,km\,s$^{-1}$. Such clusters, if evolved with direct NBODY models, may still show in-situ mergers due to a high primordial binary fraction and higher multiplicity interactions \citep{Banerjee:2018pmh, Banerjee2021, ChattopadhyayStarCluster2022buz}.

The initial mass and density of the cluster, independently from $v_\mathrm{esc}$, also govern $\mathcal{F_\mathrm{ej}}$. For the same value of initial $v_\mathrm{esc}$, a smaller-mass cluster (hence, higher-density) results in fewer ex-situ mergers, as shown in the M5 and M6 models in Fig.~\ref{fig:ejected}.  

The initial spin distribution of the BHs also determine $\mathcal{F_\mathrm{ej}}$, with high initial spin models having more ejected mergers (see the inset plot of Fig.~\ref{fig:ejected}). 
This is because high-spins translate into high recoil kicks, as explained in Sec.~\ref{subsec:initialSpin}).
This means that $\mathcal{F_\mathrm{ej}}$ increases, even though the number of ex-situ mergers does not vary much between models with different initial spin distributions. 

For the `DE' set of models, $(1/\epsilon -1)=0.2$, unlike in all other models where it is $\leq0.2$, causing more binaries to be ejected out of the cluster in the DE models. 
The effect, however, is small; $\mathcal{F_\mathrm{ej}}$ for models M7D5 and DE$_\mathrm{M7D5}$ (both with initial cluster mass of $10^7$\,M$_\odot$ and density $10^5$\,M$_\odot$\,pc$^{-3}$) are $0.04$ and $0.07$, respectively. 
For M6D5, with initial mass and density $10^6$\,M$_\odot$ and $10^5$\,M$_\odot$\,pc$^{-3}$, the ejected mergers make up a fraction of $0.46$ of all mergers, compared to a constant DE model of same initial mass and density where $\mathcal{F_\mathrm{ej}}=0.52$.
The difference disappears for clusters of high $v_\mathrm{esc}$ because there are fewer ejections. 

We compare the ejected mergers to in-cluster ones in Fig.~\ref{fig:ejgrid}, with models M6D5 and M6D6 (which have ejected merger fractions of about $0.5$ and $0.3$ respectively, with each having the ability to form an IMBH just above $100$\,M$_\odot$). 
The left panel and top middle panel show that  lower-mass binaries and equal mass-ratios are preferred within ejected mergers. 
This results in a smaller chirp mass $M_\mathrm{c}$ for these ex-situ mergers (lower middle panel of Fig.~\ref{fig:ejgrid}). Though most ex-situ mergers are first generation, a fraction of them can also belong to a higher generation, typically a second-generation merger remnant BH merging with a primordial first generation BH, as reflected in the second peak of $\sim 0.8$ in the distribution of primary spin $\chi_1$ (upper right panel of Fig.~\ref{fig:ejgrid}). The eccentricity at the time of ejection/formation ($e_\mathrm{i}$) for ex-situ/in-situ mergers is shown in the lower right panel of Fig.~\ref{fig:ejgrid}. The ex-situ binaries roughly follow a thermal distribution, albeit slightly shifted to lower values, since highly eccentric binaries tend to merge very rapidly. 

\section{Observational implications: detectability and merger rates}
\label{sec:rates}
\begin{figure}
\centering
	\includegraphics[width=0.99\columnwidth]{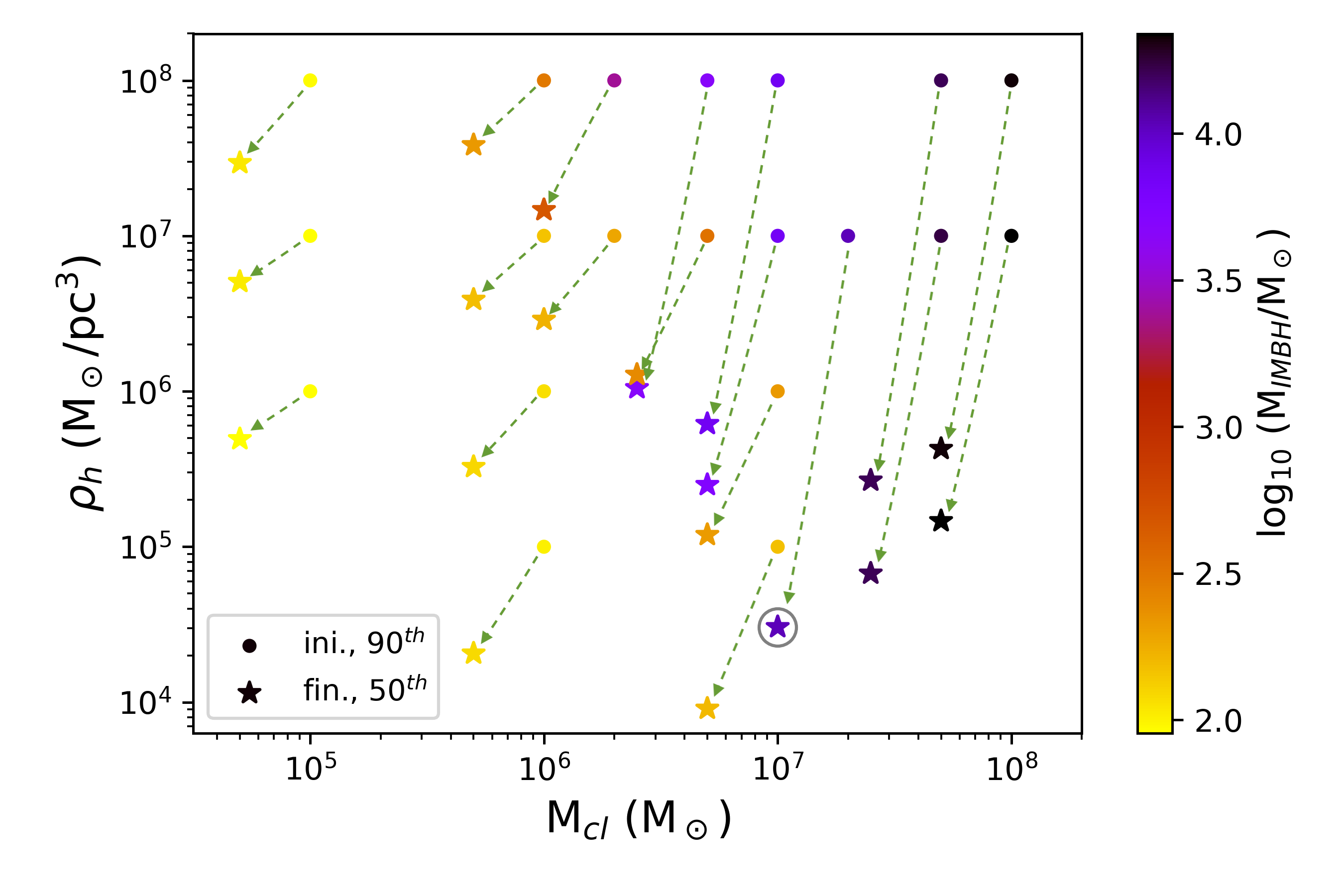}
    \caption{20 cluster models with different initial masses and densities (circles) evolved using $\tt{cBHBd}$ for a Hubble time to their final masses and densities (stars) joined by dotted green arrows. All other settings of these models are the same as the Fiducial model, which is denoted by a circle around the star (for final mass-density values, similar to that of the Milky-way nuclear cluster). The colour bar for the circular points show the $90^\mathrm{th}$ percentile of the IMBH mass that forms, while that of the stars show the $50^\mathrm{th}$ percentile. This is a visual representation of the variance in the upper IMBH mass formed solely through hierarchical mergers.} 
    \label{fig:Mcl_imbh}
\end{figure}

We face a couple of challenges in calculating the merger rates from nuclear clusters: the lack of comprehensive data on the number density evolution of nuclear clusters with redshift (also, cluster birth redshift), and the uncertainty in birth parameters (mass, half-mass density and metallicity). We therefore make a few simplifying assumptions (and then vary some of these as different models), as explained step-by-step below:
\begin{enumerate}
    \item Create $N$ clusters with different values of initial cluster mass ($M_\mathrm{NSC,i}$) and half-mass radius ($R_\mathrm{NSC,i}$). To find unique $M_\mathrm{NSC,i}$ and $R_\mathrm{NSC,i}$ for each individual cluster, there are two models groupings obtained:
    \begin{enumerate}
        \item ModelA group: Host galaxy masses ($M_\mathrm{gal}$) are drawn uniformly from a flat-in-the-log distribution between $10^8$~--~$10^{12}$\,M$_\odot$. Each host galaxy, depending on its mass, has a normalized relative weight $w_\mathrm{N}$ (such that the area under the curve of number density of galaxies per unit volume per unit dex is scaled to unity). $w_\mathrm{N}$ is calculated using the normalized weight associated with each host galaxy of the cluster, and is given by the Schechter best-fit parameters \citep{Schechter:1976} given by \cite{Mortlock2015} and \cite{Song2016}. ModelA$_{1}$ and ModelA$_{2}$ are selected from \cite{Mortlock2015} for redshifts $0.3<z<0.5$ and $2.5<z<3$, while ModelA$_{3}$ is from \cite{Song2016} for $z=5$.\footnote{Schechter function: $\phi(\mathrm{M})=\phi^{\ast}\mathrm{ln(10)}[10^\mathrm{(M-M^\ast)}]^{1+\alpha}\mathrm{exp}[-10]^\mathrm{(M-M^\ast)}$. We have a turn-over mass index in units of dex M$^\ast=10.9;11.04;10.97$ , normalization of Schechter function in logarithm log$\phi^\ast=-2.54;-4.03;-4.28$, slope $\alpha=-1.59;-1.69;-1.70$ for ModelA$_{1,2}$ \citep{Mortlock2015} and ModelA$_{3}$ \citep{Song2016} respectively.}
        These selections are made to represent three regions of the redshift parameter space. Each host galaxy is then associated with a nuclear cluster, whose $M_\mathrm{NSC,i}$ is obtained from fitting the fitting formulae for late-type galaxies through $M_\mathrm{gal}$ and then a $R_\mathrm{NSC,i}$ through $M_\mathrm{NSC,i}$, from the third and first row of Table\,1 of \cite{Georgiev:2016} (with the most likely values of the function). The nuclear clusters are all assumed to have the same metallicity of Z$=1.5\times10^{-3}$, and born uniformly between the redshifts of $0$ to $8$. 
        
        \item ModelB group: Since each fit for late-type galaxies in \cite{Georgiev:2016} (Table\,1)  has error margins, we devise two further sub-models. ModelB$_\mathrm{1}$ (with the upper error margin in obtaining $M_\mathrm{NSC,i}$ and lower error margin in obtaining $R_\mathrm{NSC,i}$, such that the most massive and most dense possible cluster, which is likely to produce more mergers, is built) and ModelB$_\mathrm{2}$ (with the lower error margin in obtaining $M_\mathrm{NSC,i}$ and upper error margin in obtaining $R_\mathrm{NSC,i}$, such that the least massive and most sparse possible cluster is built). Everything else in ModelB$_\mathrm{1,2}$ is identical to ModelA$_\mathrm{1}$.

        \item ModelC group: ModelC$_\mathrm{1}$ and ModelC$_\mathrm{2}$ are identical to ModelB$_\mathrm{1}$ and ModelB$_\mathrm{2}$ respectively, apart from the \cite{Georgiev:2016} best-fit, which for the C group of models is obtained from the sub-sample of nucleated early-type galaxies (see their Table\,1).

        \item ModelD: The James Webb Space Telescope is now confidently detecting galaxies at redshifts at high as $12$ \citep{Castellano2022}. We therefore make another variation to ModelA$_\mathrm{3}$, incorporating up to a volume of redshift $12$ (and making the uniform birth of nuclear clusters between redshifts $0$ to $12$).

        \item ModelE group: The models in the ModelE$_\mathrm{1,2}$ group are the same as ModelA$_\mathrm{1,3}$ respectively, but with metallicities Z$=1.5\times10^{-4}$.

        \item ModelF group: The masses and half-mass radii of the nuclear clusters of corresponding galaxies are obtained from observations of their present-day properties \citep{Georgiev:2016}. While for all of our previous models we stick to an assumption of steady-state, such that these present-day properties are the initial cluster properties of some other galaxies and  the mass-radius relations of the nuclear clusters are identical at each time evolution snap-shot, this assumption is likely incorrect. We therefore create two sets of nuclear clusters with all properties identical to ModelA$_\mathrm{1}$ but with initial radii of $1$\,pc for one set of clusters (ModelF$_\mathrm{1}$) and  initial radii reduced from the median \cite{Georgiev:2016} fit by $\times0.01$ for the other set (ModelF$_\mathrm{2}$).  

        \item ModelG group: In model ModelG$_\mathrm{1}$ the cluster initial properties are identical to the $228$ clusters selected from \cite{Georgiev:2016}, also utilized in the study by \cite{Antonini2016ApJ}. The clusters are equi-weighted, meaning each is given a weight of $1/228$. In this scenario, we only select the mergers at a merger time cut-off of 1\,Gyr (ModelG$_\mathrm{1}$) and 1\,Gyr post-core collapse (ModelG$_\mathrm{2}$). No cosmological evolution is accounted for in this case. The metallicity remains the same as in ModelA$_\mathrm{1}$.

        \item ModelH group: We explore the effects of a non-uniform \cite{MadauDickinson2014} cluster birth redshift distribution through ModelH$_\mathrm{1}$. ModelH$_\mathrm{2}$ uses a fixed birth redshift of 2, approximately close to the peak of the \cite{MadauDickinson2014}. All other parameters remain exactly identical to   ModelA$_\mathrm{1}$.

Once the batch of clusters is created for different models, we evolve them using $\tt{cBHBd}$ as explained below. 
\end{enumerate}
\item Each cluster is assigned a birth redshift, depending on model type --- either uniform (between 0-8 or 0-12) or \cite{MadauDickinson2014} distribution or fixed at a redshift of 2, irrespective of their initial cluster properties. 
\item Volumetric shells are created, such that they join to enclose a volume with radius with endpoints corresponding to redshift $0$ and $8$. The width of each shell is taken to be $0.2$ as step-size (although our calculation becomes independent of the step-size, as long as $N$ is large enough). 

\item A flat $\Lambda$CDM cosmology with a Hubble constant of $70$\,km\,s$^{-1} \rm Mpc^{-1}$ and $\Omega_\mathrm{o}=0.3$ is assumed.
For each merger in each cluster (at the given birth redshift), the true merger redshift and look-back times are computed. Only the mergers that occur within the lower limit of the grid redshift is taken (i.e., mergers that happen in the future of the cluster are rejected). 
Here, we also calculate the signal-to-noise ratio (SNR) of the gravitational-wave emission of the coalescing binaries for a range of different detectors. 
The SNR is a standard quantification of the detectability of a gravitational-wave signal for a given instrument, and can be calculated for a compact binary as \citep[][]{1994PhRvD..49.2658C}
\begin{equation}
    {\rm SNR}^2=\frac{5}{6}\frac{1}{\pi^{4/3}}\frac{c^2}{r^2}\left(\frac{GM_c}{c^3}\right)^{5/3}|Q(\theta,\phi;\iota)|^2\int_{f_{\rm min}}^{f_{\rm max}}{\rm d}f\,\frac{f^{-7/3}}{S_n(f)},\label{eq:SNR2}
\end{equation}
where $r$ is the luminosity distance to the binary and $M_c=(m_1m_2)^{3/5}/(m_1+m_2)^{1/5}$ its chirp mass. 
The function $Q(\theta,\phi;\iota)$ describes the antenna response of the detector to the cross and plus polarization of the gravitational wave; it  depends on the polar angles $\theta$ and $\phi$ of the binary position on the sky and the inclination $\iota$ of its orbital axis with respect to the line-of-sight. 
In this work, we marginalise $|Q(\theta,\phi;\iota)|^2$ over all angles, which yields $\langle|Q(\theta,\phi;\iota)|^2\rangle=4/25$ for an interferometric detector design \citep[][]{maggiore2008gravitational}. 
The function $S_n(f)$ is the noise power spectral density of a given instrument. 
Here, we use the noise power spectral densities for the currently operating Advanced LIGO (aLIGO) detectors and the planned Cosmic Explorer (CE) and Einstein Telescope (ET) detectors. \footnote{For aLIGO we adopt $S_n(f)$ from released data of the collaboration: \url{https://dcc.ligo.org/LIGO-T1800044/public}, last accessed 16 May 2023. For CE we adopt $S_n(f)$ from released data of the collaboration: \url{https://cosmicexplorer.org/sensitivity.html}, last accessed 16 May 2023. For ET we adopt $S_n(f)$ from released data of the collaboration: \url{https://www.et-gw.eu/index.php/etsensitivities}, last accessed 16 May 2023.(see main text).}

The frequency minimum $f_{\rm min}$ and maximum $f_{\rm max}$ of the integration in Eq.~\eqref{eq:SNR2} depend on the detector. 
Ground-based detectors like aLIGO, CE, and ET are sensitive to relatively high gravitational-wave frequencies, $\sim\mathcal{O}(10^1$~--~$10^3)\,\rm Hz$, which are emitted by binaries during their final orbits before merger. 
Hence, for ground-based detectors we set the $f_{\rm max}=c^3/[6\sqrt{6}\pi G(m_1+m_2)]$, corresponding to the frequency of the binary's Innermost Stable Circular Orbit (ISCO). 
For practical purposes, we can set the lower limit of the integration to $f_{\rm min}=0$, because for the noise power spectral densities of ground-based detectors only frequencies $f\gtrsim10\,\rm Hz$ significantly contribute to the integral in Eq.~\eqref{eq:SNR2}.

If there are ``$\Delta k$'' selected mergers within the interval of $\Delta t_\mathrm{k}$, in a particular ``$N^\mathrm{th}$'' cluster associated normalized weight ``$w_\mathrm{N}$'', the weighted contribution of that cluster in merger rate becomes
    \begin{equation}
       \left(  \frac{\Delta k}{\Delta t_\mathrm{k}} \right) w_\mathrm{N},
    \end{equation}
in units of yr$^{-1}$ (since $w_\mathrm{N}$ is dimensionless).
Averaging the contribution for $N$ cluster for each model set, we get 
    \begin{equation}
       \sum_{N} \left( \frac{\Delta k}{\Delta t_\mathrm{k}} \right) w_\mathrm{N},
    \end{equation}
since $w_\mathrm{N}$ is factorized such that $\sum w_\mathrm{N}=1$ . 
\item If the number density of galaxies is $\rho_\mathrm{s}$\,Gpc$^{-3}$ for the volume grid $v_\mathrm{s}$, we need to sum over ``$s$'' volume grids that gives the total volume $V$, such that the total merger rates $\mathcal{R}$ becomes
    \begin{equation}
      \mathcal{R}= \frac{f_\mathrm{nc}}{V}\sum_{s}^{} \rho_\mathrm{s}v_s\sum_{N}^{} \Biggl\{ \left(  \frac{\Delta k}{\Delta t_\mathrm{k}} \right) w_\mathrm{N} \Biggl\},
    \end{equation}
where $f_\mathrm{nc}$ is the fraction of galaxies in the mass range of $10^8$~--~$10^{12}$\,M$_\odot$ that have a nuclear cluster. 
Observationally, this fraction varies with galaxy mass, with $\lesssim20\%$ for galaxies with mass around $10^6$\,M$_\odot$, to as high as $90\%$ for galaxy mass of $10^9$~--~$10^{10}$\,M$_\odot$. 
However, we simplify the matter by taking 
$f_\mathrm{nc}=0.8$ for all galaxies, which is a rough estimate for Late-type galaxies in our mass range \citep[][Fig\,3]{Neumayer:2020}. 
If a more generalized condition is desired, such that $f_\mathrm{nc}$ becomes a function of the host galaxy mass ($f_\mathrm{nc},N$), this term can be added inside the summation over $N$. 
If $\rho_\mathrm{s}$ is independent of galaxy properties and redshift (and hence constant), the expression can be simplified to 
    \begin{equation}
      \mathcal{R}= f_\mathrm{nc} \rho_\mathrm{s}\sum_{N}^{} \Biggl\{ \left(  \frac{\Delta k}{\Delta t_\mathrm{k}} \right) w_\mathrm{N} \Biggl\}.
    \end{equation}
The choice of $\rho_\mathrm{s}$ is a tricky one. 
\citet{Fletcher1946MNRAS} estimated the  high value of $\rho_\mathrm{s}\approx12$\,Mpc$^{-3}$. 
More recent works have lowered this number significantly \citep{Poggianti2013,Leja2013ApJ,Ownsworth2016}, but with different studies resulting in different estimates for $\rho_\mathrm{s}$, we have taken the upper limit of $\rho_\mathrm{s}\approx0.01$\,Mpc$^{-3}$ \citep[][through the Hubble Space telescope]{Conselice2005}, as used by \cite{Antonini:2018auk}.  \cite{Conselice2016} predicts   $\approx2\times10^{12}$ galaxies within the redshift of $8$ (making $\rho_\mathrm{s}\approx0.001$\,Mpc$^{-3}$), while \cite{Lauer2021} with New Horizons shows the sky to be $10\times$ less bright. JWST data may further alter $\rho_\mathrm{s}$ in the near future.

\item Detectable rates $\mathcal{R}_\mathrm{LVK,CE,ET}$ for aLVK, CE, ET are also calculated in a similar way, but only by counting the mergers with SNR$>8$ correspondingly. 
\end{enumerate}
Finally, We note that our nuclear cluster models 
do not have a central SMBH, while observations show that at least in  some galaxies they coexist \citep[e.g.,][]{2008ApJ...678..116S,2012AdAst2012E..15N}. Thus, our merger rates should be most likely intended as upper limits.

The calculated rates for different models are tabulated in Table~\ref{tab:rates}. We find intrinsic rates between $\mathcal{R}=0.01-0.80$\,Gpc$^{-3}$yr$^{-1}$, while $\mathcal{R}_\mathrm{LVK}\lesssim0.7\mathcal{R}$, $\mathcal{R}_\mathrm{CE}\approx0.8-0.9\mathcal{R}$ and $\mathcal{R}_\mathrm{ET}\gtrsim0.9\mathcal{R}$. 
Lower-redshift, late-type host galaxies appear to have higher merger rates. 
Comparing ModelA$_{1}$ to ModelD$_{1}$, we observe that extending the redshift to $12$ from $8$ does not change $\mathcal{R}$ significantly, since peak of our detectable mergers emerge from $z\approx$1-2, a trend similar to \cite{Fragione2022ApJ}.
The lack of contribution from clusters born at higher redshifts is also apparent in models ModelH$_{1,2}$, where $\mathcal{R}$ remain identical, and negligibly lower than  ModelA$_{1}$.
As the initial cluster densities are made higher in ModelF$_{1,2}$, the rates increases by $5$ to $8$ times. 
The spread in cluster mass-radius is only partially encapsulated in ModelB$_{1,2}$ and ModelC$_{1,2}$, resulting in ModelG$_{1,2}$ with simplistic equi-weight assumptions produce slightly higher $\mathcal{R}$. The upper limit of our merger rate is about an order-of-magnitude lower than the upper limit obtained by \cite{Fragione2022ApJ} (Fig.\,1 showing random initial seed model roughly matches ours), possibly due to their assumption of a pre-existent seed BH.

Qualitatively, while the LVK and ET appear to be able to observe (SNR$>8$) for primary masses up to a few hundred\,M$_\odot$s and $q\sim\mathcal{O}$($10^{-1}$), CE pushes this threshold to about $800$\,M$_\odot$ and $q\sim\mathcal{O}$($10^{-2}$). 

Evolving the $228$ \cite{Georgiev:2016} models under two assumptions of initial conditions---(a) mass and radius are the same as in current observations, and (b) mass is the same as in current observations and radius is $1$\,pc---we find that after a Hubble Time,  in model (a) $\approx6$ of the clusters host IMBHs $>1000$\,M$_\odot$ and $\approx168$ of them host an IMBH of mass $100$ to $400$\,M$_\odot$; in model (b), $\approx20$ of the clusters host IMBHs $>1000$\,M$_\odot$ and $\approx176$ of them host an IMBH of mass $100$ to $400$\,M$_\odot$.

\section{Conclusions}
\label{sec:summary}

In this study we have studied $3,400$ massive clusters within the initial mass range of $10^6-10^8$\,M$_\odot$ and initial density range of $10^5-10^8$\,M$_\odot$\,pc$^{-3}$, which can be considered in the parameter space of  nuclear clusters  and the most massive globular clusters with the updated fast code $\tt{cBHBd}$ that incorporates initial mass function (hence metallicity) dependent probabilistic DBH pairing, binary-single encounters (including mass-dependent energy loss $\Delta\mathrm{E}/\mathrm{E}$).
In reference to the main questions we asked in Sec.~\ref{sec:intro}, we find that ---
\begin{enumerate}
    \item IMBHs ranging from $\mathcal{O}(2)-\mathcal{O}(4)$\,M$_\odot$ can be created solely through in-cluster hierarchical mergers (Fig.~\ref{fig:Mcl_imbh}). This mass range is roughly one or two magnitudes lower than the least massive SMBHs. To reach to the mass range of SMBHs from the hierarchically-created IMBHs, there must therefore be subsequent mass accretion through other processes. 
    \item The initial cluster escape velocity is the most important parameter in determining IMBH formation. The corresponding values of cluster's mass and density determine the final mass of the IMBH. For $v_\mathrm{esc}\gtrsim400$\,km\,s$^{-1}$, an  IMBH with mass up to $\mathcal{O}(4)$\,M$_\odot$ can form for sufficiently high cluster masses and densities (see the lower panel of Fig.~\ref{fig:Mimbh}) . This cut-off escape velocity results from a combination of increased BH retention post-natal kick (Fig.~\ref{fig:vkick}) and gravitational wave recoils, which average to around $400$\,km\,s$^{-1}$ for all spin magnitudes and orientations (see Fig.~\ref{fig:vreq}).   Other secondary factors that play a role in determining the mass of the IMBH are cluster metallicity (which alters the width of the initial mass function), DBH pairing prescription ($\alpha$ parameter, see Sec.~\ref{sec:methods}) and the mass-dependent functional form of $\Delta\mathrm{E}/\mathrm{E}$ (see Sec.~\ref{sec:methods}).   
On the other hand, the initial BH spin distribution, unless extremely high (i.e. $\chi_{1,2}=1;1$), does not significantly affect the final IMBH mass nor its spin.

        \item  An in initial BH seed of at least $10\times$ the most massive (stellar-evolution originated) in-cluster initial BH is required to maximise the probability of the seed BH's retention after a few hierarchical mergers (Sec.~\ref{sec:seed_models}), which can be $\sim200$\,M$_\odot$ for metal-rich clusters to $\sim400$\,M$_\odot$ for metal-poor clusters.  
  
        \item The spin of the final IMBHs shows a double peak distribution and a clear mass dependence (upper right panel of Fig.~\ref{fig:ejgrid}; and Fig.~\ref{fig:Mimbhspin}). For masses $\gtrsim 10^3\rm M_\odot$ the IMBH spin is $\chi_\mathrm{IMBH}\sim 0.15$, while for lower masses the spin distribution peaks at $\sim 0.7$.     
 
        \item $\approx6$~--~$20\%$ of all in-cluster mergers in our set of models are expected to have an eccentricity $e\geq0.1$ at $10$\,Hz. We find that eccentric mergers are particularly favoured in equal-mass binaries. This means that metal-rich and younger (age $\sim0.5$~--~$1.5$\,Gyrs) clusters, which do not form IMBHs of $>100$\,M$_\odot$, are ideal formation grounds for eccentric mergers that may be detected by the current generation of gravitational-wave detectors. 
       
        About $2$~--~$9\%$ of all in-cluster mergers are also expected to be formed at a frequency greater than $10$\,Hz (this is a subset of the eccentric merger fraction). Such extreme cases may appear in the LVK burst searches (see Fig.~\ref{fig:eccDist}).

        \item The number of mergers involving an IMBH (i.e. $\geq100$\,M$_\odot$) as a fraction of the total number of mergers is expressed by $\mathcal{F}_{100}$. Very dense clusters ($\rho_{i}=10^8$\,M$_\odot$\,pc$^{-3}$) that rapidly form a BH of $100$\,M$_\odot$ can have $0.09\lesssim\mathcal{F}_{100}\lesssim0.53$. For the Fiducial model, $\mathcal{F}_{100}\approx0.16$ (Table~\ref{tab:IMBHmass}).

        \item The fraction of ejected mergers, $\mathcal{F}_\mathrm{ej}$, is a function of the cluster mass and escape velocity, with $\approx20$~--~$100\%$ of the DBH mergers being ejected mergers in clusters with $v_\mathrm{esc}<100$\,km\,s$^{-1}$. However, $\mathcal{F}_\mathrm{ej}$ becomes negligible when $v_\mathrm{esc}>200$\,km\,s$^{-1}$ (see Sec.~\ref{sec:ejectedmer} and Fig.~\ref{fig:ejected}). Ex-situ mergers are typically less massive, more symmetric in mass, and have a longer delay time than their in-situ counterparts.

    \item The rates of DBH mergers from nuclear clusters can be rounded to $\mathcal{R}\lessapprox0.01$~--~$1$ Gpc$^{-3}$yr$^{-1}$ remembering it to be the upper limit due to not including an SMBH in our calculations. The orders-of-magnitude uncertainty on the rates arises predominantly due to the uncertain number density distribution of nucleated galaxies (we assume $\rho_\mathrm{s}=0.01$\,Mpc$^{-3}$ and $f_\mathrm{nc}=0.8$). Uncertainties in the  initial nuclear cluster mass-density distribution, the nuclear cluster mass scaling relation with respect to host galaxy mass, and the metallicity distribution of nuclear clusters have lower impact on $\mathcal{R}$. 
    
    \item At SNR$>8$, we expect ET and CE to detect about $80\%$ and at least $90\%$ respectively
    of the intrinsic DBH mergers from nuclear clusters.
    Out of the $228$ nuclear clusters with well-measured masses and radii \citep{Georgiev:2016}, we predict that up to $80\%$ of the clusters host hierarchically-formed IMBHs with masses $\lesssim400$\,M$_\odot$, and 
    that up to $20$ host IMBHs with masses $>1000$\,M$_\odot$.
    We also highlight that, while the current generation of gravitational-wave detectors can only observe IMBHs of up to a few hundred M$_\odot$, future detectors such as CE will have improved lower-frequency sensitivity, enabling detection of more massive ($\sim500$~--~$800$\,M$_\odot$) IMBHs.
    
    \end{enumerate}

Future improvements to the study presented here will include binary-binary interactions, BH mergers with other objects (e.g., neutron stars, white dwarfs, non-compact objects), both globular and nuclear clusters, mass gain through infalling globular clusters, and the wet component of gas accretion for nuclear clusters \citep{Guillard2016,Bourne2016}.

\begin{table*}
\caption{Results from {\tt cBHBd} cluster models, showing most massive IMBH properties --- mass ($M_\mathrm{IMBH}^{50,10,90}$), spin ($\chi_\mathrm{IMBH}^{50}$), generation(n$^\mathrm{th}$ G), cluster density at a Hubble time ($\rho_\mathrm{h}$/10$^5$), median formation timescales for 100\,M$_\odot$ and 1000\,M$_\odot$ IMBHs ($t_{100\,M_\odot}$ and $t_{1000\,M_\odot}$), eccentric DBH merger fraction ($\mathcal{F}_\mathrm{ecc}$), high-frequency burst merger fraction ($\mathcal{F}_\mathrm{freq}$) and fraction of DBH mergers with at-least one component $\geq$100\,M$_\odot$ ($\mathcal{F}_\mathrm{100}$).}

\centering
\begin{tabular}{llrrrrrrrrrrr}

\hline
Sl. & Model & $M_\mathrm{IMBH}^{50}$ & $M_\mathrm{IMBH}^{10}$ & $M_\mathrm{IMBH}^{90}$ & $\chi_\mathrm{IMBH}^{50}$ &  n$^\mathrm{th}$ G &   $\rho_\mathrm{h}$/10$^5$ &  $t_{100\,M_\odot}$ &  $t_{1000\,M_\odot}$ &
 $\mathcal{F}_\mathrm{ecc}$ &  $\mathcal{F}_\mathrm{freq}$ &
 $\mathcal{F}_\mathrm{100}$
\\ 

no. &  & (M$_\odot$) & (M$_\odot$) & (M$_\odot$) &  & (IMBH) &  (M$_\odot$\,pc$^{-3}$) & (Gyr) & (Gyr) &
 &  &  \\ 

\hline

 &  &  &  &  &  &   & & &  & &\\
 1. & Fiducial &  10562 & 9451 & 11742 & 0.134 & 121$^{+24}_{-20}$ & 0.31 & 0.46 & 0.61 & 17.19& 06.05 & 0.164 \\
 &  &  &  &  &  &   & & &  & & &\\
 2. & M8D8 &  21187 & 19946 & 23228 & 0.085 & 283$^{+34}_{-29}$ & 4.27 &  0.42 & 0.46 & 18.08 & 06.53 & 0.531\\
 3.& M8D7 & 21952 & 19192 & 23754 & 0.131 & 172$^{+20}_{-15}$ & 1.43 & 1.46 & 1.69 & 18.02 & 06.26 &0.273\\
  &  &  &  &  &  &   & & &  & &&\\
4. & M7D8 & 7240 & 6790 & 8068 & 0.155 & 118$^{+14}_{-18}$ & 6.27 & 0.05 & 0.06 & 19.27 & 06.67&0.410 \\
5. & M7D7 & 5310 & 409 & 7716 & 0.293 & 83$^{+37}_{-76}$ & 0.27 & 0.51 & 0.41 & 14.52 & 05.03& 0.062\\
6. & M7D6 & 209 & 164 & 259 & 0.657 & 4$^{+1}_{-1}$ & 0.93 & 1.84 & -& 10.53 & 03.54 & 0.011 \\
7. & M7D5 & 159 & 147 & 185 & 0.679 & 3$^{+0}_{-0}$ &  0.08 & 6.21 & -& 07.88 & 02.53 &0.004 \\
 &  &  &  &  &  &   & & &  & &&\\
8. & M6D8 & 212 & 170 & 367 & 0.610 & 4$^{+3}_{-0}$ & 314.05 & 0.01 & -& 18.92 & 06.93&0.091\\
9. & M6D7 & 150 & 130 & 178 & 0.716 & 3$^{+1}_{-0}$ & 25.92 & 0.05 &- & 12.69 & 04.43&0.004 \\
10. & M6D6 & 123 & 116 & 142 & 0.690 & 2$^{+1}_{-0}$ &  3.31 & 0.21 & -& 08.49 & 03.06&0.001 \\
11. & M6D5 & 121 & 113 & 124 & 0.695 & 2$^{+0}_{-0}$ &  0.22 & 0.91 &- & 04.66 & 02.05 &9e-5\\
 &  &  &  &  &  &   & & &  & &&\\
12. & Z\_10 & 10168 & 9128 & 11071 & 0.154 & 115$^{+19}_{-25}$  & 0.46 & 0.47 & 0.63 & 17.56 & 06.30&0.163 \\
13. & Z\_100 & 6472 & 5140 & 6811  & 0.166  & 113$^{+12}_{-22}$ & 2.47 & 0.86 & 1.12 & 20.16 & 08.86&0.052\\
 &  &  &  &  &  &   & & &  & &&\\
14. & SN\_D & 10862 & 9671 & 12223 & 0.163 & 125$^{+26}_{-18}$ & 0.26 & 0.41 & 0.66 & 16.81 & 06.01&0.173\\
 &  &  &  &  &  &   & & &  & &&\\
15. & Sd\_50 & 10574 & 9560 & 11572 & 0.152 & 128$^{+17}_{-20}$ &  0.31 & 0.41 & 0.81 
& 16.64 & 05.81 & 0.182\\
16. & Sd\_100 & 10556 & 9590 & 11719 & 0.141 & 125$^{+24}_{-20}$ &  0.31 & 0.44 & 0.59 & 16.93 & 05.91&0.184\\
17. & Sd\_150 & 10462 & 9485 & 11660 & 0.157 & 130$^{+20}_{-21}$ &  0.30 & 0.41 & 0.57 & 16.61 & 05.74&0.195\\
18. & Sd\_200 & 10211 & 8796 & 11232 & 0.150 & 133$^{+45}_{-26}$ &  0.30 & 0.38 & 0.59 & 16.22 & 05.64&0.204\\
 &  &  &  &  &  &   & & &  & &&\\

19.& Ml\_ev & 10456 & 9058 & 11733 & 0.184 & 125$^{+18}_{-27}$ & 0.32  & 0.44 & 0.66 & 16.62 & 05.83 & 0.152 \\
20.& Ml\_0 & 10296 & 9677 & 11466 & 0.128 & 151$^{+15}_{-20}$ &  1.53 & 0.29 & 0.33 & 16.91 & 05.90 & 0.417\\
21.& Ml\_0$_\mathrm{M7D5}$ & 219 & 185 & 276 & 0.700 & 4$^{+1}_{-1}$ & 0.39 & 4.73 &- & 16.61 & 03.34 & 0.014\\

 &  &  &  &  &  &   & & &  & &&\\
22. & Vk\_0 & 10537 & 9346 & 11512 & 0.154 & 121$^{+19}_{-20}$ &   0.31 &  0.48 & 0.64 & 16.73 & 05.78&0.154\\
23. & Vk\_0$_\mathrm{M7D7}$ & 1091 & 402 & 7041 & 0.521 & 17$^{+95}_{-10}$ &  3.21 &  0.52 & 0.56 & 14.39 & 05.05&0.053\\
24. & Vk\_0$_\mathrm{M7D5}$ & 164 & 145 & 187 & 0.704 & 3$^{+0}_{-0}$ & 0.09 & 5.97 &- & 07.90 & 02.61&0.004\\
25. & Vk\_0$_\mathrm{Z_{100}}$ & 6488 & 5340 & 6795 & 0.162 & 112$^{+14}_{-23}$ & 2.47 & 1.19 & 1.17 & 20.08 & 08.86&0.046\\
 &  &  &  &  &  &   & & &  & &&\\

26. & Sp\_01 & 10640 & 9595 & 11868 & 0.147 & 126$^{+21}_{-21}$ &  0.31 &  0.43 & 0.65 & 16.71 & 05.85&0.171\\
27.  & Sp\_33 & 11196 & 10283 & 11826 & 0.139 & 123$^{+14}_{-20}$ & 0.31 & 0.46 & 0.68 & 17.11 & 06.15&0.143\\
28. & Sp\_11 & 7148 & 423 & 7679 & 0.147 & 162$^{+20}_{-153}$ &   0.30 &  0.57 & 0.92 & 17.06 & 06.38&0.079\\
29. & Sp\_LVK & 7654 & 5517 & 8186 & 0.155 & 164$^{+13}_{-48}$ &   0.30 &  0.50 & 0.79 & 17.15 & 06.18&0.142\\
 &  &  &  &  &  &   & & &  & &&\\
30. & Ord\_BH & 7217 & 657 & 8155 & 0.161 & 162$^{+18}_{-150}$ &   0.31 &  0.57 & 0.75 & 15.99 & 05.34&0.168 \\
 &  &  &  &  &  &   & & &  & &&\\
 
31. & DE & 8902 & 8173 & 9459 & 0.157 & 141$^{+20}_{-30}$ & 0.31 & 0.43 & 0.59 & 12.18 & 04.43&0.199 \\
32. & DE$_\mathrm{M7D7}$& 667 & 346 & 6140 & 0.581 & 12$^{+106}_{-6}$ & 7.08 & 0.66 & - & 10.88 & 03.97&0.604\\
33. & DE$_\mathrm{M7D5}$ & 158 & 143 & 189 & 0.685 & 3$^{+0}_{-0}$ & 0.08 & 6.54 & -& 06.06 & 01.96&0.004\\ 
34.  & DE$_\mathrm{Z_{100}}$ & 5071 & 4582 & 5344 & 0.179 & 120$^{+18}_{-21}$ & 2.46 & 0.93  & 1.38 & 14.71 & 06.84&0.065\\
 \hline
\label{tab:IMBHmass}
\end{tabular}
\end{table*}

\begin{table*}
\centering
\begin{tabular}{lccccccc}
                  Name    &  galaxy-type & mass-radius fit & metallicity &redshift range fit & birth distribution & $\mathcal{R}$(Gpc$^{-3}$yr$^{-1}$) \\
                  \hline
ModelA$_\mathrm{1}$                 &  late            &    median     &  1.5$\times10^{-3}$&$0.3<z<0.5 ^{\ast}$   & uniform, 0-8 &   0.12   \\
ModelA$_\mathrm{2}$                 &  late            &    median     &  1.5$\times10^{-3}$&$0.5<z<3^{\ast}$  & uniform, 0-8 &   0.09   \\
ModelA$_\mathrm{3}$                 &  late            &    median     &  1.5$\times10^{-3}$&$z=5^{\dagger}$   & uniform, 0-8 &    0.06  \\

ModelB$_\mathrm{1}$                 &  late            &    high     &  1.5$\times10^{-3}$&$0.3<z<0.5$    & uniform, 0-8 &    0.15  \\
ModelB$_\mathrm{2}$                 &  late            &    low     &  1.5$\times10^{-3}$&$0.3<z<0.5$    & uniform, 0-8 &    0.08  \\
ModelC$_\mathrm{1}$                 &  early            &    high     &  1.5$\times10^{-3}$&$0.3<z<0.5$    & uniform, 0-8 &  0.04    \\
ModelC$_\mathrm{2}$                 &  early            &    low     &  1.5$\times10^{-3}$&$0.3<z<0.5$    & uniform, 0-8 &  0.03    \\
ModelC$_\mathrm{3}$                 &  early            &    median     &  1.5$\times10^{-3}$&$0.3<z<0.5$    & uniform, 0-8 &  0.01    \\
ModelD$_\mathrm{1}$                &  late            &    median     &  1.5$\times10^{-3}$&$0.3<z<0.5$    & uniform, 0-12 &  0.12    \\
ModelD$_\mathrm{2}$                &  late            &    median     &  1.5$\times10^{-3}$&$z=5$    & uniform, 0-12 &   0.10   \\
ModelE$_\mathrm{1}$                 &  late            &    median     &  1.5$\times10^{-4}$&$0.3<z<0.5$ & uniform, 0-8 &    0.13  \\
ModelE$_\mathrm{2}$                 &  late            &    median     &  1.5$\times10^{-4}$&$z=5$    & uniform, 0-8 &   0.09   \\
ModelF$_\mathrm{1}$                &  late            &    median,1\,pc     &  1.5$\times10^{-3}$&$0.3<z<0.5$    & uniform, 0-8 &   0.47 \\  
ModelF$_\mathrm{2}$                &  late            &    median,median$\times0.01$     &  1.5$\times10^{-3}$&$0.3<z<0.5$  & uniform, 0-8 &   0.75   \\
ModelG$_\mathrm{1}^\otimes$                &  -            &    -    &  1.5$\times10^{-3}$&-    & - &   0.61 \\
ModelG$_\mathrm{2}$                &  -            &    -   &  1.5$\times10^{-3}$&-& - &   0.80 \\
ModelH$_\mathrm{1}$                 &  late            &    median     &  1.5$\times10^{-3}$&$0.3<z<0.5 ^{\ast}$   & MD$^\Psi$, 0-8 &   0.11   \\
ModelH$_\mathrm{2}$                 &  late            &    median     &  1.5$\times10^{-3}$&$0.3<z<0.5 ^{\ast}$   & MD peak, 2 &   0.11 

\end{tabular}
\caption{The different models we use to explore intrinsic cosmological merger rates ($\mathcal{R}$) of DBHs from nuclear clusters. Given the uncertain nature of nuclear cluster initial parameters and their evolution, the differences in the model assumptions are aimed at highlighting the variance in the calculated rates. We note that the detector observable rates are $\mathcal{R}_\mathrm{LVK}\lesssim0.7\mathcal{R}$, $\mathcal{R}_\mathrm{CE}\approx0.8-0.9\mathcal{R}$ and $\mathcal{R}_\mathrm{ET}\gtrsim0.9\mathcal{R}$.\newline
$\ast$: \protect\cite{Mortlock2015}; $\dagger$: \protect\cite{Song2016}; $\otimes$: \protect\cite{Georgiev:2016}; $\Psi$: \protect\cite{MadauDickinson2014}.}
\label{tab:rates}
\end{table*}

\section*{Acknowledgements}

We thank Christopher Berry, Mark Gieles, Daniel Marín Pina, Simon Stevenson and Tom Wagg for useful discussions and comments. DC and JB are supported by the STFC grant ST/V005618/1, and FA is supported by an STFC Rutherford fellowship (ST/P00492X/2). IMR-S acknowledges support received from the Herchel Smith Postdoctoral Fellowship Fund. This work made  use  of  the  OzSTAR  high  performance  computer at  Swinburne  University  of  Technology. 
OzSTAR  is funded by Swinburne University of Technology and the National Collaborative Research Infrastructure Strategy (NCRIS).

\section*{Data Availability}

The data utilized for this work will be freely available  upon reasonable request to the corresponding author.



\bibliographystyle{mnras}
\bibliography{example} 

\begin{thebibliography}{}
\makeatletter
\relax
\def\mn@urlcharsother{\let\do\@makeother \do\$\do\&\do\#\do\^\do\_\do\%\do\~}
\def\mn@doi{\begingroup\mn@urlcharsother \@ifnextchar [ {\mn@doi@}
  {\mn@doi@[]}}
\def\mn@doi@[#1]#2{\def\@tempa{#1}\ifx\@tempa\@empty \href
  {http://dx.doi.org/#2} {doi:#2}\else \href {http://dx.doi.org/#2} {#1}\fi
  \endgroup}
\def\mn@eprint#1#2{\mn@eprint@#1:#2::\@nil}
\def\mn@eprint@arXiv#1{\href {http://arxiv.org/abs/#1} {{\tt arXiv:#1}}}
\def\mn@eprint@dblp#1{\href {http://dblp.uni-trier.de/rec/bibtex/#1.xml}
  {dblp:#1}}
\def\mn@eprint@#1:#2:#3:#4\@nil{\def\@tempa {#1}\def\@tempb {#2}\def\@tempc
  {#3}\ifx \@tempc \@empty \let \@tempc \@tempb \let \@tempb \@tempa \fi \ifx
  \@tempb \@empty \def\@tempb {arXiv}\fi \@ifundefined
  {mn@eprint@\@tempb}{\@tempb:\@tempc}{\expandafter \expandafter \csname
  mn@eprint@\@tempb\endcsname \expandafter{\@tempc}}}

\bibitem[\protect\citeauthoryear{Abbott et~al.}{Abbott
  et~al.}{2016}]{LIGOScientific:2016aoc}
Abbott B.~P.,  et~al., 2016, \mn@doi [Phys. Rev. Lett.]
  {10.1103/PhysRevLett.116.061102}, 116, 061102

\bibitem[\protect\citeauthoryear{Abbott et~al.}{Abbott
  et~al.}{2020}]{GW190521LIGOScientific:2020iuh}
Abbott R.,  et~al., 2020, \mn@doi [Phys. Rev. Lett.]
  {10.1103/PhysRevLett.125.101102}, 125, 101102

\bibitem[\protect\citeauthoryear{Abbott et~al.}{Abbott
  et~al.}{2021a}]{GWTC3LIGOScientific:2021djp}
Abbott R.,  et~al., 2021a, \mn@doi [arXiv e-prints]
  {10.48550/arXiv.2111.03606}, \href
  {https://ui.adsabs.harvard.edu/abs/2021arXiv211103606T} {p. arXiv:2111.03606}

\bibitem[\protect\citeauthoryear{Abbott et~al.}{Abbott
  et~al.}{2021b}]{GWTC3popLIGOScientific:2021psn}
Abbott R.,  et~al., 2021b, \mn@doi [arXiv e-prints]
  {10.48550/arXiv.2111.03634}, \href
  {https://ui.adsabs.harvard.edu/abs/2021arXiv211103634T} {p. arXiv:2111.03634}

\bibitem[\protect\citeauthoryear{Abbott et~al.}{Abbott
  et~al.}{2021c}]{GWTC2LIGOScientific:2020ibl}
Abbott R.,  et~al., 2021c, \mn@doi [Phys. Rev. X] {10.1103/PhysRevX.11.021053},
  11, 021053

\bibitem[\protect\citeauthoryear{Abbott et~al.}{Abbott
  et~al.}{2021d}]{LIGOScientific:2020kqk_ch}
Abbott R.,  et~al., 2021d, \mn@doi [Astrophys. J. Lett.]
  {10.3847/2041-8213/abe949}, 913, L7

\bibitem[\protect\citeauthoryear{Akiyama et~al.}{Akiyama
  et~al.}{2019}]{EventHorizonTelescopeM87:2019dse}
Akiyama K.,  et~al., 2019, \mn@doi [\apjl] {10.3847/2041-8213/ab0ec7}, \href
  {https://ui.adsabs.harvard.edu/abs/2019ApJ...875L...1E} {875, L1}

\bibitem[\protect\citeauthoryear{Akiyama et~al.}{Akiyama
  et~al.}{2022}]{EventHorizonTelescopeSagA:2022}
Akiyama K.,  et~al., 2022, \mn@doi [\apjl] {10.3847/2041-8213/ac6674}, \href
  {https://ui.adsabs.harvard.edu/abs/2022ApJ...930L..12E} {930, L12}

\bibitem[\protect\citeauthoryear{{Antonini} \& {Gieles}}{{Antonini} \&
  {Gieles}}{2020a}]{Antonini:2020PhRvD}
{Antonini} F.,  {Gieles} M.,  2020a, \mn@doi [\prd]
  {10.1103/PhysRevD.102.123016}, \href
  {https://ui.adsabs.harvard.edu/abs/2020PhRvD.102l3016A} {102, 123016}

\bibitem[\protect\citeauthoryear{Antonini \& Gieles}{Antonini \&
  Gieles}{2020b}]{Antonini:2019ulv}
Antonini F.,  Gieles M.,  2020b, \mn@doi [Mon. Not. Roy. Astron. Soc.]
  {10.1093/mnras/stz3584}, 492, 2936

\bibitem[\protect\citeauthoryear{{Antonini} \& {Rasio}}{{Antonini} \&
  {Rasio}}{2016}]{Antonini2016ApJ}
{Antonini} F.,  {Rasio} F.~A.,  2016, \mn@doi [\apj]
  {10.3847/0004-637X/831/2/187}, \href
  {https://ui.adsabs.harvard.edu/abs/2016ApJ...831..187A} {831, 187}

\bibitem[\protect\citeauthoryear{{Antonini}, {Murray}  \& {Mikkola}}{{Antonini}
  et~al.}{2014}]{2014ApJ...781...45A}
{Antonini} F.,  {Murray} N.,   {Mikkola} S.,  2014, \mn@doi [\apj]
  {10.1088/0004-637X/781/1/45}, \href
  {https://ui.adsabs.harvard.edu/abs/2014ApJ...781...45A} {781, 45}

\bibitem[\protect\citeauthoryear{{Antonini}, {Chatterjee}, {Rodriguez},
  {Morscher}, {Pattabiraman}, {Kalogera}  \& {Rasio}}{{Antonini}
  et~al.}{2016}]{Antonini2016}
{Antonini} F.,  {Chatterjee} S.,  {Rodriguez} C.~L.,  {Morscher} M.,
  {Pattabiraman} B.,  {Kalogera} V.,   {Rasio} F.~A.,  2016, \mn@doi [\apj]
  {10.3847/0004-637X/816/2/65}, \href
  {https://ui.adsabs.harvard.edu/abs/2016ApJ...816...65A} {816, 65}

\bibitem[\protect\citeauthoryear{Antonini, Gieles  \& Gualandris}{Antonini
  et~al.}{2019}]{Antonini:2018auk}
Antonini F.,  Gieles M.,   Gualandris A.,  2019, \mn@doi [Mon. Not. Roy.
  Astron. Soc.] {10.1093/mnras/stz1149}, 486, 5008

\bibitem[\protect\citeauthoryear{{Antonini}, {Gieles}, {Dosopoulou}  \&
  {Chattopadhyay}}{{Antonini} et~al.}{2023}]{Antonini:2022vib}
{Antonini} F.,  {Gieles} M.,  {Dosopoulou} F.,   {Chattopadhyay} D.,  2023,
  \mn@doi [\mnras] {10.1093/mnras/stad972}, \href
  {https://ui.adsabs.harvard.edu/abs/2023MNRAS.522..466A} {522, 466}

\bibitem[\protect\citeauthoryear{{Arca Sedda}, {Kamlah}, {Spurzem}, {Rizzuto},
  {Giersz}, {Naab}  \& {Berczik}}{{Arca Sedda} et~al.}{2023}]{ArcaSedda2023}
{Arca Sedda} M.,  {Kamlah} A. W.~H.,  {Spurzem} R.,  {Rizzuto} F.~P.,  {Giersz}
  M.,  {Naab} T.,   {Berczik} P.,  2023, \mn@doi [arXiv e-prints]
  {10.48550/arXiv.2307.04806}, \href
  {https://ui.adsabs.harvard.edu/abs/2023arXiv230704806A} {p. arXiv:2307.04806}

\bibitem[\protect\citeauthoryear{{Askar}, {Szkudlarek}, {Gondek-Rosi{\'n}ska},
  {Giersz}  \& {Bulik}}{{Askar} et~al.}{2017}]{Askar2017}
{Askar} A.,  {Szkudlarek} M.,  {Gondek-Rosi{\'n}ska} D.,  {Giersz} M.,
  {Bulik} T.,  2017, \mn@doi [\mnras] {10.1093/mnrasl/slw177}, \href
  {https://ui.adsabs.harvard.edu/abs/2017MNRAS.464L..36A} {464, L36}

\bibitem[\protect\citeauthoryear{{Asplund}, {Grevesse}, {Sauval}  \&
  {Scott}}{{Asplund} et~al.}{2009}]{Asplund:2009}
{Asplund} M.,  {Grevesse} N.,  {Sauval} A.~J.,   {Scott} P.,  2009, \mn@doi
  [\araa] {10.1146/annurev.astro.46.060407.145222}, \href
  {https://ui.adsabs.harvard.edu/abs/2009ARA&A..47..481A} {47, 481}

\bibitem[\protect\citeauthoryear{{Atallah}, {Trani}, {Kremer}, {Weatherford},
  {Fragione}, {Spera}  \& {Rasio}}{{Atallah} et~al.}{2023}]{Atallah2023MNRAS}
{Atallah} D.,  {Trani} A.~A.,  {Kremer} K.,  {Weatherford} N.~C.,  {Fragione}
  G.,  {Spera} M.,   {Rasio} F.~A.,  2023, \mn@doi [\mnras]
  {10.1093/mnras/stad1634}, \href
  {https://ui.adsabs.harvard.edu/abs/2023MNRAS.523.4227A} {523, 4227}

\bibitem[\protect\citeauthoryear{Banerjee}{Banerjee}{2018}]{Banerjee:2018pmh}
Banerjee S.,  2018, \mn@doi [Mon. Not. Roy. Astron. Soc.]
  {10.1093/mnras/sty2608}, 481, 5123

\bibitem[\protect\citeauthoryear{{Banerjee}}{{Banerjee}}{2021}]{Banerjee2021}
{Banerjee} S.,  2021, \mn@doi [\mnras] {10.1093/mnras/staa2392}, \href
  {https://ui.adsabs.harvard.edu/abs/2021MNRAS.500.3002B} {500, 3002}

\bibitem[\protect\citeauthoryear{{Banerjee}, {Baumgardt}  \&
  {Kroupa}}{{Banerjee} et~al.}{2010}]{Banerjee2010}
{Banerjee} S.,  {Baumgardt} H.,   {Kroupa} P.,  2010, \mn@doi [\mnras]
  {10.1111/j.1365-2966.2009.15880.x}, \href
  {https://ui.adsabs.harvard.edu/abs/2010MNRAS.402..371B} {402, 371}

\bibitem[\protect\citeauthoryear{{Bavera} et~al.,}{{Bavera}
  et~al.}{2020}]{Bavera2020}
{Bavera} S.~S.,  et~al., 2020, \mn@doi [\aap] {10.1051/0004-6361/201936204},
  \href {https://ui.adsabs.harvard.edu/abs/2020A&A...635A..97B} {635, A97}

\bibitem[\protect\citeauthoryear{{Belczynski}}{{Belczynski}}{2020}]{Belczynski:2020:0521}
{Belczynski} K.,  2020, \mn@doi [\apjl] {10.3847/2041-8213/abcbf1}, \href
  {https://ui.adsabs.harvard.edu/abs/2020ApJ...905L..15B} {905, L15}

\bibitem[\protect\citeauthoryear{{Belczynski}, {Taam}, {Kalogera}, {Rasio}  \&
  {Bulik}}{{Belczynski} et~al.}{2007}]{Belczynski2007A}
{Belczynski} K.,  {Taam} R.~E.,  {Kalogera} V.,  {Rasio} F.~A.,   {Bulik} T.,
  2007, \mn@doi [\apj] {10.1086/513562}, \href
  {https://ui.adsabs.harvard.edu/abs/2007ApJ...662..504B} {662, 504}

\bibitem[\protect\citeauthoryear{{Belczynski}, {Bulik}, {Fryer}, {Ruiter},
  {Valsecchi}, {Vink}  \& {Hurley}}{{Belczynski}
  et~al.}{2010}]{Belczynski:2010}
{Belczynski} K.,  {Bulik} T.,  {Fryer} C.~L.,  {Ruiter} A.,  {Valsecchi} F.,
  {Vink} J.~S.,   {Hurley} J.~R.,  2010, \mn@doi [\apj]
  {10.1088/0004-637X/714/2/1217}, \href
  {https://ui.adsabs.harvard.edu/abs/2010ApJ...714.1217B} {714, 1217}

\bibitem[\protect\citeauthoryear{Belczynski, Holz, Bulik  \&
  O'Shaughnessy}{Belczynski et~al.}{2016a}]{Belczynski:2016obo}
Belczynski K.,  Holz D.~E.,  Bulik T.,   O'Shaughnessy R.,  2016a, \mn@doi
  [Nature] {10.1038/nature18322}, 534, 512

\bibitem[\protect\citeauthoryear{{Belczynski} et~al.,}{{Belczynski}
  et~al.}{2016b}]{Belczynski2016A}
{Belczynski} K.,  et~al., 2016b, \mn@doi [\aap] {10.1051/0004-6361/201628980},
  \href {https://ui.adsabs.harvard.edu/abs/2016A&A...594A..97B} {594, A97}

\bibitem[\protect\citeauthoryear{{Belczynski} et~al.,}{{Belczynski}
  et~al.}{2020}]{Belczynski:2019fed}
{Belczynski} K.,  et~al., 2020, \mn@doi [\apj] {10.3847/1538-4357/ab6d77},
  \href {https://ui.adsabs.harvard.edu/abs/2020ApJ...890..113B} {890, 113}

\bibitem[\protect\citeauthoryear{{Belkus}, {Van Bever}  \&
  {Vanbeveren}}{{Belkus} et~al.}{2007}]{Belkus2007}
{Belkus} H.,  {Van Bever} J.,   {Vanbeveren} D.,  2007, \mn@doi [\apj]
  {10.1086/512181}, \href
  {https://ui.adsabs.harvard.edu/abs/2007ApJ...659.1576B} {659, 1576}

\bibitem[\protect\citeauthoryear{Binney \& Tremaine}{Binney \&
  Tremaine}{2008}]{10.2307/j.ctvc778ff}
Binney J.,  Tremaine S.,  2008, Galactic Dynamics: Second Edition, rev -
  revised, 2 edn.
Princeton University Press, \url {http://www.jstor.org/stable/j.ctvc778ff}

\bibitem[\protect\citeauthoryear{{B{\"o}ker}, {Sarzi}, {McLaughlin}, {van der
  Marel}, {Rix}, {Ho}  \& {Shields}}{{B{\"o}ker}
  et~al.}{2004}]{2004AJ....127..105B}
{B{\"o}ker} T.,  {Sarzi} M.,  {McLaughlin} D.~E.,  {van der Marel} R.~P.,
  {Rix} H.-W.,  {Ho} L.~C.,   {Shields} J.~C.,  2004, \mn@doi [\aj]
  {10.1086/380231}, \href
  {https://ui.adsabs.harvard.edu/abs/2004AJ....127..105B} {127, 105}

\bibitem[\protect\citeauthoryear{{Bonson} \& {Gallo}}{{Bonson} \&
  {Gallo}}{2016}]{Bonson2016}
{Bonson} K.,  {Gallo} L.~C.,  2016, \mn@doi [\mnras] {10.1093/mnras/stw466},
  \href {https://ui.adsabs.harvard.edu/abs/2016MNRAS.458.1927B} {458, 1927}

\bibitem[\protect\citeauthoryear{{Bourne} \& {Power}}{{Bourne} \&
  {Power}}{2016}]{Bourne2016}
{Bourne} M.~A.,  {Power} C.,  2016, \mn@doi [\mnras] {10.1093/mnrasl/slv162},
  \href {https://ui.adsabs.harvard.edu/abs/2016MNRAS.456L..20B} {456, L20}

\bibitem[\protect\citeauthoryear{{Breen} \& {Heggie}}{{Breen} \&
  {Heggie}}{2013}]{Breen_Heggie:2013}
{Breen} P.~G.,  {Heggie} D.~C.,  2013, \mn@doi [\mnras] {10.1093/mnras/stt628},
  \href {https://ui.adsabs.harvard.edu/abs/2013MNRAS.432.2779B} {432, 2779}

\bibitem[\protect\citeauthoryear{{Broekgaarden}, {Stevenson}  \&
  {Thrane}}{{Broekgaarden} et~al.}{2022}]{Broekgaarden2022spin}
{Broekgaarden} F.~S.,  {Stevenson} S.,   {Thrane} E.,  2022, \mn@doi [\apj]
  {10.3847/1538-4357/ac8879}, \href
  {https://ui.adsabs.harvard.edu/abs/2022ApJ...938...45B} {938, 45}

\bibitem[\protect\citeauthoryear{{Castellano} et~al.,}{{Castellano}
  et~al.}{2022}]{Castellano2022}
{Castellano} M.,  et~al., 2022, \mn@doi [\apjl] {10.3847/2041-8213/ac94d0},
  \href {https://ui.adsabs.harvard.edu/abs/2022ApJ...938L..15C} {938, L15}

\bibitem[\protect\citeauthoryear{{Chakraborty}, {Navale}, {Ratheesh}  \&
  {Bhattacharyya}}{{Chakraborty} et~al.}{2020}]{xrayChakraborty2020}
{Chakraborty} S.,  {Navale} N.,  {Ratheesh} A.,   {Bhattacharyya} S.,  2020,
  \mn@doi [\mnras] {10.1093/mnras/staa2711}, \href
  {https://ui.adsabs.harvard.edu/abs/2020MNRAS.498.5873C} {498, 5873}

\bibitem[\protect\citeauthoryear{{Charles} et~al.,}{{Charles}
  et~al.}{2022}]{xrayCharles2022}
{Charles} P.,  et~al., 2022, in High Energy Astrophysics in Southern Africa
  2021. p.~24 (\mn@eprint {arXiv} {2201.11442}), \mn@doi{10.22323/1.401.0024}

\bibitem[\protect\citeauthoryear{{Chattopadhyay}, {Stevenson}, {Hurley},
  {Bailes}  \& {Broekgaarden}}{{Chattopadhyay}
  et~al.}{2021}]{Chattopadhyay2021MNRAS}
{Chattopadhyay} D.,  {Stevenson} S.,  {Hurley} J.~R.,  {Bailes} M.,
  {Broekgaarden} F.,  2021, \mn@doi [\mnras] {10.1093/mnras/stab973}, \href
  {https://ui.adsabs.harvard.edu/abs/2021MNRAS.504.3682C} {504, 3682}

\bibitem[\protect\citeauthoryear{Chattopadhyay, Hurley, Stevenson  \&
  Raidani}{Chattopadhyay et~al.}{2022a}]{ChattopadhyayStarCluster2022buz}
Chattopadhyay D.,  Hurley J.,  Stevenson S.,   Raidani A.,  2022a, \mn@doi
  [Mon. Not. Roy. Astron. Soc.] {10.1093/mnras/stac1163}, 513, 4527

\bibitem[\protect\citeauthoryear{{Chattopadhyay}, {Stevenson}, {Broekgaarden},
  {Antonini}  \& {Belczynski}}{{Chattopadhyay}
  et~al.}{2022b}]{Chattopadhyay2022}
{Chattopadhyay} D.,  {Stevenson} S.,  {Broekgaarden} F.,  {Antonini} F.,
  {Belczynski} K.,  2022b, \mn@doi [\mnras] {10.1093/mnras/stac1283}, \href
  {https://ui.adsabs.harvard.edu/abs/2022MNRAS.513.5780C} {513, 5780}

\bibitem[\protect\citeauthoryear{{Chia}}{{Chia}}{2021}]{Chia2021}
{Chia} H.~S.,  2021, \mn@doi [\prd] {10.1103/PhysRevD.104.024013}, \href
  {https://ui.adsabs.harvard.edu/abs/2021PhRvD.104b4013C} {104, 024013}

\bibitem[\protect\citeauthoryear{{Conselice}, {Blackburne}  \&
  {Papovich}}{{Conselice} et~al.}{2005}]{Conselice2005}
{Conselice} C.~J.,  {Blackburne} J.~A.,   {Papovich} C.,  2005, \mn@doi [\apj]
  {10.1086/426102}, \href
  {https://ui.adsabs.harvard.edu/abs/2005ApJ...620..564C} {620, 564}

\bibitem[\protect\citeauthoryear{{Conselice}, {Wilkinson}, {Duncan}  \&
  {Mortlock}}{{Conselice} et~al.}{2016}]{Conselice2016}
{Conselice} C.~J.,  {Wilkinson} A.,  {Duncan} K.,   {Mortlock} A.,  2016,
  \mn@doi [\apj] {10.3847/0004-637X/830/2/83}, \href
  {https://ui.adsabs.harvard.edu/abs/2016ApJ...830...83C} {830, 83}

\bibitem[\protect\citeauthoryear{{C{\^o}t{\'e}} et~al.,}{{C{\^o}t{\'e}}
  et~al.}{2006}]{2006ApJS..165...57C}
{C{\^o}t{\'e}} P.,  et~al., 2006, \mn@doi [\apjs] {10.1086/504042}, \href
  {https://ui.adsabs.harvard.edu/abs/2006ApJS..165...57C} {165, 57}

\bibitem[\protect\citeauthoryear{{Cutler} \& {Flanagan}}{{Cutler} \&
  {Flanagan}}{1994}]{1994PhRvD..49.2658C}
{Cutler} C.,  {Flanagan} {\'E}.~E.,  1994, \mn@doi [\prd]
  {10.1103/PhysRevD.49.2658}, \href
  {https://ui.adsabs.harvard.edu/abs/1994PhRvD..49.2658C} {49, 2658}

\bibitem[\protect\citeauthoryear{{Dall'Amico}, {Mapelli}, {Torniamenti}  \&
  {Arca Sedda}}{{Dall'Amico} et~al.}{2023}]{DallAmico2023}
{Dall'Amico} M.,  {Mapelli} M.,  {Torniamenti} S.,   {Arca Sedda} M.,  2023,
  \mn@doi [arXiv e-prints] {10.48550/arXiv.2303.07421}, \href
  {https://ui.adsabs.harvard.edu/abs/2023arXiv230307421D} {p. arXiv:2303.07421}

\bibitem[\protect\citeauthoryear{{Di Carlo}, {Giacobbo}, {Mapelli}, {Pasquato},
  {Spera}, {Wang}  \& {Haardt}}{{Di Carlo} et~al.}{2019}]{DiCarlo2019}
{Di Carlo} U.~N.,  {Giacobbo} N.,  {Mapelli} M.,  {Pasquato} M.,  {Spera} M.,
  {Wang} L.,   {Haardt} F.,  2019, \mn@doi [\mnras] {10.1093/mnras/stz1453},
  \href {https://ui.adsabs.harvard.edu/abs/2019MNRAS.487.2947D} {487, 2947}

\bibitem[\protect\citeauthoryear{{Di Carlo} et~al.,}{{Di Carlo}
  et~al.}{2021}]{DiCarlo2021}
{Di Carlo} U.~N.,  et~al., 2021, \mn@doi [\mnras] {10.1093/mnras/stab2390},
  \href {https://ui.adsabs.harvard.edu/abs/2021MNRAS.507.5132D} {507, 5132}

\bibitem[\protect\citeauthoryear{{Doctor}, {Farr}  \& {Holz}}{{Doctor}
  et~al.}{2021}]{Doctor:2021:SpinRemnants}
{Doctor} Z.,  {Farr} B.,   {Holz} D.~E.,  2021, \mn@doi [\apjl]
  {10.3847/2041-8213/ac0334}, \href
  {https://ui.adsabs.harvard.edu/abs/2021ApJ...914L..18D} {914, L18}

\bibitem[\protect\citeauthoryear{{Dolgov} \& {Postnov}}{{Dolgov} \&
  {Postnov}}{2017}]{Dolgov2017}
{Dolgov} A.,  {Postnov} K.,  2017, \mn@doi [\jcap]
  {10.1088/1475-7516/2017/04/036}, \href
  {https://ui.adsabs.harvard.edu/abs/2017JCAP...04..036D} {2017, 036}

\bibitem[\protect\citeauthoryear{Farmer, Renzo, de Mink, Marchant  \&
  Justham}{Farmer et~al.}{2019a}]{Farmer2019}
Farmer R.,  Renzo M.,  de Mink S.~E.,  Marchant P.,   Justham S.,  2019a, ApJ,
  887, 53

\bibitem[\protect\citeauthoryear{{Farmer}, {Renzo}, {de Mink}, {Marchant}  \&
  {Justham}}{{Farmer} et~al.}{2019b}]{Farmer:2019jed}
{Farmer} R.,  {Renzo} M.,  {de Mink} S.~E.,  {Marchant} P.,   {Justham} S.,
  2019b, \mn@doi [\apj] {10.3847/1538-4357/ab518b}, \href
  {https://ui.adsabs.harvard.edu/abs/2019ApJ...887...53F} {887, 53}

\bibitem[\protect\citeauthoryear{{Fletcher}}{{Fletcher}}{1946}]{Fletcher1946MNRAS}
{Fletcher} A.,  1946, \mn@doi [\mnras] {10.1093/mnras/106.2.121}, \href
  {https://ui.adsabs.harvard.edu/abs/1946MNRAS.106..121F} {106, 121}

\bibitem[\protect\citeauthoryear{Fragione \& Loeb}{Fragione \&
  Loeb}{2020}]{Fragione:2020khu}
Fragione G.,  Loeb A.,  2020, \mn@doi [Astrophys. J. Lett.]
  {10.3847/2041-8213/abb9b4}, 901, L32

\bibitem[\protect\citeauthoryear{Fragione \& Silk}{Fragione \&
  Silk}{2020}]{Fragione:2020nib}
Fragione G.,  Silk J.,  2020, \mn@doi [Mon. Not. Roy. Astron. Soc.]
  {10.1093/mnras/staa2629}, 498, 4591

\bibitem[\protect\citeauthoryear{{Fragione}, {Loeb}, {Kocsis}  \&
  {Rasio}}{{Fragione} et~al.}{2022}]{Fragione2022ApJ}
{Fragione} G.,  {Loeb} A.,  {Kocsis} B.,   {Rasio} F.~A.,  2022, \mn@doi [\apj]
  {10.3847/1538-4357/ac75d0}, \href
  {https://ui.adsabs.harvard.edu/abs/2022ApJ...933..170F} {933, 170}

\bibitem[\protect\citeauthoryear{{Fryer}, {Belczynski}, {Wiktorowicz},
  {Dominik}, {Kalogera}  \& {Holz}}{{Fryer} et~al.}{2012}]{Fryer:2012}
{Fryer} C.~L.,  {Belczynski} K.,  {Wiktorowicz} G.,  {Dominik} M.,  {Kalogera}
  V.,   {Holz} D.~E.,  2012, \mn@doi [\apj] {10.1088/0004-637X/749/1/91}, \href
  {https://ui.adsabs.harvard.edu/abs/2012ApJ...749...91F} {749, 91}

\bibitem[\protect\citeauthoryear{{Garofalo}}{{Garofalo}}{2013}]{Garofalo2013}
{Garofalo} D.,  2013, \mn@doi [Advances in Astronomy] {10.1155/2013/213105},
  \href {https://ui.adsabs.harvard.edu/abs/2013AdAst2013E..13G} {2013, 213105}

\bibitem[\protect\citeauthoryear{{Gayathri} et~al.,}{{Gayathri}
  et~al.}{2020}]{Gayathri2020}
{Gayathri} V.,  et~al., 2020, \mn@doi [arXiv e-prints]
  {10.48550/arXiv.2009.05461}, \href
  {https://ui.adsabs.harvard.edu/abs/2020arXiv200905461G} {p. arXiv:2009.05461}

\bibitem[\protect\citeauthoryear{{Generozov}, {Stone}, {Metzger}  \&
  {Ostriker}}{{Generozov} et~al.}{2018}]{xrayGenerozov2018}
{Generozov} A.,  {Stone} N.~C.,  {Metzger} B.~D.,   {Ostriker} J.~P.,  2018,
  \mn@doi [\mnras] {10.1093/mnras/sty1262}, \href
  {https://ui.adsabs.harvard.edu/abs/2018MNRAS.478.4030G} {478, 4030}

\bibitem[\protect\citeauthoryear{{Georgiev}, {B{\"o}ker}, {Leigh},
  {L{\"u}tzgendorf}  \& {Neumayer}}{{Georgiev} et~al.}{2016}]{Georgiev:2016}
{Georgiev} I.~Y.,  {B{\"o}ker} T.,  {Leigh} N.,  {L{\"u}tzgendorf} N.,
  {Neumayer} N.,  2016, \mn@doi [\mnras] {10.1093/mnras/stw093}, \href
  {https://ui.adsabs.harvard.edu/abs/2016MNRAS.457.2122G} {457, 2122}

\bibitem[\protect\citeauthoryear{{Giacobbo} \& {Mapelli}}{{Giacobbo} \&
  {Mapelli}}{2018}]{Giacobbo2018}
{Giacobbo} N.,  {Mapelli} M.,  2018, \mn@doi [\mnras] {10.1093/mnras/sty1999},
  \href {https://ui.adsabs.harvard.edu/abs/2018MNRAS.480.2011G} {480, 2011}

\bibitem[\protect\citeauthoryear{{Gond{\'a}n}, {Kocsis}, {Raffai}  \&
  {Frei}}{{Gond{\'a}n} et~al.}{2018}]{Gond2018}
{Gond{\'a}n} L.,  {Kocsis} B.,  {Raffai} P.,   {Frei} Z.,  2018, \mn@doi [\apj]
  {10.3847/1538-4357/aaad0e}, \href
  {https://ui.adsabs.harvard.edu/abs/2018ApJ...855...34G} {855, 34}

\bibitem[\protect\citeauthoryear{{Gonz{\'a}lez}, {Kremer}, {Chatterjee},
  {Fragione}, {Rodriguez}, {Weatherford}, {Ye}  \& {Rasio}}{{Gonz{\'a}lez}
  et~al.}{2021}]{Gonzalez2021ApJ}
{Gonz{\'a}lez} E.,  {Kremer} K.,  {Chatterjee} S.,  {Fragione} G.,  {Rodriguez}
  C.~L.,  {Weatherford} N.~C.,  {Ye} C.~S.,   {Rasio} F.~A.,  2021, \mn@doi
  [\apjl] {10.3847/2041-8213/abdf5b}, \href
  {https://ui.adsabs.harvard.edu/abs/2021ApJ...908L..29G} {908, L29}

\bibitem[\protect\citeauthoryear{{Guillard}, {Emsellem}  \&
  {Renaud}}{{Guillard} et~al.}{2016}]{Guillard2016}
{Guillard} N.,  {Emsellem} E.,   {Renaud} F.,  2016, \mn@doi [\mnras]
  {10.1093/mnras/stw1570}, \href
  {https://ui.adsabs.harvard.edu/abs/2016MNRAS.461.3620G} {461, 3620}

\bibitem[\protect\citeauthoryear{Hannam et~al.}{Hannam
  et~al.}{2022}]{Hannam:2021pit}
Hannam M.,  et~al., 2022, \mn@doi [Nature] {10.1038/s41586-022-05212-z}, 610,
  652

\bibitem[\protect\citeauthoryear{{Heggie} \& {Hut}}{{Heggie} \&
  {Hut}}{1993}]{1993ApJS...85..347H}
{Heggie} D.~C.,  {Hut} P.,  1993, \mn@doi [\apjs] {10.1086/191768}, \href
  {https://ui.adsabs.harvard.edu/abs/1993ApJS...85..347H} {85, 347}

\bibitem[\protect\citeauthoryear{Heggie \& Hut}{Heggie \&
  Hut}{2003}]{Heggie_Hut:2003}
Heggie D.,  Hut P.,  2003, The Gravitational Million–Body Problem: A
  Multidisciplinary Approach to Star Cluster Dynamics.
Cambridge University Press, \mn@doi{10.1017/CBO9781139164535}

\bibitem[\protect\citeauthoryear{H{\'e}non}{H{\'e}non}{1972}]{Henon:1972}
H{\'e}non M.,  1972, in Lecar M.,  ed., Gravitational N-Body Problem. Springer
  Netherlands, Dordrecht, pp 44--59

\bibitem[\protect\citeauthoryear{{Hills} \& {Fullerton}}{{Hills} \&
  {Fullerton}}{1980}]{Hills_Fullerton:1980}
{Hills} J.~G.,  {Fullerton} L.~W.,  1980, \mn@doi [\aj] {10.1086/112798}, \href
  {https://ui.adsabs.harvard.edu/abs/1980AJ.....85.1281H} {85, 1281}

\bibitem[\protect\citeauthoryear{{Hobbs}, {Lorimer}, {Lyne}  \&
  {Kramer}}{{Hobbs} et~al.}{2005}]{Hobbs:2005}
{Hobbs} G.,  {Lorimer} D.~R.,  {Lyne} A.~G.,   {Kramer} M.,  2005, \mn@doi
  [\mnras] {10.1111/j.1365-2966.2005.09087.x}, \href
  {http://adsabs.harvard.edu/abs/2005MNRAS.360..974H} {360, 974}

\bibitem[\protect\citeauthoryear{Hong, Askar, Giersz, Hypki  \& Yoon}{Hong
  et~al.}{2020}]{Hong:2020dsl}
Hong J.,  Askar A.,  Giersz M.,  Hypki A.,   Yoon S.-J.,  2020, \mn@doi [Mon.
  Not. Roy. Astron. Soc.] {10.1093/mnras/staa2677}, 498, 4287

\bibitem[\protect\citeauthoryear{{Hurley}, {Pols}  \& {Tout}}{{Hurley}
  et~al.}{2000}]{HurleySSE:2000pk}
{Hurley} J.~R.,  {Pols} O.~R.,   {Tout} C.~A.,  2000, \mn@doi [\mnras]
  {10.1046/j.1365-8711.2000.03426.x}, \href
  {http://cdsads.u-strasbg.fr/abs/2000MNRAS.315..543H} {315, 543}

\bibitem[\protect\citeauthoryear{{Hurley}, {Tout}  \& {Pols}}{{Hurley}
  et~al.}{2002}]{Hurley2002MNRAS}
{Hurley} J.~R.,  {Tout} C.~A.,   {Pols} O.~R.,  2002, \mn@doi [\mnras]
  {10.1046/j.1365-8711.2002.05038.x}, \href
  {https://ui.adsabs.harvard.edu/abs/2002MNRAS.329..897H} {329, 897}

\bibitem[\protect\citeauthoryear{Janka}{Janka}{2013}]{Janka:2013hfa}
Janka H.~T.,  2013, \mn@doi [Mon. Not. Roy. Astron. Soc.]
  {10.1093/mnras/stt1106}, 434, 1355

\bibitem[\protect\citeauthoryear{{Kamlah} et~al.,}{{Kamlah}
  et~al.}{2022}]{Kamlah:2022}
{Kamlah} A.~W.~H.,  et~al., 2022, \mn@doi [\mnras] {10.1093/mnras/stab3748},
  \href {https://ui.adsabs.harvard.edu/abs/2022MNRAS.511.4060K} {511, 4060}

\bibitem[\protect\citeauthoryear{{Kroupa}}{{Kroupa}}{2001}]{Kroupa2001}
{Kroupa} P.,  2001, \mn@doi [\mnras] {10.1046/j.1365-8711.2001.04022.x}, \href
  {https://ui.adsabs.harvard.edu/abs/2001MNRAS.322..231K} {322, 231}

\bibitem[\protect\citeauthoryear{{Lauer} et~al.,}{{Lauer}
  et~al.}{2021}]{Lauer2021}
{Lauer} T.~R.,  et~al., 2021, \mn@doi [\apj] {10.3847/1538-4357/abc881}, \href
  {https://ui.adsabs.harvard.edu/abs/2021ApJ...906...77L} {906, 77}

\bibitem[\protect\citeauthoryear{{Le Tiec} \& {Casals}}{{Le Tiec} \&
  {Casals}}{2021}]{LeTiec2021L}
{Le Tiec} A.,  {Casals} M.,  2021, \mn@doi [\prl]
  {10.1103/PhysRevLett.126.131102}, \href
  {https://ui.adsabs.harvard.edu/abs/2021PhRvL.126m1102L} {126, 131102}

\bibitem[\protect\citeauthoryear{Le~Tiec, Blanchet  \& Will}{Le~Tiec
  et~al.}{2010}]{LeTiec_spinkick:2009yg}
Le~Tiec A.,  Blanchet L.,   Will C.~M.,  2010, \mn@doi [Class. Quant. Grav.]
  {10.1088/0264-9381/27/1/012001}, 27, 012001

\bibitem[\protect\citeauthoryear{{Leja}, {van Dokkum}  \& {Franx}}{{Leja}
  et~al.}{2013}]{Leja2013ApJ}
{Leja} J.,  {van Dokkum} P.,   {Franx} M.,  2013, \mn@doi [\apj]
  {10.1088/0004-637X/766/1/33}, \href
  {https://ui.adsabs.harvard.edu/abs/2013ApJ...766...33L} {766, 33}

\bibitem[\protect\citeauthoryear{{Lipunov}, {Postnov}  \&
  {Prokhorov}}{{Lipunov} et~al.}{1997}]{Lipunov1997}
{Lipunov} V.~M.,  {Postnov} K.~A.,   {Prokhorov} M.~E.,  1997, \mn@doi [\mnras]
  {10.1093/mnras/288.1.245}, \href
  {https://ui.adsabs.harvard.edu/abs/1997MNRAS.288..245L} {288, 245}

\bibitem[\protect\citeauthoryear{{Lousto} \& {Zlochower}}{{Lousto} \&
  {Zlochower}}{2009}]{Lousto:2009}
{Lousto} C.~O.,  {Zlochower} Y.,  2009, \mn@doi [\prd]
  {10.1103/PhysRevD.79.064018}, \href
  {https://ui.adsabs.harvard.edu/abs/2009PhRvD..79f4018L} {79, 064018}

\bibitem[\protect\citeauthoryear{{Lousto}, {Campanelli}, {Zlochower}  \&
  {Nakano}}{{Lousto} et~al.}{2010}]{Lousto:2010}
{Lousto} C.~O.,  {Campanelli} M.,  {Zlochower} Y.,   {Nakano} H.,  2010,
  \mn@doi [Classical and Quantum Gravity] {10.1088/0264-9381/27/11/114006},
  \href {https://ui.adsabs.harvard.edu/abs/2010CQGra..27k4006L} {27, 114006}

\bibitem[\protect\citeauthoryear{{Loutrel}}{{Loutrel}}{2020}]{Loutrel2020}
{Loutrel} N.,  2020, \mn@doi [arXiv e-prints] {10.48550/arXiv.2009.11332},
  \href {https://ui.adsabs.harvard.edu/abs/2020arXiv200911332L} {p.
  arXiv:2009.11332}

\bibitem[\protect\citeauthoryear{Lower et~al.}{Lower et~al.}{2018}]{Lower18}
Lower M.,  et~al., 2018, \mn@doi [Phys. Rev. D] {10.1103/PhysRevD.98.083028},
  98

\bibitem[\protect\citeauthoryear{{Lucy} \& {Solomon}}{{Lucy} \&
  {Solomon}}{1970}]{Lucy1970}
{Lucy} L.~B.,  {Solomon} P.~M.,  1970, \mn@doi [\apj] {10.1086/150365}, \href
  {https://ui.adsabs.harvard.edu/abs/1970ApJ...159..879L} {159, 879}

\bibitem[\protect\citeauthoryear{{Ma} \& {Fuller}}{{Ma} \&
  {Fuller}}{2023}]{Ma2023}
{Ma} L.,  {Fuller} J.,  2023, \mn@doi [arXiv e-prints]
  {10.48550/arXiv.2305.08356}, \href
  {https://ui.adsabs.harvard.edu/abs/2023arXiv230508356M} {p. arXiv:2305.08356}

\bibitem[\protect\citeauthoryear{{Madau} \& {Dickinson}}{{Madau} \&
  {Dickinson}}{2014}]{MadauDickinson2014}
{Madau} P.,  {Dickinson} M.,  2014, \mn@doi [\araa]
  {10.1146/annurev-astro-081811-125615}, \href
  {https://ui.adsabs.harvard.edu/abs/2014ARA&A..52..415M} {52, 415}

\bibitem[\protect\citeauthoryear{Maggiore}{Maggiore}{2008}]{maggiore2008gravitational}
Maggiore M.,  2008, Gravitational Waves: Volume 1: Theory and Experiments.
Gravitational Waves, OUP Oxford, \url
  {https://books.google.de/books?id=AqVpQgAACAAJ}

\bibitem[\protect\citeauthoryear{{Mapelli}}{{Mapelli}}{2016}]{Mapelli2016MNRAS_ch}
{Mapelli} M.,  2016, \mn@doi [\mnras] {10.1093/mnras/stw869}, \href
  {https://ui.adsabs.harvard.edu/abs/2016MNRAS.459.3432M} {459, 3432}

\bibitem[\protect\citeauthoryear{{Mapelli} et~al.,}{{Mapelli}
  et~al.}{2021}]{Mapelli:2021MNRAS}
{Mapelli} M.,  et~al., 2021, \mn@doi [\mnras] {10.1093/mnras/stab1334}, \href
  {https://ui.adsabs.harvard.edu/abs/2021MNRAS.505..339M} {505, 339}

\bibitem[\protect\citeauthoryear{{Marchant} \& {Moriya}}{{Marchant} \&
  {Moriya}}{2020a}]{Marchant2020}
{Marchant} P.,  {Moriya} T.~J.,  2020a, \mn@doi [\aap]
  {10.1051/0004-6361/202038902}, \href
  {https://ui.adsabs.harvard.edu/abs/2020A&A...640L..18M} {640, L18}

\bibitem[\protect\citeauthoryear{Marchant \& Moriya}{Marchant \&
  Moriya}{2020b}]{Marchant:2020haw}
Marchant P.,  Moriya T.,  2020b, \mn@doi [Astron. Astrophys.]
  {10.1051/0004-6361/202038902}, 640, L18

\bibitem[\protect\citeauthoryear{{Miller} \& {Hamilton}}{{Miller} \&
  {Hamilton}}{2002}]{Miller2002}
{Miller} M.~C.,  {Hamilton} D.~P.,  2002, \mn@doi [\mnras]
  {10.1046/j.1365-8711.2002.05112.x}, \href
  {https://ui.adsabs.harvard.edu/abs/2002MNRAS.330..232C} {330, 232}

\bibitem[\protect\citeauthoryear{{Mortlock} et~al.,}{{Mortlock}
  et~al.}{2015}]{Mortlock2015}
{Mortlock} A.,  et~al., 2015, \mn@doi [\mnras] {10.1093/mnras/stu2403}, \href
  {https://ui.adsabs.harvard.edu/abs/2015MNRAS.447....2M} {447, 2}

\bibitem[\protect\citeauthoryear{{Neumayer} \& {Walcher}}{{Neumayer} \&
  {Walcher}}{2012}]{2012AdAst2012E..15N}
{Neumayer} N.,  {Walcher} C.~J.,  2012, \mn@doi [Advances in Astronomy]
  {10.1155/2012/709038}, \href
  {https://ui.adsabs.harvard.edu/abs/2012AdAst2012E..15N} {2012, 709038}

\bibitem[\protect\citeauthoryear{{Neumayer}, {Seth}  \& {B{\"o}ker}}{{Neumayer}
  et~al.}{2020}]{Neumayer:2020}
{Neumayer} N.,  {Seth} A.,   {B{\"o}ker} T.,  2020, \mn@doi [\aapr]
  {10.1007/s00159-020-00125-0}, \href
  {https://ui.adsabs.harvard.edu/abs/2020A&ARv..28....4N} {28, 4}

\bibitem[\protect\citeauthoryear{{Nokhrina}, {Gurvits}, {Beskin}, {Nakamura},
  {Asada}  \& {Hada}}{{Nokhrina} et~al.}{2019}]{Nokhrina2019}
{Nokhrina} E.~E.,  {Gurvits} L.~I.,  {Beskin} V.~S.,  {Nakamura} M.,  {Asada}
  K.,   {Hada} K.,  2019, \mn@doi [\mnras] {10.1093/mnras/stz2116}, \href
  {https://ui.adsabs.harvard.edu/abs/2019MNRAS.489.1197N} {489, 1197}

\bibitem[\protect\citeauthoryear{{Ownsworth}, {Conselice}, {Mundy}, {Mortlock},
  {Hartley}, {Duncan}  \& {Almaini}}{{Ownsworth} et~al.}{2016}]{Ownsworth2016}
{Ownsworth} J.~R.,  {Conselice} C.~J.,  {Mundy} C.~J.,  {Mortlock} A.,
  {Hartley} W.~G.,  {Duncan} K.,   {Almaini} O.,  2016, \mn@doi [\mnras]
  {10.1093/mnras/stw1207}, \href
  {https://ui.adsabs.harvard.edu/abs/2016MNRAS.461.1112O} {461, 1112}

\bibitem[\protect\citeauthoryear{{Peters}}{{Peters}}{1964}]{Peters:1964}
{Peters} P.~C.,  1964, \mn@doi [Physical Review] {10.1103/PhysRev.136.B1224},
  \href {http://cdsads.u-strasbg.fr/abs/1964PhRv..136.1224P} {136, 1224}

\bibitem[\protect\citeauthoryear{{Piotrovich}, {Buliga}  \&
  {Natsvlishvili}}{{Piotrovich} et~al.}{2022}]{Piotrovich2022P}
{Piotrovich} M.~Y.,  {Buliga} S.~D.,   {Natsvlishvili} T.~M.,  2022, \mn@doi
  [Astronomische Nachrichten] {10.1002/asna.20210020}, \href
  {https://ui.adsabs.harvard.edu/abs/2022AN....34310020P} {343, e10020}

\bibitem[\protect\citeauthoryear{{Poggianti}, {Moretti}, {Calvi}, {D'Onofrio},
  {Valentinuzzi}, {Fritz}  \& {Renzini}}{{Poggianti}
  et~al.}{2013}]{Poggianti2013}
{Poggianti} B.~M.,  {Moretti} A.,  {Calvi} R.,  {D'Onofrio} M.,  {Valentinuzzi}
  T.,  {Fritz} J.,   {Renzini} A.,  2013, \mn@doi [\apj]
  {10.1088/0004-637X/777/2/125}, \href
  {https://ui.adsabs.harvard.edu/abs/2013ApJ...777..125P} {777, 125}

\bibitem[\protect\citeauthoryear{{Qin}, {Fragos}, {Meynet}, {Andrews},
  {S{\o}rensen}  \& {Song}}{{Qin} et~al.}{2018}]{Qin2018}
{Qin} Y.,  {Fragos} T.,  {Meynet} G.,  {Andrews} J.,  {S{\o}rensen} M.,
  {Song} H.~F.,  2018, \mn@doi [\aap] {10.1051/0004-6361/201832839}, \href
  {https://ui.adsabs.harvard.edu/abs/2018A&A...616A..28Q} {616, A28}

\bibitem[\protect\citeauthoryear{{Quinlan}}{{Quinlan}}{1996}]{Quinlan:1996}
{Quinlan} G.~D.,  1996, \mn@doi [\na] {10.1016/S1384-1076(96)00003-6}, \href
  {https://ui.adsabs.harvard.edu/abs/1996NewA....1...35Q} {1, 35}

\bibitem[\protect\citeauthoryear{Reynolds}{Reynolds}{2013}]{Reynolds:2013rva}
Reynolds C.~S.,  2013, \mn@doi [Class. Quant. Grav.]
  {10.1088/0264-9381/30/24/244004}, 30, 244004

\bibitem[\protect\citeauthoryear{{Reynolds}}{{Reynolds}}{2021}]{Reynolds:2021}
{Reynolds} C.~S.,  2021, \mn@doi [\araa] {10.1146/annurev-astro-112420-035022},
  \href {https://ui.adsabs.harvard.edu/abs/2021ARA&A..59..117R} {59, 117}

\bibitem[\protect\citeauthoryear{{Rizzuto} et~al.,}{{Rizzuto}
  et~al.}{2021}]{Rizzuto2021}
{Rizzuto} F.~P.,  et~al., 2021, \mn@doi [\mnras] {10.1093/mnras/staa3634},
  \href {https://ui.adsabs.harvard.edu/abs/2021MNRAS.501.5257R} {501, 5257}

\bibitem[\protect\citeauthoryear{{Rodriguez}, {Chatterjee}  \&
  {Rasio}}{{Rodriguez} et~al.}{2016a}]{RodriguezChatterjee2016}
{Rodriguez} C.~L.,  {Chatterjee} S.,   {Rasio} F.~A.,  2016a, \mn@doi [\prd]
  {10.1103/PhysRevD.93.084029}, \href
  {https://ui.adsabs.harvard.edu/abs/2016PhRvD..93h4029R} {93, 084029}

\bibitem[\protect\citeauthoryear{Rodriguez, Zevin, Pankow, Kalogera  \&
  Rasio}{Rodriguez et~al.}{2016b}]{RodriguezSpin:2016vmx}
Rodriguez C.~L.,  Zevin M.,  Pankow C.,  Kalogera V.,   Rasio F.~A.,  2016b,
  \mn@doi [Astrophys. J. Lett.] {10.3847/2041-8205/832/1/L2}, 832, L2

\bibitem[\protect\citeauthoryear{Rodriguez, Amaro-Seoane, Chatterjee, Kremer,
  Rasio, Samsing, Ye  \& Zevin}{Rodriguez
  et~al.}{2018a}]{RodriguezEccen:2018pss}
Rodriguez C.~L.,  Amaro-Seoane P.,  Chatterjee S.,  Kremer K.,  Rasio F.~A.,
  Samsing J.,  Ye C.~S.,   Zevin M.,  2018a, \mn@doi [Phys. Rev. D]
  {10.1103/PhysRevD.98.123005}, 98, 123005

\bibitem[\protect\citeauthoryear{Rodriguez, Amaro-Seoane, Chatterjee  \&
  Rasio}{Rodriguez et~al.}{2018b}]{Rodriguez18a}
Rodriguez C.~L.,  Amaro-Seoane P.,  Chatterjee S.,   Rasio F.~A.,  2018b,
  \mn@doi [Phys. Rev. Lett.] {10.1103/PhysRevLett.120.151101}, 120, 151101

\bibitem[\protect\citeauthoryear{{Rodriguez}, {Zevin}, {Amaro-Seoane},
  {Chatterjee}, {Kremer}, {Rasio}  \& {Ye}}{{Rodriguez}
  et~al.}{2019}]{Rodriguez2019PhRvD_ch}
{Rodriguez} C.~L.,  {Zevin} M.,  {Amaro-Seoane} P.,  {Chatterjee} S.,  {Kremer}
  K.,  {Rasio} F.~A.,   {Ye} C.~S.,  2019, \mn@doi [\prd]
  {10.1103/PhysRevD.100.043027}, \href
  {https://ui.adsabs.harvard.edu/abs/2019PhRvD.100d3027R} {100, 043027}

\bibitem[\protect\citeauthoryear{{Romero-Shaw}, {Lasky}, {Thrane}  \&
  {Calder{\'o}n Bustillo}}{{Romero-Shaw} et~al.}{2020}]{RomeroShaw2020ApJ}
{Romero-Shaw} I.,  {Lasky} P.~D.,  {Thrane} E.,   {Calder{\'o}n Bustillo} J.,
  2020, \mn@doi [\apjl] {10.3847/2041-8213/abbe26}, \href
  {https://ui.adsabs.harvard.edu/abs/2020ApJ...903L...5R} {903, L5}

\bibitem[\protect\citeauthoryear{{Romero-Shaw}, {Lasky}  \&
  {Thrane}}{{Romero-Shaw} et~al.}{2021}]{Romero-Shaw2021}
{Romero-Shaw} I.,  {Lasky} P.~D.,   {Thrane} E.,  2021, \mn@doi [\apjl]
  {10.3847/2041-8213/ac3138}, \href
  {https://ui.adsabs.harvard.edu/abs/2021ApJ...921L..31R} {921, L31}

\bibitem[\protect\citeauthoryear{{Romero-Shaw}, {Loutrel}  \&
  {Zevin}}{{Romero-Shaw} et~al.}{2022a}]{Romero-Shaw2022arXiv}
{Romero-Shaw} I.~M.,  {Loutrel} N.,   {Zevin} M.,  2022a, \mn@doi [arXiv
  e-prints] {10.48550/arXiv.2211.07278}, \href
  {https://ui.adsabs.harvard.edu/abs/2022arXiv221107278R} {p. arXiv:2211.07278}

\bibitem[\protect\citeauthoryear{{Romero-Shaw}, {Lasky}  \&
  {Thrane}}{{Romero-Shaw}
  et~al.}{2022b}]{Romero-Shaw:2022:FourEccentricMergers}
{Romero-Shaw} I.,  {Lasky} P.~D.,   {Thrane} E.,  2022b, \mn@doi [\apj]
  {10.3847/1538-4357/ac9798}, \href
  {https://ui.adsabs.harvard.edu/abs/2022ApJ...940..171R} {940, 171}

\bibitem[\protect\citeauthoryear{{Romero-Shaw}, {Gerosa}  \&
  {Loutrel}}{{Romero-Shaw} et~al.}{2023}]{Romero-Shaw2023}
{Romero-Shaw} I.~M.,  {Gerosa} D.,   {Loutrel} N.,  2023, \mn@doi [\mnras]
  {10.1093/mnras/stad031}, \href
  {https://ui.adsabs.harvard.edu/abs/2023MNRAS.519.5352R} {519, 5352}

\bibitem[\protect\citeauthoryear{{Sabhahit}, {Vink}, {Sander}  \&
  {Higgins}}{{Sabhahit} et~al.}{2023}]{Sabhahit2023}
{Sabhahit} G.~N.,  {Vink} J.~S.,  {Sander} A. A.~C.,   {Higgins} E.~R.,  2023,
  \mn@doi [\mnras] {10.1093/mnras/stad1888}, \href
  {https://ui.adsabs.harvard.edu/abs/2023MNRAS.524.1529S} {524, 1529}

\bibitem[\protect\citeauthoryear{{Sakstein}, {Croon}, {McDermott}, {Straight}
  \& {Baxter}}{{Sakstein} et~al.}{2020}]{Sakstein2020PhRvL}
{Sakstein} J.,  {Croon} D.,  {McDermott} S.~D.,  {Straight} M.~C.,   {Baxter}
  E.~J.,  2020, \mn@doi [\prl] {10.1103/PhysRevLett.125.261105}, \href
  {https://ui.adsabs.harvard.edu/abs/2020PhRvL.125z1105S} {125, 261105}

\bibitem[\protect\citeauthoryear{Samsing}{Samsing}{2018}]{Samsing:2017xmd}
Samsing J.,  2018, \mn@doi [Phys. Rev. D] {10.1103/PhysRevD.97.103014}, 97,
  103014

\bibitem[\protect\citeauthoryear{Samsing, Askar  \& Giersz}{Samsing
  et~al.}{2018}]{Samsing:2017oij}
Samsing J.,  Askar A.,   Giersz M.,  2018, \mn@doi [Astrophys. J.]
  {10.3847/1538-4357/aaab52}, 855, 124

\bibitem[\protect\citeauthoryear{{Schechter}}{{Schechter}}{1976}]{Schechter:1976}
{Schechter} P.,  1976, \mn@doi [\apj] {10.1086/154079}, \href
  {https://ui.adsabs.harvard.edu/abs/1976ApJ...203..297S} {203, 297}

\bibitem[\protect\citeauthoryear{{Sch{\"o}del} et~al.,}{{Sch{\"o}del}
  et~al.}{2007}]{2007A&A...469..125S}
{Sch{\"o}del} R.,  et~al., 2007, \mn@doi [\aap] {10.1051/0004-6361:20065089},
  \href {https://ui.adsabs.harvard.edu/abs/2007A&A...469..125S} {469, 125}

\bibitem[\protect\citeauthoryear{{Sch{\"o}del}, {Merritt}  \&
  {Eckart}}{{Sch{\"o}del} et~al.}{2009}]{Schodel2009}
{Sch{\"o}del} R.,  {Merritt} D.,   {Eckart} A.,  2009, \mn@doi [\aap]
  {10.1051/0004-6361/200810922}, \href
  {https://ui.adsabs.harvard.edu/abs/2009A&A...502...91S} {502, 91}

\bibitem[\protect\citeauthoryear{{Sch{\"o}del}, {Nogueras-Lara},
  {Gallego-Cano}, {Shahzamanian}, {Gallego-Calvente}  \&
  {Gardini}}{{Sch{\"o}del} et~al.}{2020}]{Schodel2020}
{Sch{\"o}del} R.,  {Nogueras-Lara} F.,  {Gallego-Cano} E.,  {Shahzamanian} B.,
  {Gallego-Calvente} A.~T.,   {Gardini} A.,  2020, \mn@doi [\aap]
  {10.1051/0004-6361/201936688}, \href
  {https://ui.adsabs.harvard.edu/abs/2020A&A...641A.102S} {641, A102}

\bibitem[\protect\citeauthoryear{{Seth}, {Ag{\"u}eros}, {Lee}  \&
  {Basu-Zych}}{{Seth} et~al.}{2008}]{2008ApJ...678..116S}
{Seth} A.,  {Ag{\"u}eros} M.,  {Lee} D.,   {Basu-Zych} A.,  2008, \mn@doi
  [\apj] {10.1086/528955}, \href
  {https://ui.adsabs.harvard.edu/abs/2008ApJ...678..116S} {678, 116}

\bibitem[\protect\citeauthoryear{{Song} et~al.,}{{Song}
  et~al.}{2016}]{Song2016}
{Song} M.,  et~al., 2016, \mn@doi [\apj] {10.3847/0004-637X/825/1/5}, \href
  {https://ui.adsabs.harvard.edu/abs/2016ApJ...825....5S} {825, 5}

\bibitem[\protect\citeauthoryear{{Sopuerta}, {Yunes}  \& {Laguna}}{{Sopuerta}
  et~al.}{2007}]{Sopuerta_spinkickecc:2007}
{Sopuerta} C.~F.,  {Yunes} N.,   {Laguna} P.,  2007, \mn@doi [\apjl]
  {10.1086/512067}, \href
  {https://ui.adsabs.harvard.edu/abs/2007ApJ...656L...9S} {656, L9}

\bibitem[\protect\citeauthoryear{{Spera} \& {Mapelli}}{{Spera} \&
  {Mapelli}}{2017}]{Spera2017}
{Spera} M.,  {Mapelli} M.,  2017, \mn@doi [\mnras] {10.1093/mnras/stx1576},
  \href {https://ui.adsabs.harvard.edu/abs/2017MNRAS.470.4739S} {470, 4739}

\bibitem[\protect\citeauthoryear{{Spera}, {Trani}  \& {Mencagli}}{{Spera}
  et~al.}{2022}]{Spera2022Galax}
{Spera} M.,  {Trani} A.~A.,   {Mencagli} M.,  2022, \mn@doi [Galaxies]
  {10.3390/galaxies10040076}, \href
  {https://ui.adsabs.harvard.edu/abs/2022Galax..10...76S} {10, 76}

\bibitem[\protect\citeauthoryear{{Stegmann}, {Antonini}, {Schneider}, {Tiwari}
  \& {Chattopadhyay}}{{Stegmann} et~al.}{2022}]{Stegmann2022PhRvD}
{Stegmann} J.,  {Antonini} F.,  {Schneider} F. R.~N.,  {Tiwari} V.,
  {Chattopadhyay} D.,  2022, \mn@doi [\prd] {10.1103/PhysRevD.106.023014},
  \href {https://ui.adsabs.harvard.edu/abs/2022PhRvD.106b3014S} {106, 023014}

\bibitem[\protect\citeauthoryear{Stevenson, Vigna-G\'omez, Mandel, Barrett,
  Neijssel, Perkins  \& de Mink}{Stevenson et~al.}{2017a}]{Stevenson:2017tfq}
Stevenson S.,  Vigna-G\'omez A.,  Mandel I.,  Barrett J.~W.,  Neijssel C.~J.,
  Perkins D.,   de Mink S.~E.,  2017a, \mn@doi [Nature Commun.]
  {10.1038/ncomms14906}, 8, 14906

\bibitem[\protect\citeauthoryear{{Stevenson}, {Berry}  \& {Mandel}}{{Stevenson}
  et~al.}{2017b}]{Stevenson2017MNRASS}
{Stevenson} S.,  {Berry} C. P.~L.,   {Mandel} I.,  2017b, \mn@doi [\mnras]
  {10.1093/mnras/stx1764}, \href
  {https://ui.adsabs.harvard.edu/abs/2017MNRAS.471.2801S} {471, 2801}

\bibitem[\protect\citeauthoryear{{Stone}, {K{\"u}pper}  \& {Ostriker}}{{Stone}
  et~al.}{2017}]{Stone2017}
{Stone} N.~C.,  {K{\"u}pper} A. H.~W.,   {Ostriker} J.~P.,  2017, \mn@doi
  [\mnras] {10.1093/mnras/stx097}, \href
  {https://ui.adsabs.harvard.edu/abs/2017MNRAS.467.4180S} {467, 4180}

\bibitem[\protect\citeauthoryear{{Tetarenko} et~al.,}{{Tetarenko}
  et~al.}{2016}]{xrayTetarenko2016}
{Tetarenko} B.~E.,  et~al., 2016, \mn@doi [\apj] {10.3847/0004-637X/825/1/10},
  \href {https://ui.adsabs.harvard.edu/abs/2016ApJ...825...10T} {825, 10}

\bibitem[\protect\citeauthoryear{{Tiwari}}{{Tiwari}}{2023}]{Tiwari2023}
{Tiwari} V.,  2023, \mn@doi [arXiv e-prints] {10.48550/arXiv.2304.03498}, \href
  {https://ui.adsabs.harvard.edu/abs/2023arXiv230403498T} {p. arXiv:2304.03498}

\bibitem[\protect\citeauthoryear{{Tiwari} \& {Fairhurst}}{{Tiwari} \&
  {Fairhurst}}{2021}]{Tiwari:2021}
{Tiwari} V.,  {Fairhurst} S.,  2021, \mn@doi [\apjl]
  {10.3847/2041-8213/abfbe7}, \href
  {https://ui.adsabs.harvard.edu/abs/2021ApJ...913L..19T} {913, L19}

\bibitem[\protect\citeauthoryear{{Valcin}, {Jimenez}, {Verde}, {Bernal}  \&
  {Wandelt}}{{Valcin} et~al.}{2021}]{Valcin2021}
{Valcin} D.,  {Jimenez} R.,  {Verde} L.,  {Bernal} J.~L.,   {Wandelt} B.~D.,
  2021, \mn@doi [\jcap] {10.1088/1475-7516/2021/08/017}, \href
  {https://ui.adsabs.harvard.edu/abs/2021JCAP...08..017V} {2021, 017}

\bibitem[\protect\citeauthoryear{{Vink}, {de Koter}  \& {Lamers}}{{Vink}
  et~al.}{2001}]{Vink:2001}
{Vink} J.~S.,  {de Koter} A.,   {Lamers} H.~J.~G.~L.~M.,  2001, \mn@doi [\aap]
  {10.1051/0004-6361:20010127}, \href
  {https://ui.adsabs.harvard.edu/abs/2001A&A...369..574V} {369, 574}

\bibitem[\protect\citeauthoryear{{Vink}, {Higgins}, {Sander}  \&
  {Sabhahit}}{{Vink} et~al.}{2021}]{Vink2021MNRAS}
{Vink} J.~S.,  {Higgins} E.~R.,  {Sander} A. A.~C.,   {Sabhahit} G.~N.,  2021,
  \mn@doi [\mnras] {10.1093/mnras/stab842}, \href
  {https://ui.adsabs.harvard.edu/abs/2021MNRAS.504..146V} {504, 146}

\bibitem[\protect\citeauthoryear{{Wang}, {Huang}  \& {Wang}}{{Wang}
  et~al.}{2019}]{Wang2019}
{Wang} K.,  {Huang} Z.-p.,   {Wang} J.-m.,  2019, \mn@doi [\caa]
  {10.1016/j.chinastron.2019.04.005}, \href
  {https://ui.adsabs.harvard.edu/abs/2019ChA&A..43..217W} {43, 217}

\bibitem[\protect\citeauthoryear{{Wang}, {Iwasawa}, {Nitadori}  \&
  {Makino}}{{Wang} et~al.}{2020}]{Wang2020MNRAS}
{Wang} L.,  {Iwasawa} M.,  {Nitadori} K.,   {Makino} J.,  2020, \mn@doi
  [\mnras] {10.1093/mnras/staa1915}, \href
  {https://ui.adsabs.harvard.edu/abs/2020MNRAS.497..536W} {497, 536}

\bibitem[\protect\citeauthoryear{{Wen}}{{Wen}}{2003}]{Wen2003}
{Wen} L.,  2003, \mn@doi [\apj] {10.1086/378794}, \href
  {https://ui.adsabs.harvard.edu/abs/2003ApJ...598..419W} {598, 419}

\bibitem[\protect\citeauthoryear{Woosley}{Woosley}{2017}]{Woosley:2016hmi}
Woosley S.~E.,  2017, \mn@doi [Astrophys. J.] {10.3847/1538-4357/836/2/244},
  836, 244

\bibitem[\protect\citeauthoryear{{Woosley} \& {Heger}}{{Woosley} \&
  {Heger}}{2021}]{Woosley2021}
{Woosley} S.~E.,  {Heger} A.,  2021, \mn@doi [\apjl]
  {10.3847/2041-8213/abf2c4}, \href
  {https://ui.adsabs.harvard.edu/abs/2021ApJ...912L..31W} {912, L31}

\bibitem[\protect\citeauthoryear{{Yuan} et~al.,}{{Yuan}
  et~al.}{2023}]{Yuan2023}
{Yuan} G.-W.,  et~al., 2023, \mn@doi [arXiv e-prints]
  {10.48550/arXiv.2303.09391}, \href
  {https://ui.adsabs.harvard.edu/abs/2023arXiv230309391Y} {p. arXiv:2303.09391}

\bibitem[\protect\citeauthoryear{{Yungelson}, {van den Heuvel}, {Vink},
  {Portegies Zwart}  \& {de Koter}}{{Yungelson} et~al.}{2008}]{Yungelson2008}
{Yungelson} L.~R.,  {van den Heuvel} E.~P.~J.,  {Vink} J.~S.,  {Portegies
  Zwart} S.~F.,   {de Koter} A.,  2008, \mn@doi [\aap]
  {10.1051/0004-6361:20078345}, \href
  {https://ui.adsabs.harvard.edu/abs/2008A&A...477..223Y} {477, 223}

\bibitem[\protect\citeauthoryear{{Zevin}, {Romero-Shaw}, {Kremer}, {Thrane}  \&
  {Lasky}}{{Zevin} et~al.}{2021}]{Zevin:2021}
{Zevin} M.,  {Romero-Shaw} I.~M.,  {Kremer} K.,  {Thrane} E.,   {Lasky} P.~D.,
  2021, \mn@doi [\apjl] {10.3847/2041-8213/ac32dc}, \href
  {https://ui.adsabs.harvard.edu/abs/2021ApJ...921L..43Z} {921, L43}

\makeatother
\end{thebibliography}







\bsp	
\label{lastpage}
\end{document}